\documentclass[twocolumn,a4paper,times,preprint2]{aastex631}
\usepackage{url}
\usepackage{epstopdf}
\usepackage{ulem}

\shorttitle{The $Insight$-$HXMT$ GPSS}
\shortauthors{Wang et al.}
\graphicspath{{./}{figures/}}

\begin{document}

\title{The Long-term Monitoring Results of \textit{Insight-HXMT} in the First 4 Yr Galactic Plane Scanning Survey}

\author[0000-0002-3206-5293]{Chen Wang}
\affiliation{Key Laboratory of Space Astronomy and Technology, National Astronomical Observatories, Chinese Academy of Sciences, Beijing 100101, People’s Republic of China}
\affiliation{University of Chinese Academy of Sciences, Beijing 100049, People’s Republic of China}

\author[0000-0001-8277-6133]{Jin-Yuan Liao}
\affiliation{Key Laboratory of Particle Astrophysics, Institute of High Energy Physics, Chinese Academy of Sciences, 19B Yuquan Road, Beijing 100049, People’s Republic of China}

\author{Ju Guan}
\affiliation{Key Laboratory of Particle Astrophysics, Institute of High Energy Physics, Chinese Academy of Sciences, 19B Yuquan Road, Beijing 100049, People’s Republic of China}

\author{Yuan Liu}
\affiliation{Key Laboratory of Space Astronomy and Technology, National Astronomical Observatories, Chinese Academy of Sciences, Beijing 100101, People’s Republic of China}

\author[0000-0001-5798-4491]{Cheng-Kui Li}
\affiliation{Key Laboratory of Particle Astrophysics, Institute of High Energy Physics, Chinese Academy of Sciences, 19B Yuquan Road, Beijing 100049, People’s Republic of China}

\author[0000-0001-8378-5904]{Na Sai}
\affiliation{Department of Astronomy, School of Physics and Technology, Wuhan University, Wuhan 430072, China}
\affiliation{WHU-NAOC Joint Center for Astronomy, Wuhan University, Wuhan 430072, China}

\author{Qi Luo}
\author[0000-0002-7085-6328]{Jing Jin}
\affiliation{Key Laboratory of Particle Astrophysics, Institute of High Energy Physics, Chinese Academy of Sciences, 19B Yuquan Road, Beijing 100049, People’s Republic of China}

\author{Yi Nang}
\affiliation{Key Laboratory of Particle Astrophysics, Institute of High Energy Physics, Chinese Academy of Sciences, 19B Yuquan Road, Beijing 100049, People’s Republic of China}

\author[0000-0001-5586-1017]{Shuang-Nan Zhang}
\affiliation{Key Laboratory of Particle Astrophysics, Institute of High Energy Physics, Chinese Academy of Sciences, 19B Yuquan Road, Beijing 100049, People’s Republic of China}
\affiliation{University of Chinese Academy of Sciences, Beijing 100049, People’s Republic of China}
\affiliation{Key Laboratory of Space Astronomy and Technology, National Astronomical Observatories, Chinese Academy of Sciences, Beijing 100101, People’s Republic of China}

\correspondingauthor{Chen Wang}
\email{cwang@bao.ac.cn}

\correspondingauthor{Jin-Yuan Liao}
\email{liaojinyuan@ihep.ac.cn}



\begin{abstract}
    The first X-ray source catalog of \textit{Insight-HXMT} Galactic Plane ($|b|<10^{\circ}$) Scanning Survey (GPSS) is presented based on the data accumulated from June 2017 to August 2021. The 4 yr limit sensitivities at main energy bands can reach $\mathrm{8.2 \times 10^{-12}\,erg\,s^{-1}\,cm^{2}}$ (2$-$6\,keV), $\mathrm{4.21 \times 10^{-11}\,erg\,s^{-1}\,cm^{2}}$ (7$-$40\,keV) and $\mathrm{2.78 \times 10^{-11}\,erg\,s^{-1}\,cm^{2}}$ (25$-$100\,keV). More than 1300 sources have been monitored at a wide band (1$-$100\,keV), of which 223 sources have a signal-to-noise ratio greater than 5. We combined the GPSS data of \textit{Insight-HXMT} and MAXI and found it is feasible to obtain more complete long-term light curves from their scanning results. The flux variabilities at different energy bands of the 223 bright sources are analyzed based on the excess variances. It is found that the fluxes of X-ray binaries are more active than those of supernova remnants and isolated pulsars. Different types of binaries, e.g., low-mass X-ray binaries (LMXBs), high-mass X-ray binaries (HMXBs), neutron star binaries, and black hole binaries, also distinctively show different regularities. In addition, the relations between the hardness ratio (HR) and excess variances, and HR and source types are analyzed. It is obvious that the HRs of HMXBs tend to be harder than those of LMXBs and HMXBs tend to be more active than those of LMXBs.
\end{abstract}


\section{Introduction} \label{sec:intro}
There are various X-ray sources in the universe, such as stars, X-ray binaries, gamma-ray bursts, active galactic nuclei, galaxy clusters, etc. They exhibit not only spatial distribution characteristics from inside to outside the Galaxy but also morphological characteristics from point to diffuse sources of various scales. Observing and studying these sources in the X-ray band is of great significance for the development of astronomy. For example, understanding the formation and evolutionary history of stars, galaxies, and the universe, as well as the accretion and radiation processes of compact objects, etc. 

As the influence of the absorption effect of the Earth's atmosphere, astronomical observations in the X-ray band must be carried out in space observations. A variety of X-ray astronomical satellites have been launched for decades, and they have advanced astronomy by conducting large numbers of observations of sources in different energy bands from different regions to different depths of the sky. The first all-sky X-ray survey was performed by Uhuru, which was launched on Dec. 12th, 1970 \citep{1971ApJ...165L..27G}. It found 339 X-ray sources at 2$-$6\,keV band \citep{1978ApJS...38..357F}, and identified radiation resulted from accretion process of compact objects \citep{1973ApJ...182L.103F}. The first Wolter-I type imaging telescope HEAO$-$2 was launched in 1978, which has a higher sensitivity than any previous telescopes \citep{1979ApJ...230..540G}. It demonstrated that X-ray emission could be produced in all types of sources, including stars \citep{1990ixra.conf...61V} and supernova remnants \citep[SNRs,][]{1982AdSpR...2i.153S}, and also took our study of X-ray astronomy from the Galaxy to nearby galaxies \citep{1979ApJ...234L..45V,1981ApJ...248..925L}. ROSAT is an all-sky survey imaging telescope in the soft X-ray band (0.1$-$2.4\,keV), which has further increased the sensitivity with lower instrument noise and better background shielding technology \citep{1982AdSpR...2d.241T}. A total of 145,060 sources had been found during this survey \citep{1999A&A...349..389V}, increasing the number of known sources by two orders of magnitude. MAXI was launched in 2009, which is an all-sky monitor onboard the International Space Station. It covers the energy band 2$-$30\,keV with high survey efficiency of one all-sky scan per 92 minutes \citep{2009PASJ...61..999M,2018ApJS..235....7H}. A large quantity of scientific data has been accumulated during its orbiting period, with monitoring of long-term variabilities of X-ray sources, and discovering over 30 new sources. There are also many other satellites, which have their own advantages and complement with each other. For example, the coded-mask telescope INTEGRAL/IBIS has been observing the sky in hard X-ray bands (above 20 keV) since 2002 \citep{2021arXiv211102996K}, while Swift has the capability of rapid positioning to newly discovered GRBs and making multiwavelength observations \citep{2005SSRv..120..165B}. The SRG/ART-XC observes the sky with subarcminute angular resolution and excellent sensitivity above 2\,keV. It has detected 867 sources at 4$-$12\,keV in its first-year all-sky survey and expects to discover a significant number of new X-ray objects \citep{2022AA...661A..38P}. These satellites have discovered new X-ray sources or published catalogs during their operation, which have contributed to the development of astronomy. Some large-area surveys (e.g., all-sky or Galactic Plane or serendipitous surveys with sky-coverage larger than 3000\,deg$^2$) are listed in Table \ref{tab:satellites}.

\begin{figure*}
    \gridline{\fig{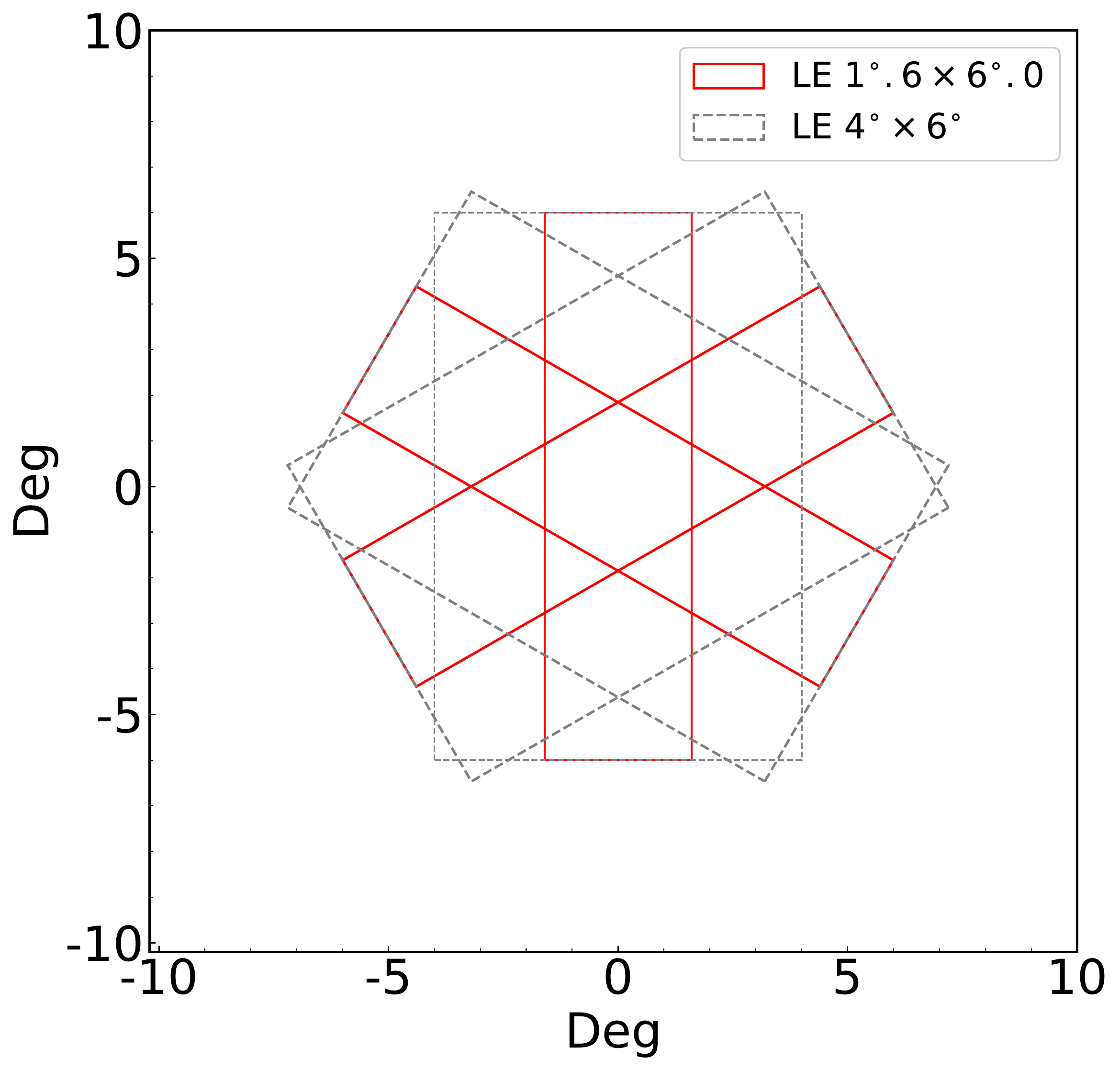}{0.3\textwidth}{(a) LE}
             \fig{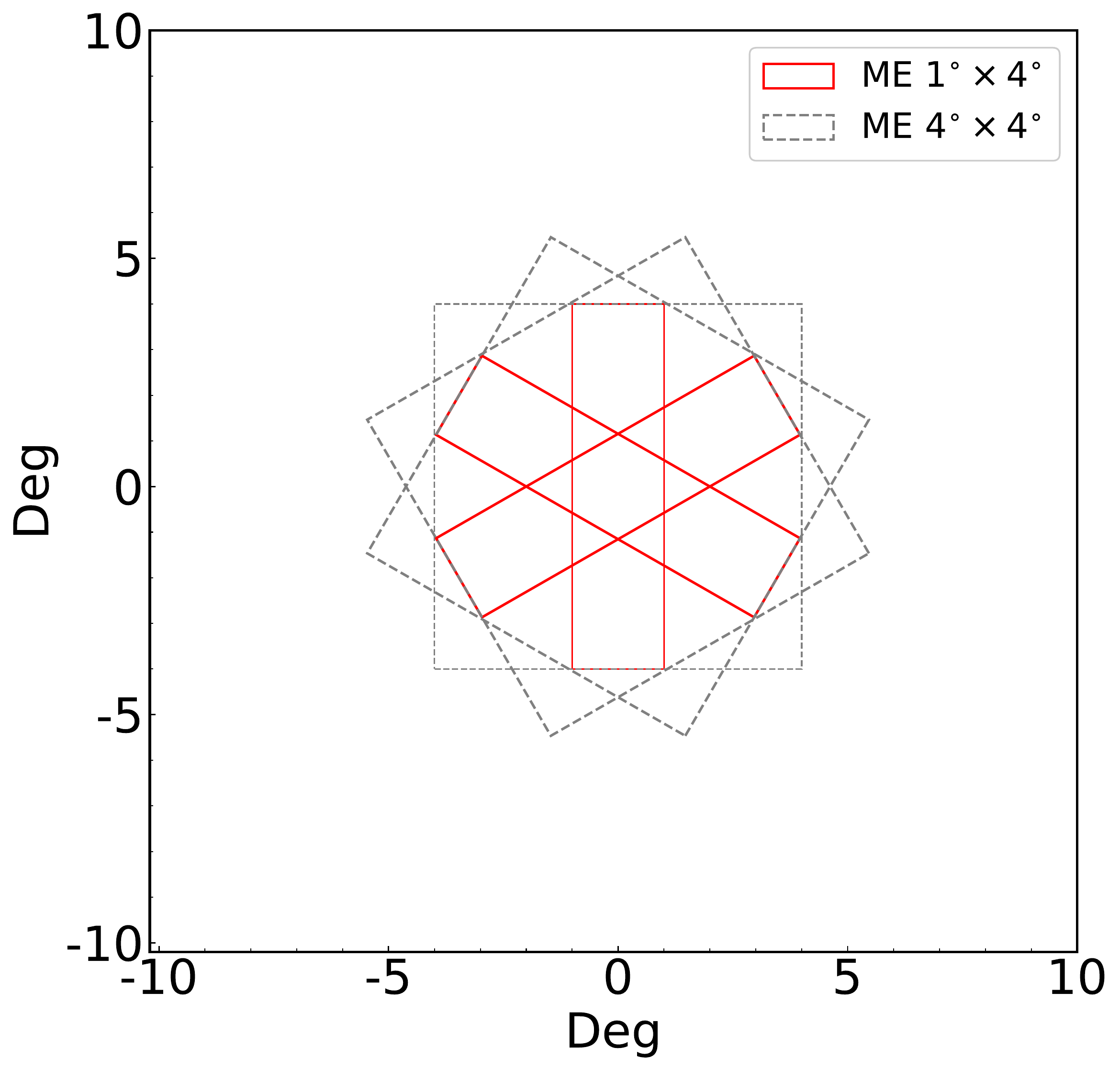}{0.3\textwidth}{(b) ME}
             \fig{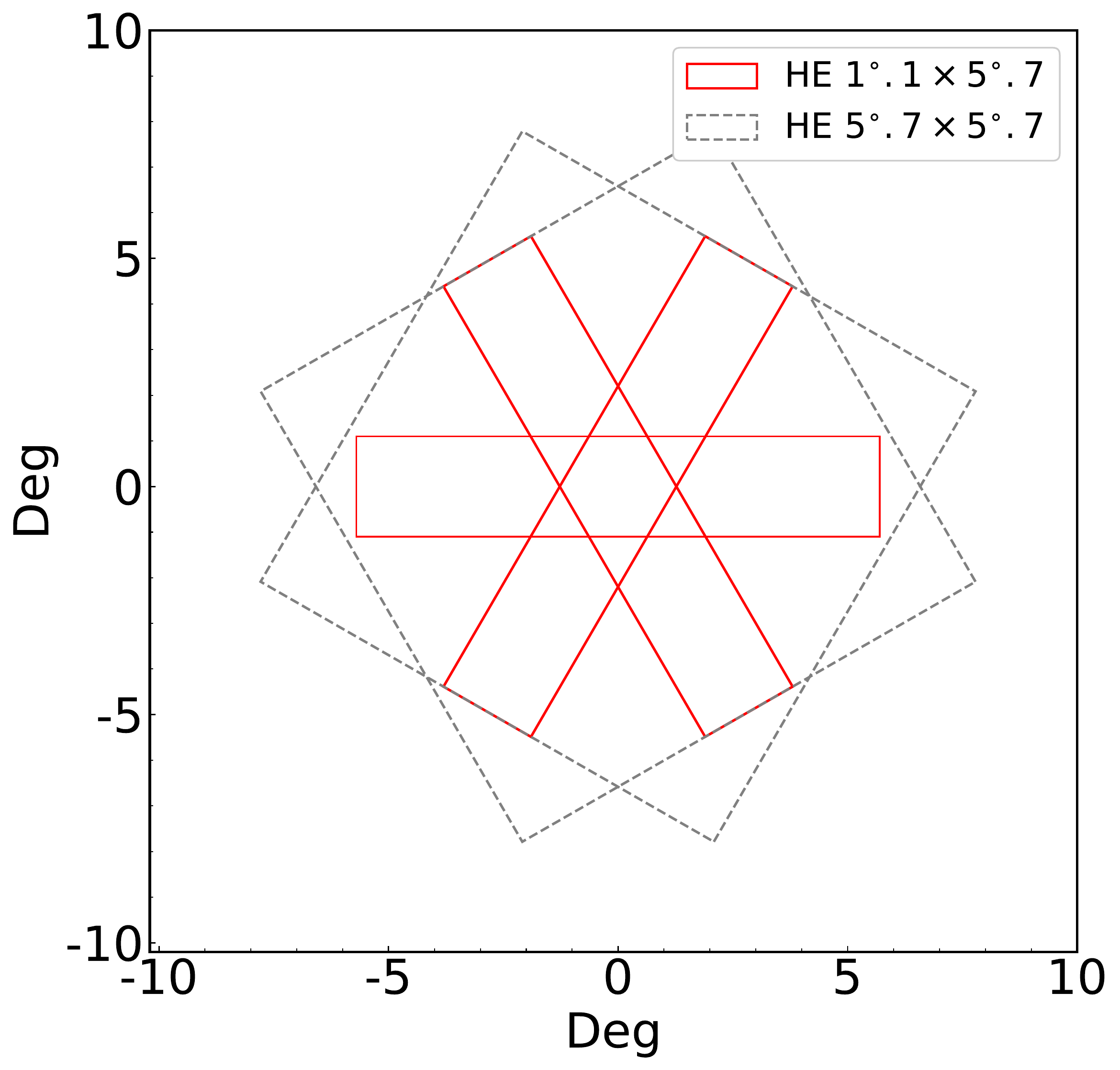}{0.3\textwidth}{(c) HE}
    }   
    \caption{The FOVs of LE, ME and HE, respectivvely.}
    \label{fig:fov}
\end{figure*}

The \textit{Hard X-ray Modulation Telescope} (dubbed as \textit{Insight-HXMT}) is an X-ray astronomy satellite covering a wide energy band (1$-$250\,keV) with large Fields of Views \citep[FOVs,][]{2020SCPMA..6349502Z}. It was launched on June 15th, 2017, with three main scientific payloads: the High Energy X-ray Telescope (HE), the Medium Energy X-ray Telescope (ME), and the Low Energy X-ray Telescope (LE). Details of the three payloads are listed in Table \ref{tab:payloads}. The three telescopes are composed of 18 NaI(Tl)/CsI(Na) crystal detector units \citep[HE,][]{2020SCPMA..6349503L}, 1728 pixels of Si-PIN detectors \citep[ME,][]{2019arXiv191004451C}, and Swept Charge Device (SCD) sensor arrays \citep[LE,][]{2020SCPMA..6349505C}, respectively. They are collimating telescopes, and each contains three detector groups with an intersecting orientation angle of $60^{\circ}$. The FOVs during scanning observations are $5^{\circ}.7 \times 1^{\circ}.1$ (HE), $4^{\circ} \times 1^{\circ}$ (ME), and $1^{\circ}.6 \times 6^{\circ}$ (LE), respectively. The FOVs of the three telescopes are shown in Figure \ref{fig:fov}.

\begin{figure}
    \centering
    \includegraphics[scale=0.45]{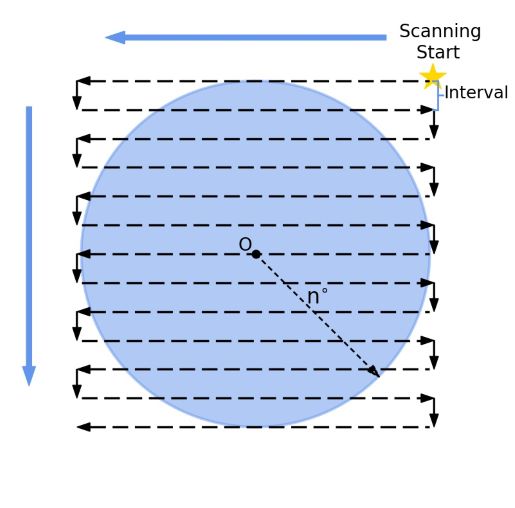}
    \caption{Schematic diagram of the scanning method of a small sky area. The circle is the observed area, and `O' is the center of the circle. The arrow points to the scanning direction.}
    \label{fig:scanning-schematic}
\end{figure}

\begin{figure*}
    \gridline{\fig{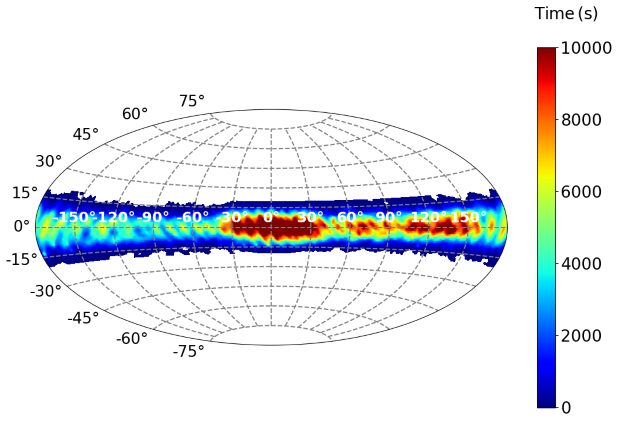}{0.49\textwidth}{LE}\label{fig:expoure_LE}
              \fig{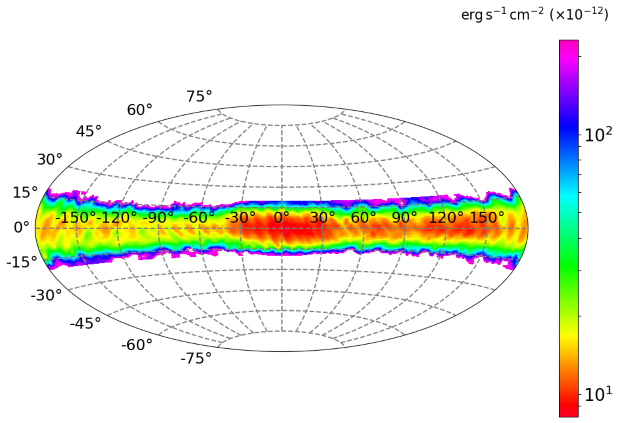}{0.49\textwidth}{LE (2$-$6\,keV)}}
    \gridline{\fig{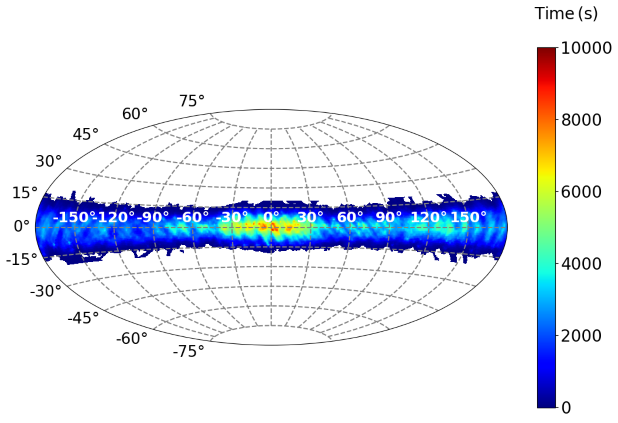}{0.49\textwidth}{ME}\label{fig:expoure_LE}
              \fig{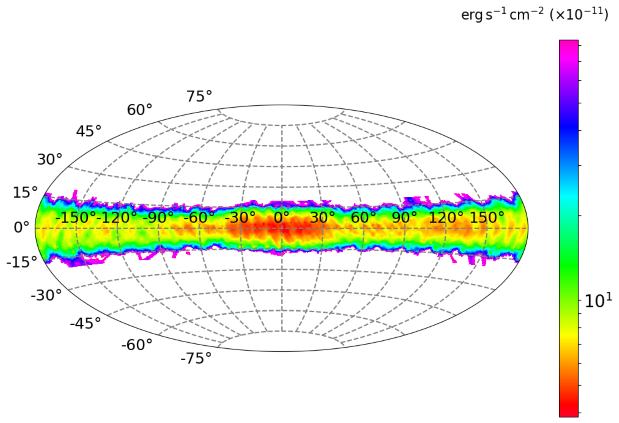}{0.49\textwidth}{ME (7$-$40\,keV)}}
    \gridline{\fig{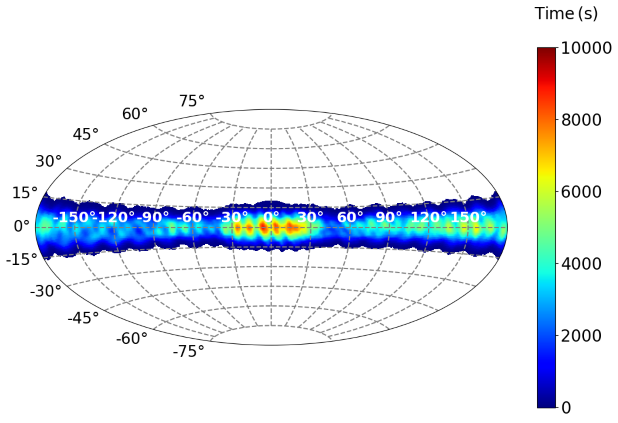}{0.49\textwidth}{HE}\label{fig:expoure_LE}
              \fig{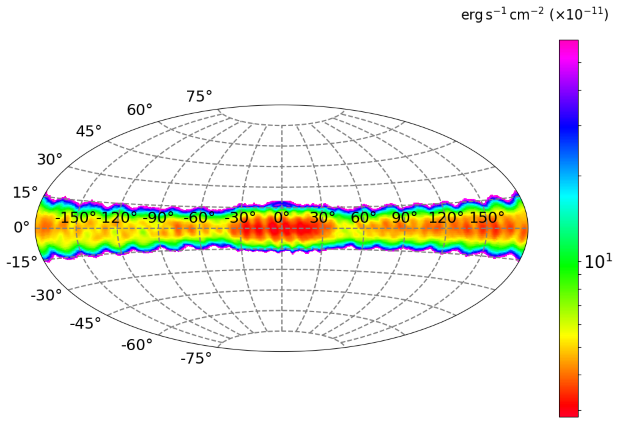}{0.49\textwidth}{HE (25$-$100\,keV)}}
    \caption{The exposure and sensitivity maps in the Galactic Plane for \textit{Insight-HXMT} are shown in the left and right columns, respectively. In the left column, the colors correspond to the exposure time from 2017-06-29 to 2021-08-27. The difference in exposures between the three telescopes comes mainly from the difference in Good Time Intervals and FOV sizes. In addition, HE has less GPSS data than LE and ME because it spends some observation time on gamma-ray burst monitoring.}
    \label{fig:expourse}
\end{figure*}

\textit{Insight-HXMT} provides us with scientific data to statistically study the timescales, duty cycles and spatial distribution of different types of sources, etc. \textit{Insight-HXMT} has three types of observations: pointing observation of X-ray sources, monitoring GRBs, and Galactic Plane Scanning Survey (GPSS). It spends about one-third of observation time on scanning the Galactic plane to monitor known sources and find new X-ray transients. We adopt a small-sky-area scanning strategy in GPSS, which divides the entire Galactic plane ($0^{\circ}<l<360^{\circ}, -10^{\circ}<b<10^{\circ}$) into several small regions, and performs a line-by-line scanning of each region. The schematic diagram of a small-sky-area scanning is shown in Figure \ref{fig:scanning-schematic}.
The early phase strategy, before April 2019, was designed with a scanning speed of $0^{\circ}.06\,\mathrm{s^{-1}}$ and a interval of $0^{\circ}.8$. It divided the Galactic Plane into 22 small regions, each of which has a radius of 10$^{\circ}$. For finer scanning, the current strategy has 50 small regions with a radius of $7^{\circ}$, a speed of $0^{\circ}.06\,\mathrm{s^{-1}}$ and a interval of $0^{\circ}.4$.
Thanks to the design of large effective area and wide-band coverage, \textit{Insight-HXMT} not only has a high scanning efficiency ($14^{\circ} \times 14^{\circ}$ coverage every 2.3 hours) but also can monitor the radiation intensity of X-ray sources from soft to hard X-ray energy bands at the same time. 
Meanwhile, the large amount of observation time ensures intensive monitoring of X-ray sources within its visible sky areas. 
In addition, with the background shielding technology, LE can reach 5$\sigma$ sensitivity of $\sim 5.7 \times 10^{-11} \mathrm{erg\,s^{-1}\,cm^{-2}}$ (1$-$6\,keV, 2.5\,mCrab) for each scanning observation ($\sim$2.3 h). 
Up to August 2021, \textit{Insight-HXMT} has conducted over 2000 scanning observations, covering the entire Galactic plane. Exposure maps for the three telescopes are shown in the left column of Figure \ref{fig:expourse}. Sensitivity maps for three main energybands are shown in the right column, with the best sensitivities up to $8.2 \times 10^{-12} \mathrm{erg\,s^{-1}\,cm^{-2}}$ (2$-$6\,keV, 0.61\,mCrab), $4.2 \times 10^{-11} \mathrm{erg\,s^{-1}\,cm^{-2}}$ (7$-$40\,keV, 2.35\,mCrab), and $2.8 \times 10^{-11} \mathrm{erg\,s^{-1}\,cm^{-2}}$ (25$-$100\,keV, 2.2\,mCrab), which are listed in Table \ref{tab:Information}. 
Long-term light curves for 1345, 957, and 935 X-ray sources are given by LE, ME, and HE, respectively.

In this paper, we report the GPSS results of \textit{Insight-HXMT}, using the 4 yr data from June 2017 to August 2021. The data reduction and light-curve fitting methods are described in Section \ref{sec:data_light}. The GPSS results and error analyses are presented in Section \ref{sec:GPSSresults}. We combine the long-term light curves from \textit{Insight-HXMT} monitored at 2$-$4\,keV with those from  MAXI in Section \ref{sec:compMAXI} and perform a series of statistical analysis on the bright sources in Section \ref{sec:Analysis}. The conclusion and summary are presented in Section \ref{sec:conclusion}.

\begin{deluxetable*}{LRRRRRRR}
    \setcounter{table}{0}
    \tablecaption{Information of some X-ray missions. \label{tab:satellites}}
    \tablewidth{0pt}
    \tablehead{
        \colhead{Mission$^{a}$}    &
        \colhead{Launch Date}      &
        \colhead{Catalog}          &
        \colhead{Energy Band}&
        \colhead{Scan Region}      &
        \colhead{Sky Coverage}     &
        \colhead{Release Date}     &
        \colhead{Reference}
        \\
        \colhead{} &
        \colhead{} &
        \colhead{} &
        \colhead{(keV)} &
        \colhead{} &
        \colhead{(deg$^2$)} &
        \colhead{} &
        \colhead{}
                }
    \startdata
        \mathrm{Uhuru}       & \mathrm{Dec.\,12th,\,1970}     & 339     & 2$-$6       & \mathrm{All-sky}              &        & \mathrm{1978}  & \text{\cite{1978ApJS...38..357F}} \\ 
        \mathrm{Ariel\,V}    & \mathrm{Oct.\,15th,\,1974}     & 109     & 2$-$10      & $|b|<10^{\circ}$              &        & \mathrm{1981}  & \text{\cite{1981MNRAS.197..865W}} \\ 
        \mathrm{HEAO-1}      & \mathrm{Aug.\,12th,\, 1977}    & 842     & 0.5$-$25    & \mathrm{All-sky}              &        & \mathrm{1983}  & \text{\cite{1984ApJS...56..507W}}\\ 
        \mathrm{EXOSAT}      & \mathrm{May.\,26th,\,1983}     & 1210    & 1$-$8       & \mathrm{All-sky}              &        & \mathrm{1998}  & \text{\cite{1999AAS..134..287R}} \\ 
        \mathrm{ROSAT}       & \mathrm{Jun.\,1th,\,1990}      & 135,000 & 0.1$-$2.4   & \mathrm{All-sky}              &        & \mathrm{2016}  & \text{\cite{2016AA...588A.103B}} \\ 
        \mathrm{RXTE}        & \mathrm{Dec.\,30th,\, 1995}    & 294     & 3$-$20      & \mathrm{serendipitous$^{b}$}  & 34,090 & \mathrm{2004}  & \text{\cite{2004AA...418..927R}} \\
        \mathrm{XMM-Newton}  & \mathrm{Dec.\,10th,\,1999}     & 72,352  & 0.2$-$12    & \mathrm{serendipitous}        & 65,000 & \mathrm{2017}  & \text{\cite{XMMSL2}}\\
        INTEGRAL             & \mathrm{Oct.\,17th,\,2002}     & 929     & 17$-$60     & \mathrm{All-sky}              &        & \mathrm{2021}  & \text{\cite{2021arXiv211102996K}} \\
        Swift                & \mathrm{Nov.\,20th,\,2004}     & 206,335 & 0.3$-$10    & \mathrm{serendipitous}        & 3,790  & \mathrm{2020}  & \text{\cite{2020ApJS..247...54E}}\\
        \mathrm{MAXI}        & \mathrm{Aug.\,15th,\,2009}     & 221     & 4$-$10      & $|b|^{c}<10^{\circ}$          &        & \mathrm{2018}  & \text{\cite{2018ApJS..235....7H}} \\         
        \mathrm{SRG/ART-XC}  & \mathrm{Jul.\,19th,\,2019}     & 867     & 4$-$12      & \mathrm{All-sky}              &        & \mathrm{2022}  & \text{\cite{2022AA...661A..38P}}\\
    \enddata
\tablecomments{\\$^{a}$ Here in the table we briefly list one of the catalogs for each mission, even though some missions have released several catalogs at various regions, different energy bands, or observation times. For example, HEAO-1 had provided catalogs at different energy bands, such as 842 sources at 0.5$-$25\,keV \citep{1984ApJS...56..507W}, 114 sources at 0.18$-$0.44\,keV \citep{1983ApJS...51....1N}, 44 sources at 40$-$80\,keV, and 14 sources at 80$-$180\,keV \citep{1984ApJS...54..581L}.\\
$^{b}$ The survey is based on slew observations or a compilation of pointed observations.\\
$^{c}$ Outside the Galactic center region ($|b|<5^{\circ}$, $l<30^{\circ}$, and $l>330^{\circ}$).\\
}
\end{deluxetable*}

\begin{deluxetable}{LLLL}
    \setcounter{table}{1}
    \tablecaption{Payloads of \textit{Insight-HXMT}. \label{tab:payloads}}
    \tablewidth{0pt}
    \tablehead{
        \colhead{ }            &
        \colhead{HE} &
        \colhead{ME} &
        \colhead{LE} 
        }
    \startdata
        \mathrm{Geometrical~Area~(cm$^{2}$)} & 5100 & 952 & 384\\
        \mathrm{Small~ FOV~(FWHM)} & $1^{\circ}.1 \times 5^{\circ}.7$ & $1^{\circ} \times 4^{\circ}$ & $1^{\circ}.6 \times 6^{\circ}$\\
        \mathrm{Large~ FOV~(FWHM)} & $5^{\circ}.7 \times 5^{\circ}.7$ & $4^{\circ} \times 4^{\circ}$ & $4^{\circ} \times 6^{\circ}$\\
        \mathrm{Energy~Band1$^{a}$~ (keV)}    &  20$-$250    & 5$-$40    & 0.7$-$13  \\
        \mathrm{Energy~Band$2^{b}$~ (keV)}    &  25$-$100    & 7$-$40    & 1$-$7$^{c}$\\
    \enddata
\tablecomments{\\
               $\mathrm{^a}$ The energy bands that three instruments covered.\\
               $\mathrm{^b}$ The energy bands that used in \textit{Insight-HXMT} GPSS.\\
               $\mathrm{^c}$ This energy band is sub-divided into 1$-$2\,keV, 2$-$6\,keV, 2$-$4\,keV, 4$-$6\,keV, 3$-$5\,keV, 5$-$7\,keV and 1$-6$\,keV.}
\end{deluxetable}

\section{Data Reduction And Light Curve Fitting} \label{sec:data_light}
\subsection{Data Reduction} \label{sec:data}
\begin{figure*}
    \centering
    \includegraphics[scale=0.45]{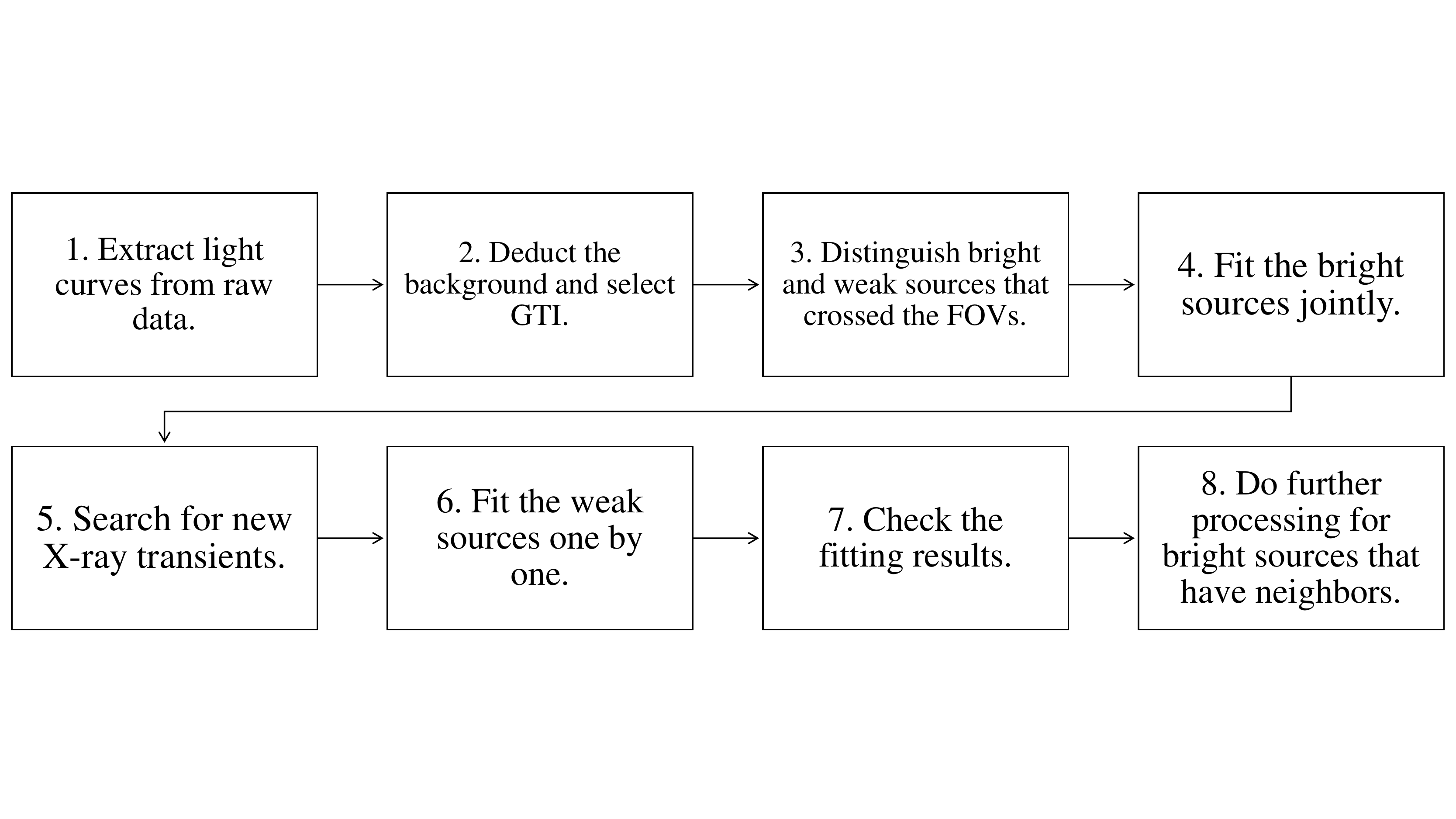}
    \caption{GPSS data analysis process.}
    \label{fig:flowchart}
\end{figure*}

We have developed a set of pipelines to reduce the GPSS data. The general flowchart is shown in Figure \ref{fig:flowchart}.
Step 1 is the extraction process of light curves from the raw data with \textit{Insight-HMXT} data analysis software HXMTDAS v2.0 \citep{HXMTusrManual}. In the next step, we estimate the background and further select Good Time Interval (GTI). GTI is generally related to the orbital environment of the satellite and can be selected by the data analysis software. However, sometimes the LE data require extra selection, because the data at 1$-$6\,keV band are vulnerable to contamination of particle events. The details are described in \cite{2020JHEAp..26....1S}. The backgrounds of HE and ME are characterized by the fluctuations that come from the modulation of geomagnetic field \citep{2009ChA&A..33..333L, 2015Ap&SS.360...47X}, and a polynomial fit is sufficient enough to estimate them \citep{2020JHEAp..25...39N}. The background of LE is generally weak and of low count rates during GTI. However, it often rises to a dramatically high level when particle events accumulate on the detectors or \textit{Insight-HXMT} passes through the South Atlantic Anomaly (SAA) region \citep{2020JHEAp..27...24L}. Consequently, it can be reliably estimated by the Statistics-sensitive Nonlinear Iterative Peak (SNIP) method \citep{1997NIMPA.401..113M, 1988NIMPB..34..396R}. 

To obtain high-quality scientific data to the maximum extent, the mean GTI coverage rates in each scan area of HE, ME, and LE are chosen to be $39.1\,\%$, $31.5\,\%$, and $27.7\,\%$, respectively. Each telescope has covered the entire Galactic plane after completion of the first-year observation. They are all collimating telescopes and hence only the X-ray sources that cross the FOVs can be recorded. That the detection efficiency increases with decreasing orientation angle of the collimator induces a triangular peak on the light curve as the FOVs sweep through a source. The peak will change when the telescope sweeps different areas on account of the various brightness of X-ray sources. This can be analogized as Point-Spread Function (PSF). The source position and flux can be reconstructed by fitting the corresponding light curves via PSF models. During the PSF fitting, the flux is assumed as a constant and hence the time-averaged value is obtained for each observation. It is accurate in most cases because an exposure for one target generally lasts within a few hundred seconds per observation, during which the flux variabilities of most sources are insignificant. The details of PSF fitting are given in \cite{2020JHEAp..25...39N}. To improve the positioning accuracy, we generate light curves according to the three collimator orientations in HE, ME, and LE, respectively.

\subsection{Light Curve Fitting} \label{sec:light}
To ensure the completeness of the X-ray sources in GPSS, the source catalog we adopted is a merge of catalogs from the websites of $Swift$ \footnote{\url{https://swift.gsfc.nasa.gov/results/transients/}}, $INTEGRAL$ \footnote{\url{http://isdc.unige.ch/integral/catalog/43/gnrl_refr_cat_0043.html}} and MAXI \footnote{\url{http://134.160.243.88/top/slist.html}}. It contains 2881 sources in total, whereas some sources are located too close to be distinguished because of the limited angular resolution of \textit{Insight-HXMT}. To improve the fitting efficiency, the sources that locate within $0^{\circ}.6$, $0^{\circ}.6$, and $0^{\circ}.1$ in HE, ME, and LE, are merged in our catalog, respectively. After the scanning data are screened, we fit the corresponding light curves with steps 3$-$7 illustrated in Figure \ref{fig:flowchart}. The details are described in \cite{2020JHEAp..26....1S}, and we outline the processes as follows:
\begin{itemize}
    \item[1.] Extract the known sources covered by the scan area from our catalog and find which of them are crossed by the FOVs via calculating their effective areas.
    \item[2.] Fix the positions of the sources obtained in the previous step one after another and use the corresponding PSF models to estimate their fluxes. 
    \item[3.] Distinguish bright and faint sources based on the fluxes and signal-to-noise ratios (S/N). The source with flux higher than the detection limit and S/N $>$ 2 is defined as a bright source. Otherwise, it is a faint source, and the S/N is defined as
        \begin{equation}
            S/N = \frac{F}{\sigma} \label{equ:snr1},    
        \end{equation}
    where $F$ is the best-fitted flux, and $\sigma$ the statistical uncertainty. In other words, bright sources mainly contribute to signal peaks in the light curves, and faint ones have no obvious contribution.
    \item[4.] Freeze all positions and then fit the fluxes of bright sources simultaneously. The fluxes with the optimal chi-squared values are considered to be the final results. Hereafter, the jointly fitting method and the corresponding results are called as method 1 and results 1, respectively.
    \item[5.] Search for the location and flux of a new source candidate based on the residuals of the previous step.
    \item[6.] Fit the faint sources with a single PSF model based on the residuals after fitting bright sources and searching for a new candidate. To improve the efficiency and accuracy, fitting the PSF models of the faint sources is put as the final step.
    \item[7.] Check the fitting results. First, each bright source is determined whether it is in quiescence. Second, the information of a candidate is checked. If there is a candidate with S/N $>$ 3 and the flux is higher than the sensitivity, a visual inspection will be carefully performed to determine the credibility.  
\end{itemize}

\begin{figure*}
    \gridline{\fig{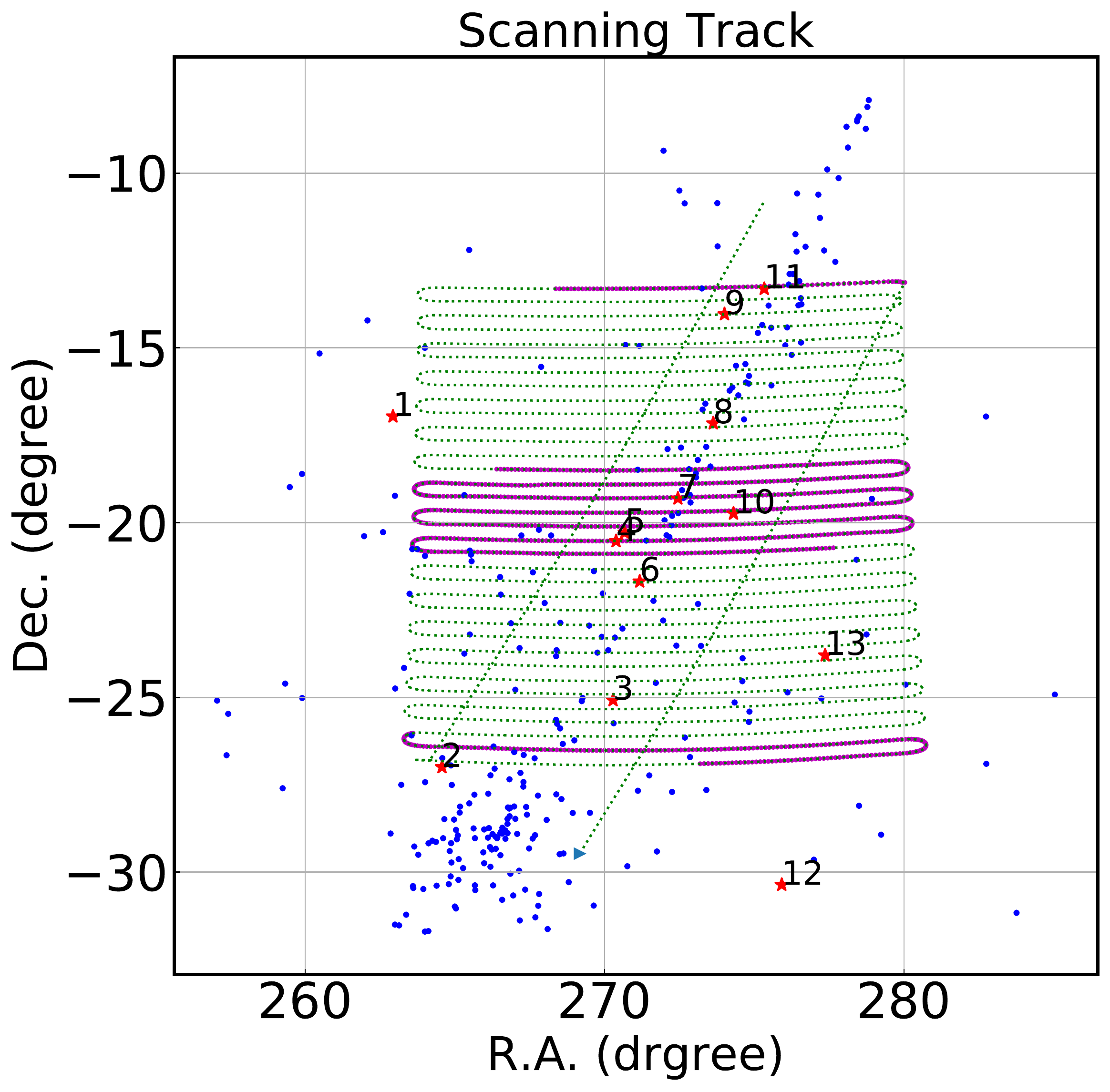}{0.43\textwidth}{(a)}\label{fig:scan}
              \fig{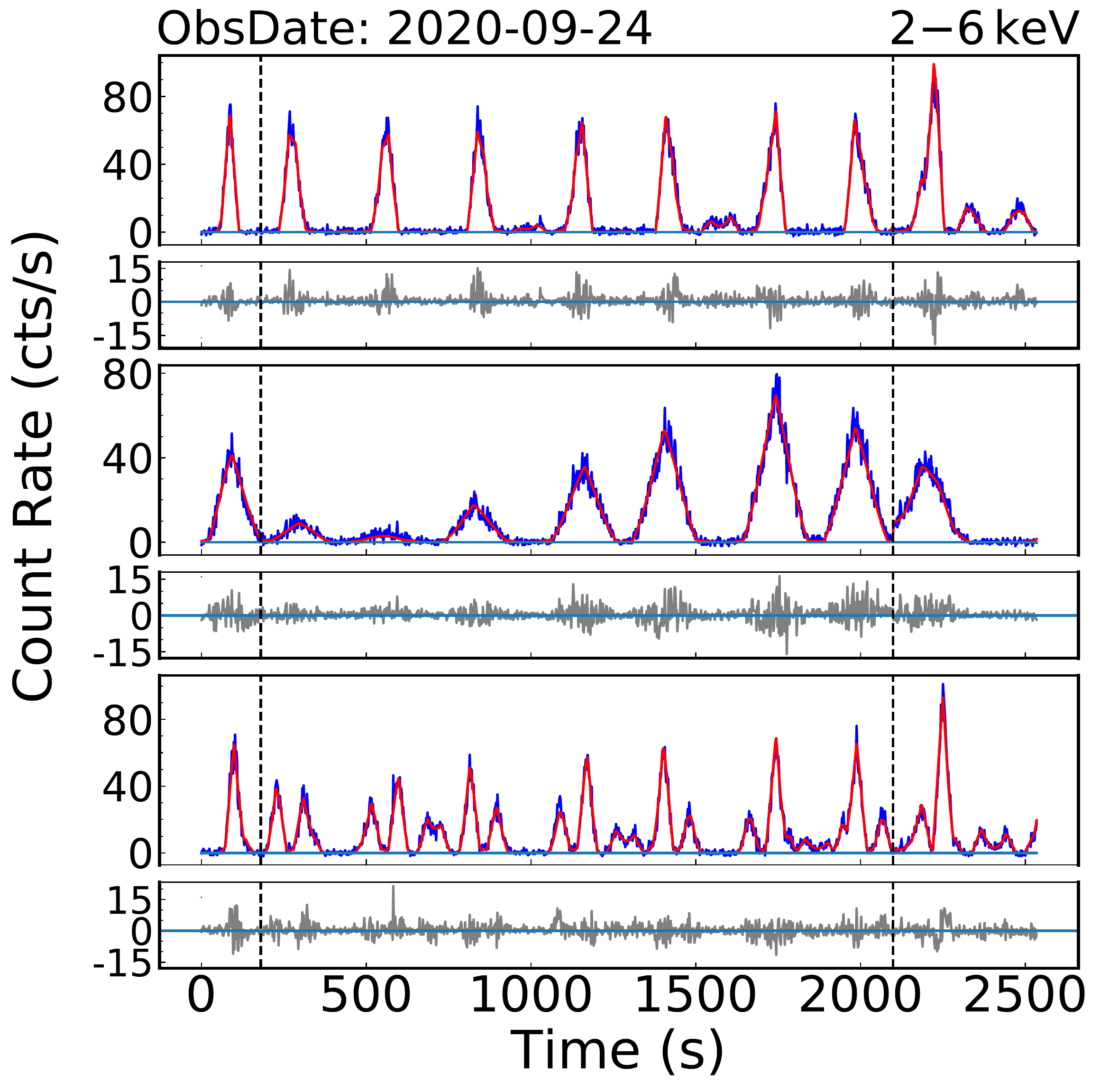}{0.43\textwidth}{(b)}\label{fig:lightcurve}
             }
    \caption{Panel (a) shows the corresponding scanned area: The green and purple dashed lines are scanning tracks in the whole observation and those with GTI, respectively. The blue points and red stars denote faint and bright sources in this area. Panel (b) shows an example of light curves at 2$-$6\,keV obtained with LE: The black dashed and red solid lines are segmentations of GTI and the best-fit PSF models of bright sources in the three groups, respectively. The blue and grey solid lines represent clean light curves in this area and residuals, respectively.}
    \label{fig:example}
\end{figure*}

Figure \ref{fig:example} displays an example of the final light curve fitting result of LE (2$-$6\,keV) based on method 1. The left panel is the corresponding scanning tracks. Thirteen bright sources are monitored in this observation, and their information is listed in Table \ref{tab:Result}. In the right panel, the blue lines represent light curves after the background is subtracted and GTI is selected. The red solid lines refer to the final fitting results of the bright sources in the three groups, and the grey lines are the corresponding residuals, respectively. Statistically, our fitting result is acceptable.

\begin{deluxetable*}{LLLLRR}
    \setcounter{table}{2}
    \tablecaption{Information of the bright sources in the scanning area shown in Fig \ref{fig:example}. \label{tab:Result}}
    \tablewidth{0pt}
    \tablehead{
        \colhead{Number} &
        \colhead{Source Name} &
        \colhead{R.A.} &
        \colhead{Dec.} &
        \colhead{Rate} &
        \colhead{S/N}
        \\
        \colhead{} &
        \colhead{} &
        \colhead{(deg)} &
        \colhead{(deg)} &
        \colhead{($\mathrm{cts\,s^{-1}}$)} &
        \colhead{}
                }
\decimalcolnumbers
    \startdata
        1	& \mathrm{3A\,1728$-$169}	    &   262.93	& -16.96	&   30.3	&   23.9  \\  
        2	& \mathrm{SLX\,1735$-$269}	    &   264.57	& -26.99	&   2.7	    &   5.1   \\  
        3	& \mathrm{GX\,5$-$1}	        &   270.28	& -25.08	&   127.2	&   56.3  \\     
        4	& \mathrm{GX\,9+1}	            &   270.38	& -20.53	&   58.1	&   42.1  \\  
        5	& \mathrm{IGR\,J18027$-$2016}	&   270.67	& -20.29	&   8.8	    &   6.3   \\  
        6	& \mathrm{HESS\,J1804$-$216}	&   271.17	& -21.68	&   5.1	    &   4.8   \\  
        7	& \mathrm{AX\,J1809.8$-$1918}	&   272.45	& -19.31	&   1.4	    &   4.1   \\  
        8	& \mathrm{GX\,13+1}	            &   273.63	& -17.16	&   41.0	&   14.8  \\      
        9	& \mathrm{GX\,17+2}	            &   274.01	& -14.04	&   74.1	&   51.5  \\      
        10	& \mathrm{IGR\,J18172$-$1944}	&   274.31	& -19.74	&   1.5	    &   6.5   \\  
        11	& \mathrm{IGR\,J18214$-$1318}	&   275.33	& -13.31	&   3.1	    &   4.8   \\  
        12	& \mathrm{H\,1820$-$303}	    &   275.92	& -30.36	&   41.2	&   30.9  \\      
        13	& \mathrm{Ginga\,1826$-$24}	    &   277.37	& -23.80	&   30.8	&   28.9  \\      
    \enddata
\end{deluxetable*}

For most cases, flux information can be obtained unambiguously through method 1. However the X-ray sources are extremely crowded in some regions (such as the galactic center). For these sources, it is sometimes difficult to determine whether the peaks on the light curves are contributed by one or several neighboring sources. Method 1 jointly fits all possible contributing sources simultaneously and acquires the final results based on the returned optimal chi-squared values, which may cause the fluxes of the neighboring sources to interfere with each other. The distance between the sources is the primary interference factor. In addition, the satellite scanning tracks, the GTI coverage, and the fluxes of the target sources during each scanning observation can cause various interference situations. Therefore, further processing is required. The following steps are taken if some neighbors may influence with each other (The target source is denoted as `A'.):
\begin{itemize}
    \item Figure out the contaminating neighbors of `A' based on the integral overlap rates of PSF models. To accomplish this, we assume that the complicated case can be simplified by comparing the overlap of PSF models of two sources in turn, which is defined as:
    \begin{equation}
        R_{\mathrm{AB}} = \frac{\int {M_{\mathrm{A}}(t) \bigcap M_{\mathrm{B}}(t)}dt}{\int M_\mathrm{A}(t) dt} \times 100\% \label{equ:overlap1},
    \end{equation}
    and
    \begin{equation}
        R_{\mathrm{BA}} = \frac{\int {M_{\mathrm{B}}(t) \bigcap M_{\mathrm{A}}(t)dt}}{\int M_\mathrm{B}(t) dt} \times 100\% \label{equ:overlap2},
    \end{equation}
    where $M_{\mathrm{A}}(t)$ and $M_{\mathrm{B}}(t)$ are the PSF models of `A' and `B', respectively. The numerator is the integration of the overlapped PSF models of `A' and `B' over time. If both $R_{\mathrm{AB}}$ and $R_{\mathrm{BA}}$ of each group are higher than 85\%, the two sources are considered as coupled with each other in this observation.
    \item Find the contaminators of `A' in each observational data according to step 1 and exclude the contributions of other sources. Then, fit the flux of `A' individually based on the residuals (This is called method 2).
    \item Publish the two results and mark the contaminators for the sources in each observational data: result 1 and result 2 (based on method 2).
\end{itemize}

\begin{figure*}
    \gridline{\fig{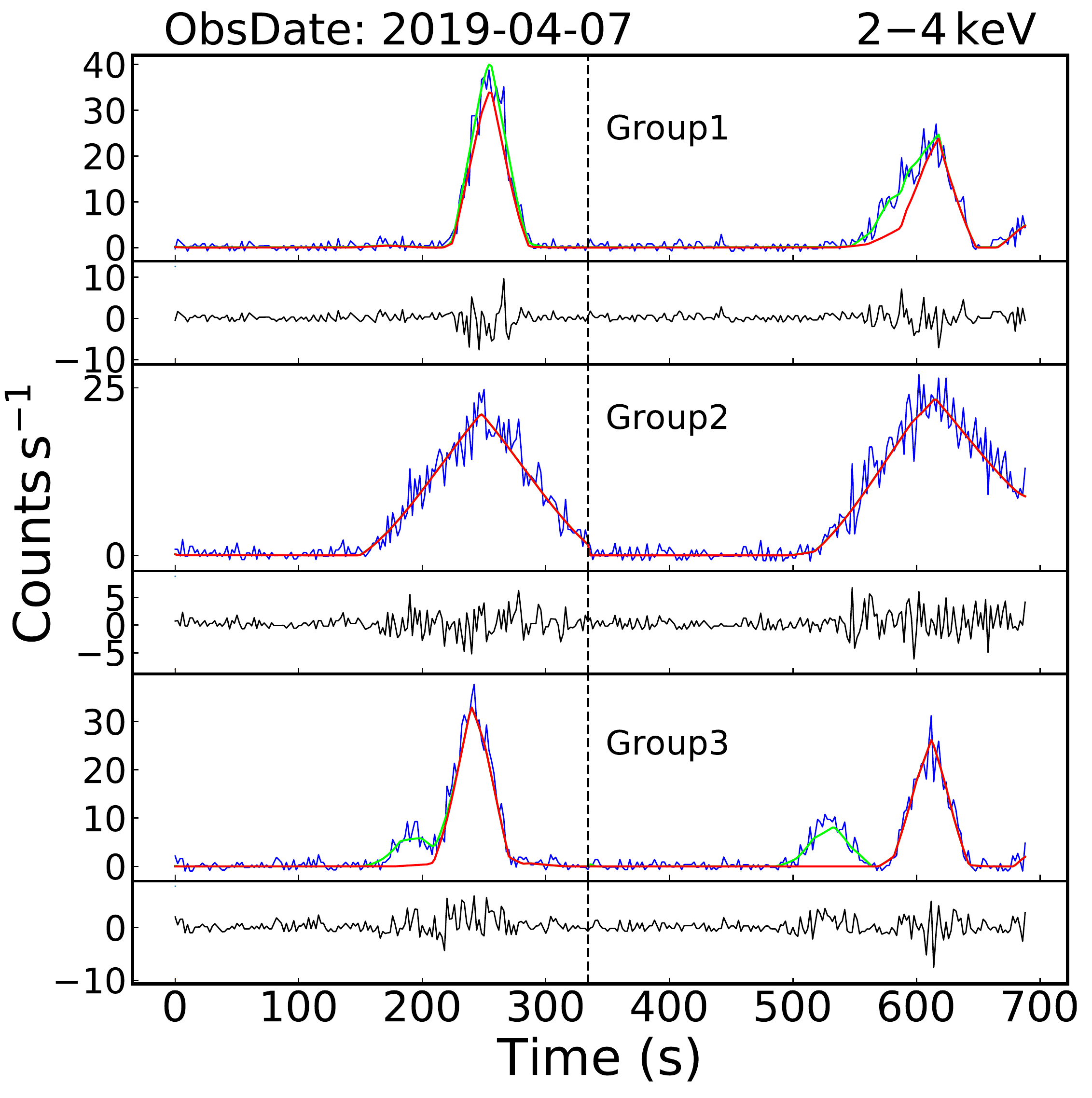}{0.4\textwidth}{(a)}
              \fig{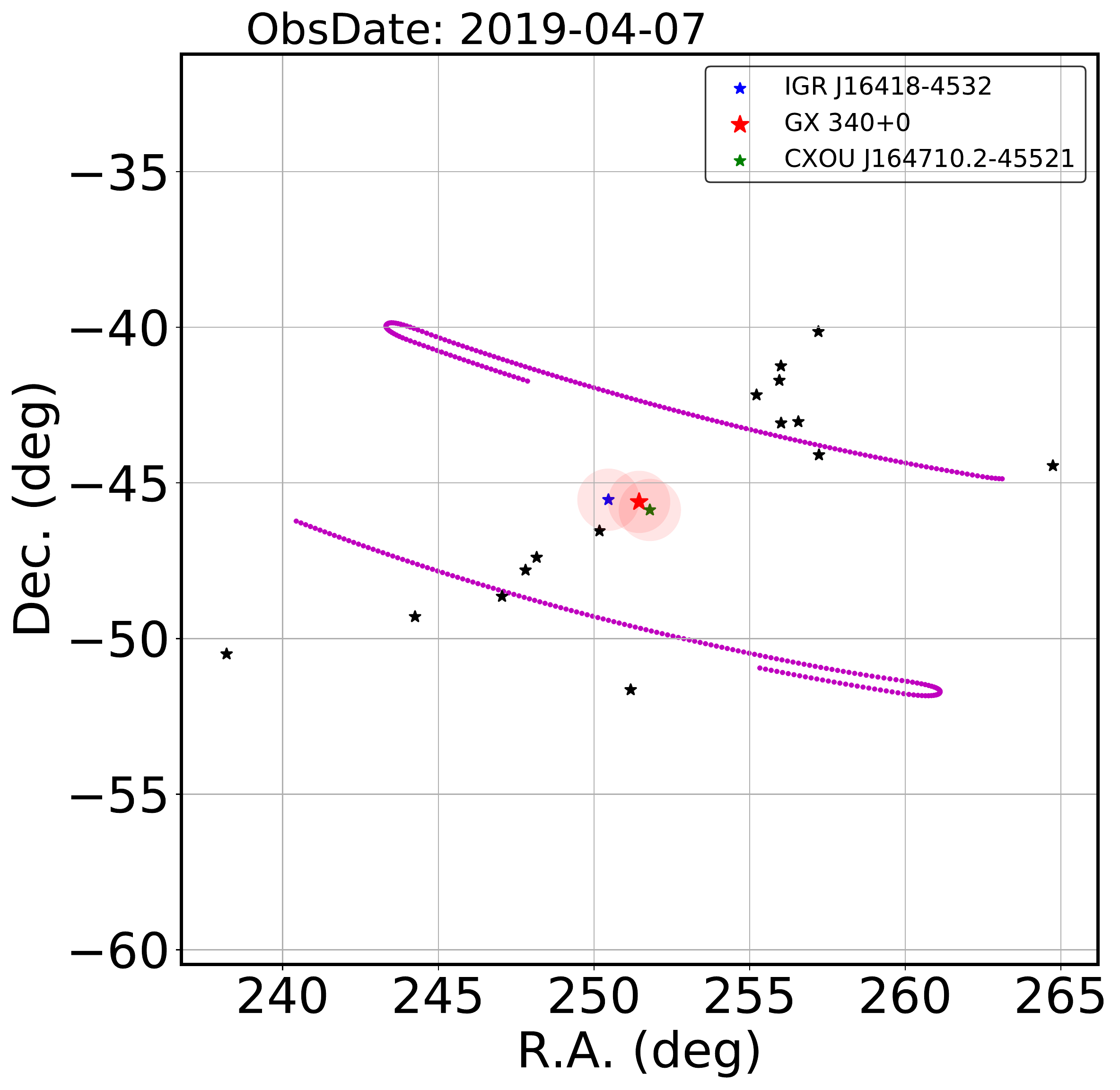}{0.4\textwidth}{(b)}
             }
    \gridline{\fig{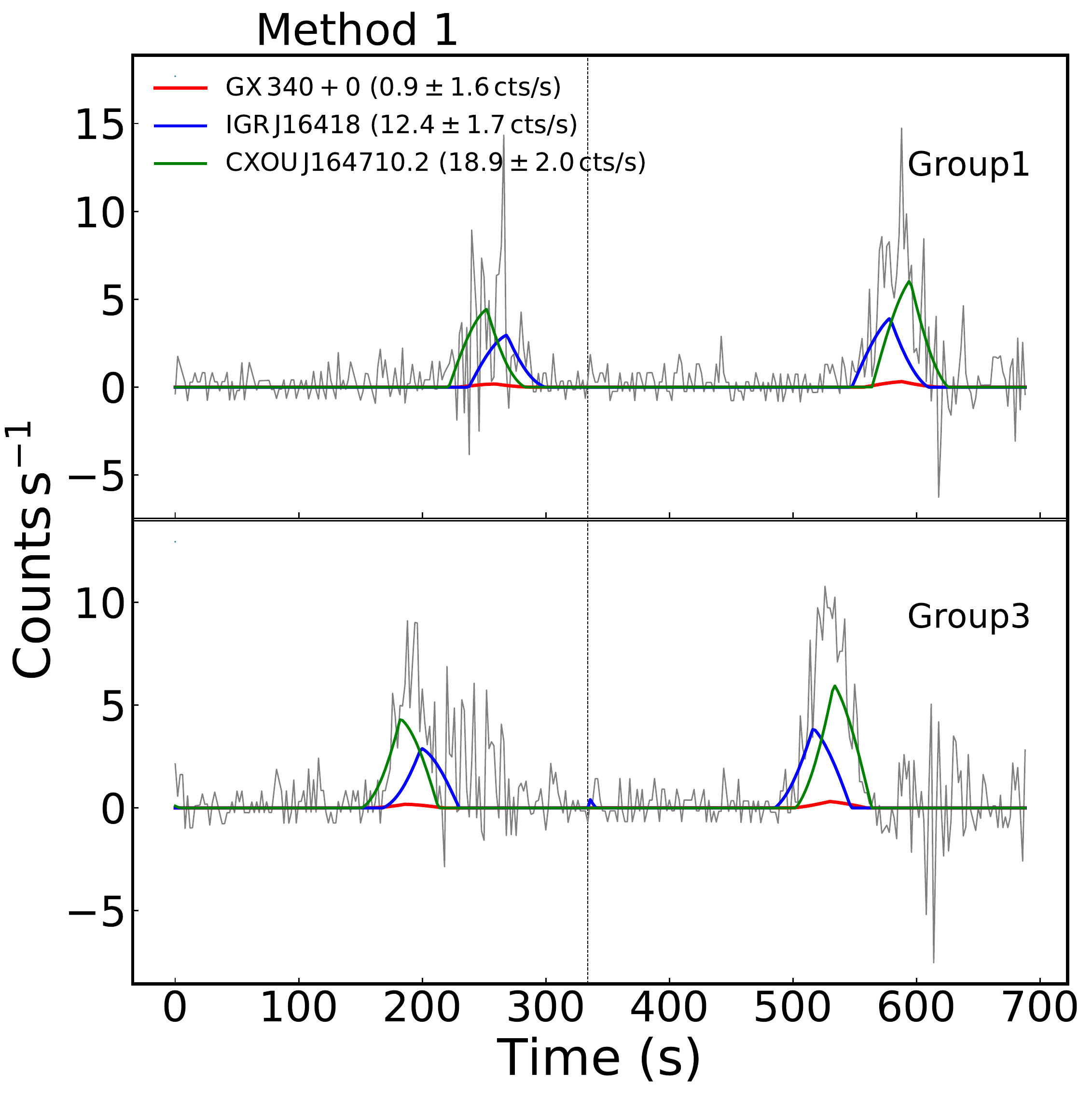}{0.4\textwidth}{(c)}
              \fig{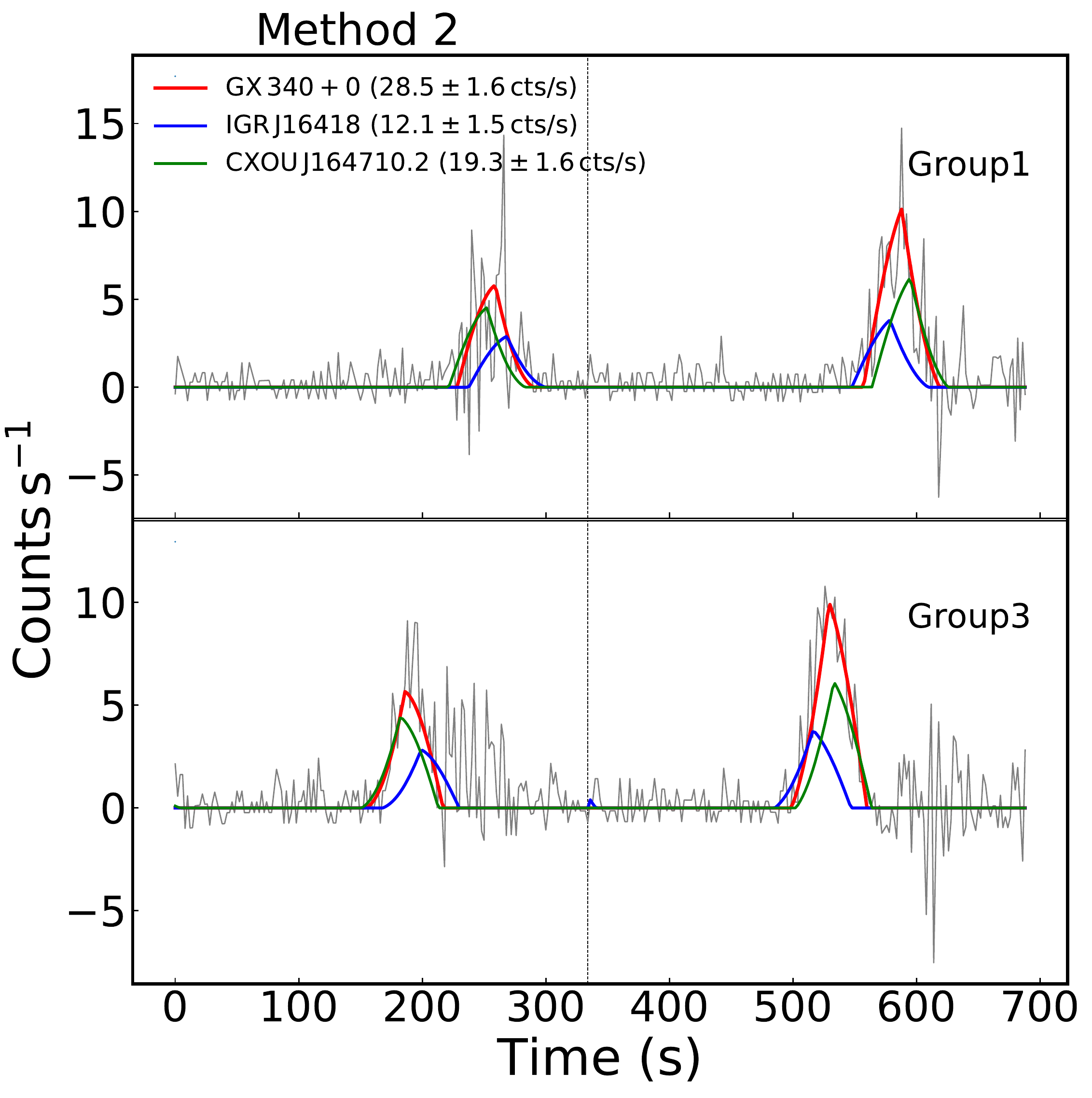}{0.4\textwidth}{(d)}}
    \caption{The example shows the observational data that GX\,340$+$0 is influenced by its neighbors.\\
    Panel (a) shows the example of light curves that the FOVs crossed GX\,340$+$0. The three blue and green lines are the light curves and the final fitting results (based on method 1) in the three groups of collimator orientation, respectively. The bottom of each group shows the corresponding residuals, and the red lines represent the contributions of sources that do not influence GX\,340$+$0. The vertical black dashed line is the segmentation of GTI. It can be seen that the red and the green lines are consistent in group2, which means the second FOV did not cross GX\,340$+$0 and its contaminators. \\ Panel (b) shows the corresponding scanning tracks, the location of GX\,340$+$0 and its contaminators in this observation. Each red circle radius is 1$^{\circ}$. Colored stars are assigned to sources as follows: Red is GX\,340$+$0. Blue and green are interference sources. Black stars denote bright sources. \\ Panels (c) and (d) are the fitting results of GX\,340$+$0 and its contaminators based on methods 1 and 2, respectively. The grey lines are the residuals of light curves and contributions of sources that do not influence GX\,340$+$0. Only the results of group1 and group3 are shown because the second FOV did not cross GX\,340$+$0. The fitting results of each source that interferes with the GX\,340$+$0 are marked with colors. The fluxes of the three sources obtained by the two methods are shown with the corresponding legends.}
    \label{fig:gx340+0lightcurve}
\end{figure*}

\begin{figure*}
    \gridline{\fig{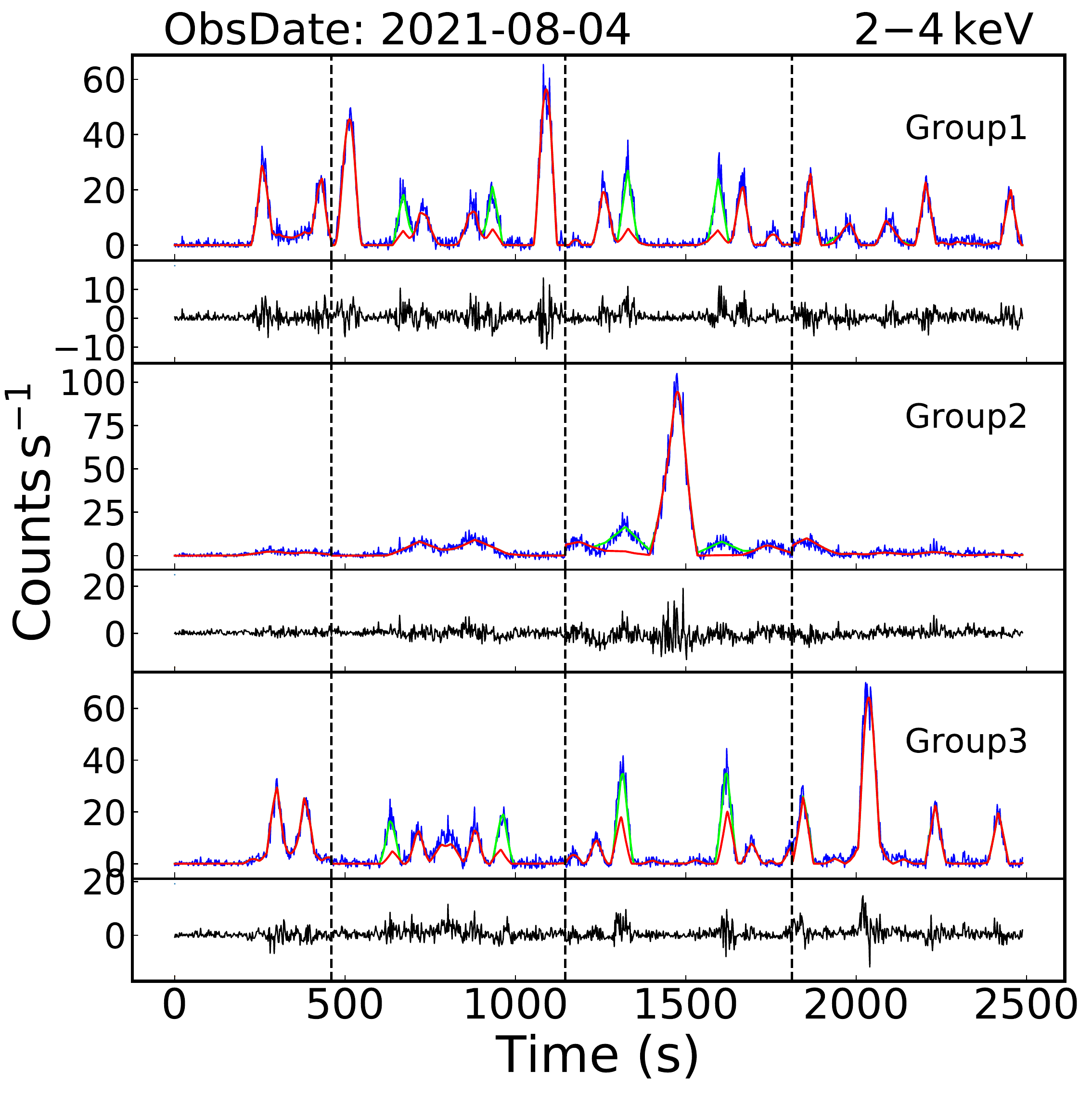}{0.4\textwidth}{(a)}
              \fig{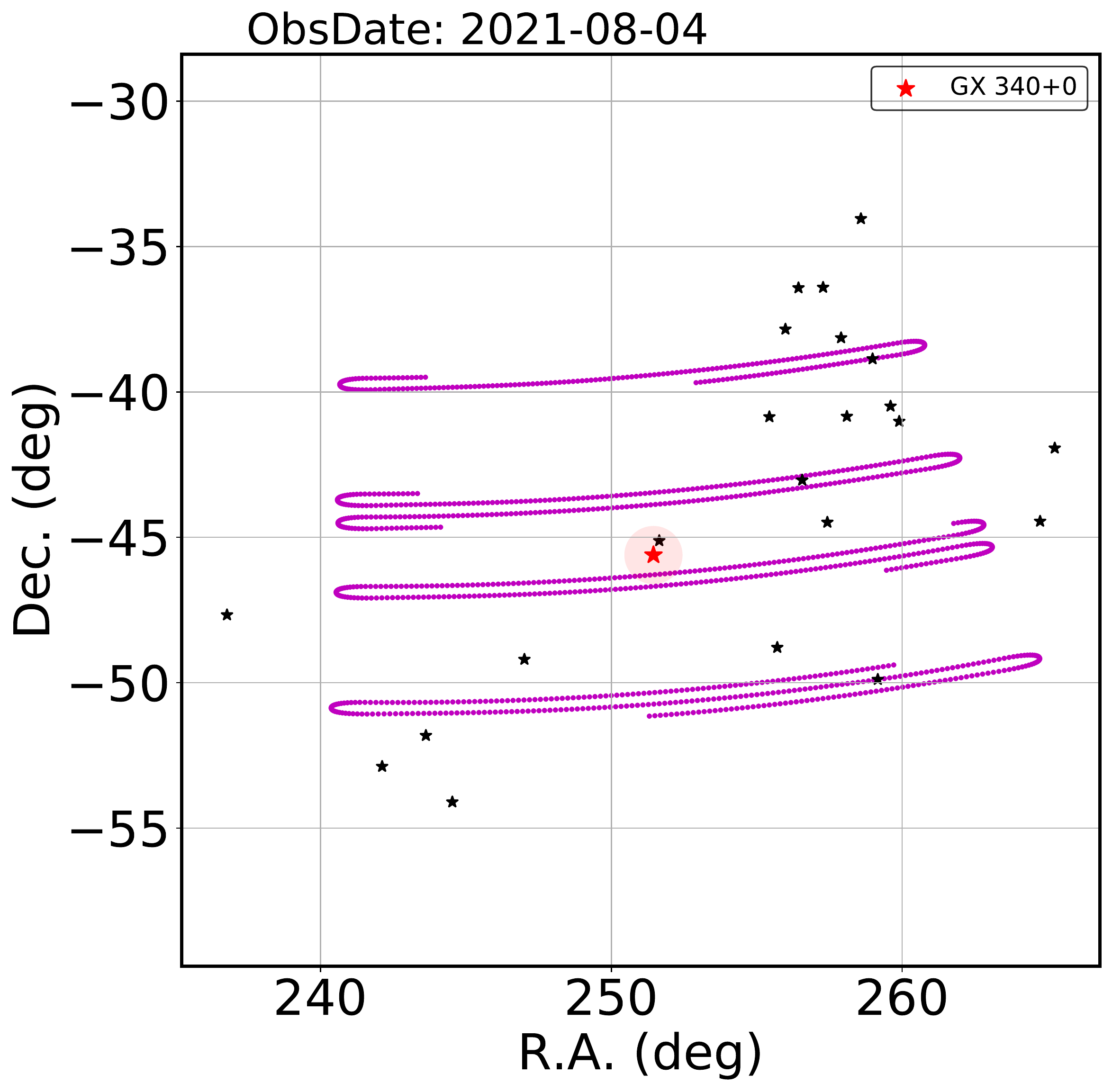}{0.4\textwidth}{(b)}
             }
    \gridline{\fig{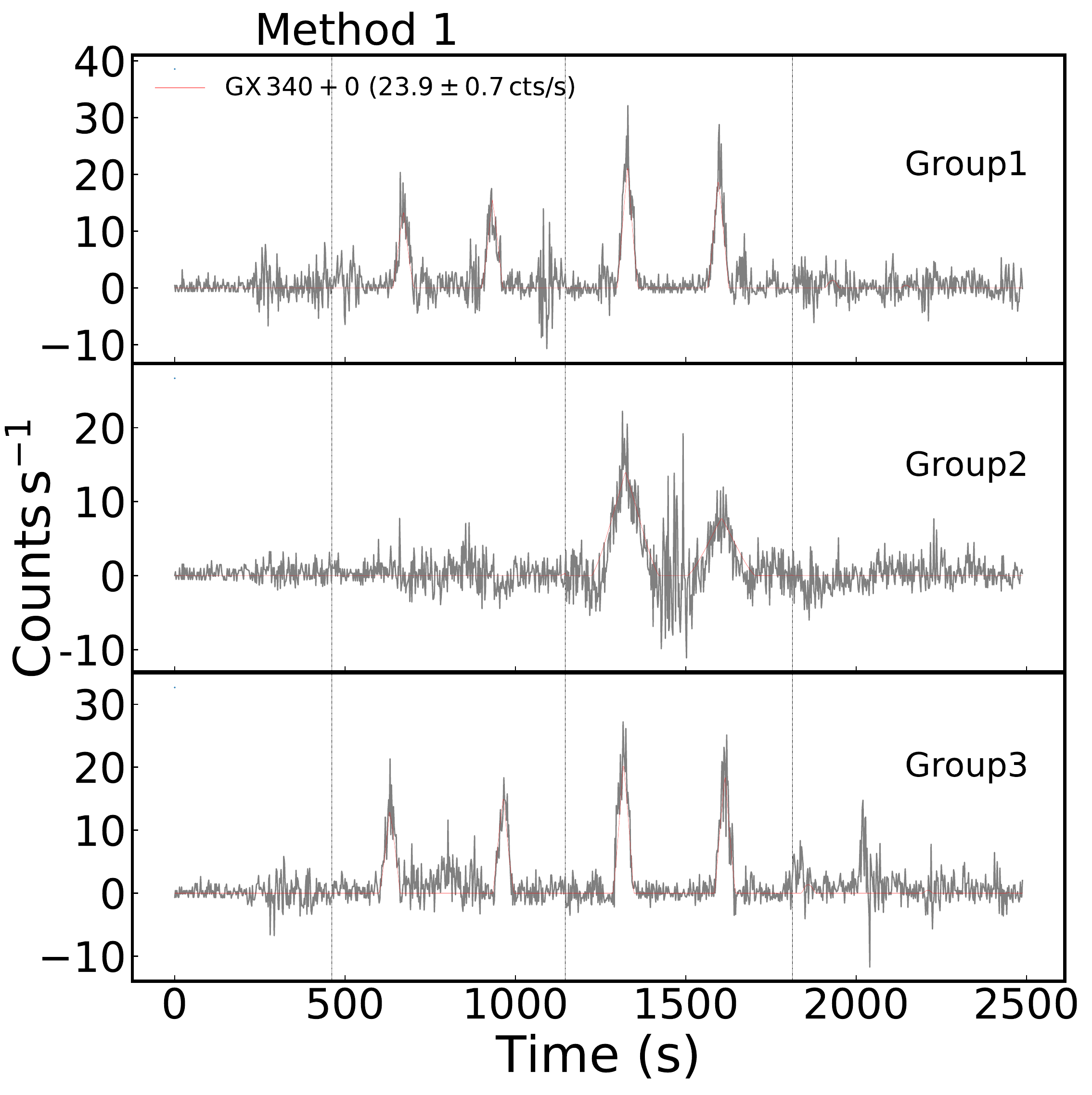}{0.4\textwidth}{(c)}
              \fig{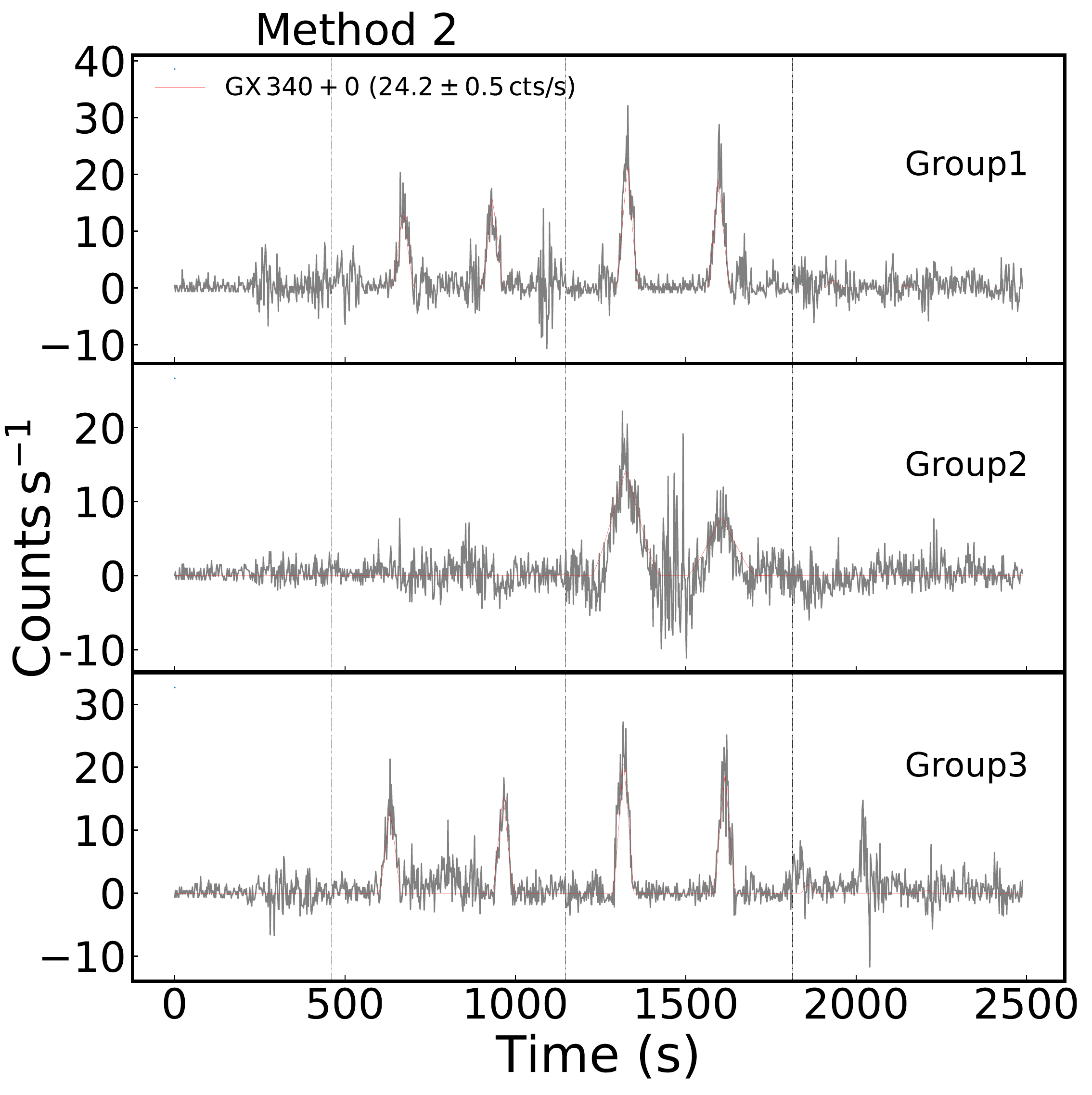}{0.4\textwidth}{(d)}}
    \caption{The example shows the observational data that GX\,340$+$0 has no contaminators. The figure contents are the same as Figure \ref{fig:gx340+0lightcurve}. The obtained fluxes are consistent within errors in panels (c) and (d).}
    \label{fig:gx340+0lightcurve2}
\end{figure*}

\begin{figure}
    \centering
    \includegraphics[scale=0.32]{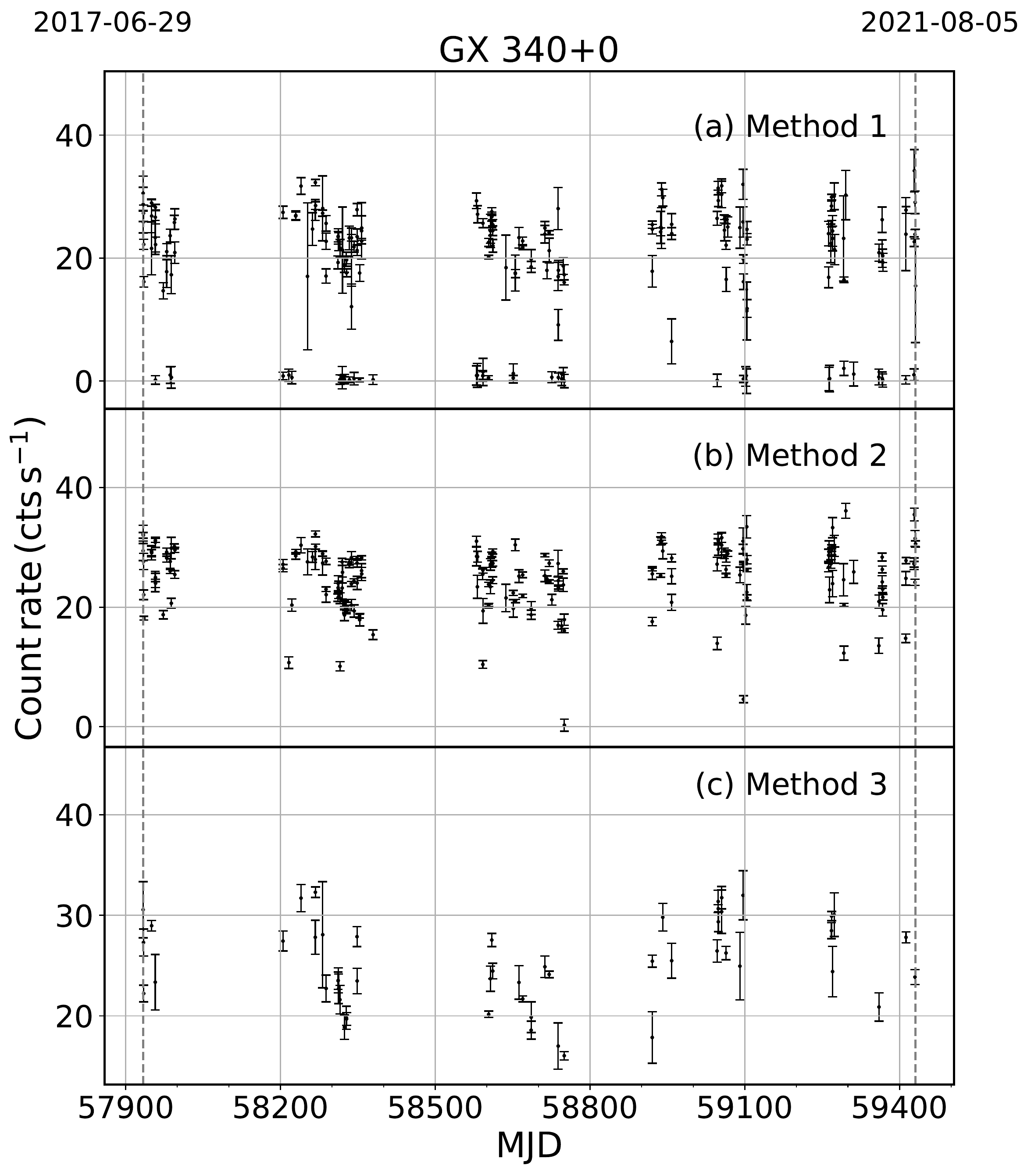}
    \caption{The long-term light curves of GX\,340+0 at 2$-$4\,keV based on different methods.}
    \label{fig:GX340+0-longterm}
\end{figure}

Taking GX\,340$+$0 as an example, two observational data are selected to illustrate the different mutual interference situations, which are shown in Figures \ref{fig:gx340+0lightcurve} and \ref{fig:gx340+0lightcurve2}, respectively. In each of the two figures, panel (a) shows the lightcurves and jointly fitting results, and panle (b) represents the corresponding scanning tracks. Panels (c) and (d) are fitting results of GX\,340$+$0 and its contaminators based on methods 1 and 2, respectivrly. Observably, GX\,340$+$0 are swept by the edges of FOVs during the first observation, while by the center of FOVs during the second one. Meanwhile, the exposures are 242 and 758\,s, respectively. Thus, the contaminators of GX340+0 are different in the two observations: the first one has two contaminators while the second one has no contaminators. This is mainly due to the more accurate positioning accuracy when the three FOVs centers cross a source with longer exposure. It is difficult to accurately know the specific contribution of CXOU\,J164710.2$-$45521, IGR\,J16418$-$4532, and GX\,340$+$0 to the peaks in Figure \ref{fig:gx340+0lightcurve} (c). In this observation, the fluxes of GX\,340$+$0 obtained by methods 1 and 2 are 0.9$\pm$1.6\,cts/s and 28.5$\pm$1.6\,cts/s, which are shown in the corresponding panel. The final long-term light curves of GX\,340$+$0 based on the two methods are shown in Figure \ref{fig:GX340+0-longterm} (a) and (b). Visibly, considerable data points at panel (a) are close to zero, which is probably because the neighboring sources shared the fluxes of GX\,340$+$0. The reason why most of the data points at panel (b) are larger than zero is that method 2 excludes all neighbors of GX\,340$+$0. Panel (c) shows the long-term light curves of GX\,340$+$0 after all uncertain \footnote{When the target source interferes with its neighbors, it is impossible to determine whether the flux peaks are contributed by one or several sources. We call the flux of the target source in this observed data `uncertain flux'. Such as the fitting result of GX\,340$+$0 in the observation of Figure \ref{fig:gx340+0lightcurve}.} data points are discarded (this is called method 3). In this example, the advantages and disadvantages of the three methods can be seen: 1) method 1 may underestimate the fluxes of some data points; 2) method 2 may overestimate the fluxes of some data points; 3) method 3 retains all convincing results (of which result 1 is consistent with result 2.) but may have fewer data points.

Considering that the radiation of most X-ray sources is mainly concentrated below 6\,keV, we perform further processing at 2$-$4\,keV, 4$-$6\,keV, and 2$-$6\,keV for all X-ray sources with neighbors according to method 2.

\section{GPSS Results} \label{sec:GPSSresults}
\textit{Insight-HXMT} has performed over 2000 observations since 2017. In GPSS, 1336, 957, and 935 sources are monitored by LE (1$-$6\,keV\footnote{Due to the instrument aging of LE, the data at 1$-$2\,keV became inaccurate after August 2020, so all data covering 1$-$2\,keV in this article are the monitored results before August 2020.}), ME (7$-$40\,keV) and HE (25$-$100\,keV), respectively.
In this section, we present the results of \textit{Insight-HXMT} GPSS and quantify the properties of the three telescopes.

\subsection{Catalog}
Long-term light curves are essential for understanding the nature of X-ray objects \citep{2018ApJS..235....7H}. They are generally obtained by monitoring and recording fluxes of sources. For each observation, \textit{Insight-HXMT} provides the following information, which is contained in light curves and can be obtained by PSF fitting:\\
\begin{itemize}
    \item For known X-ray objects that have crossed the FOVs, their fluxes and S/N at different energy bands are obtained.
    \item For a new X-ray candidate, its location, flux and S/N are obtained.
\end{itemize}

\begin{figure}
    \centering
    \includegraphics[scale=0.32]{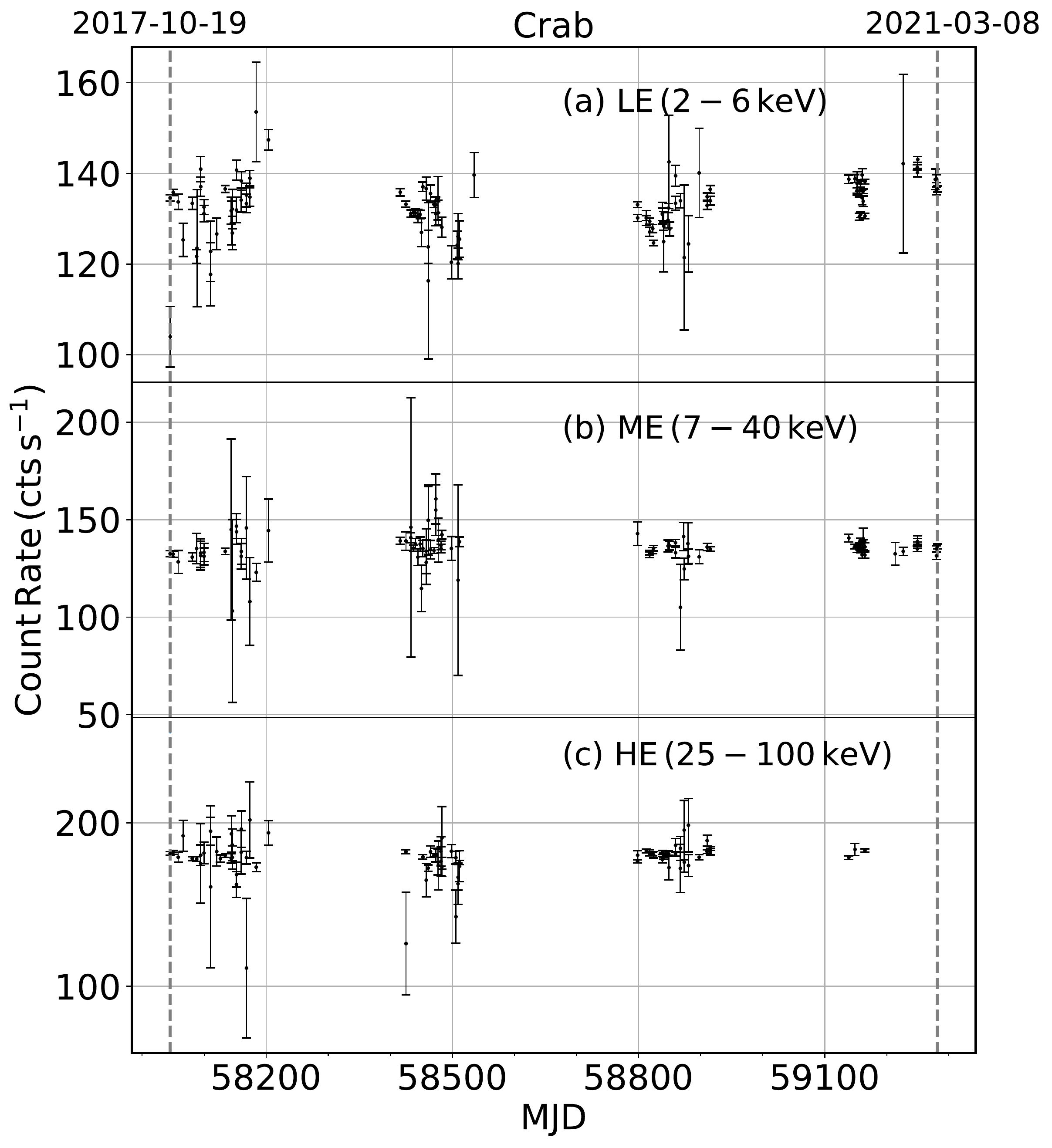}
    \caption{The long-term light curves of Crab at 2$-$6\,keV, 7$-$40\,keV, 25$-$100\,keV  obtained with LE, ME and HE, respectively.}
    \label{fig:crab-long-term2}
\end{figure}

\noindent Therefore, for \textit{Insight-HXMT}, a source's long-term variability can be derived by fitting all light curves corresponding to the scanning tracks covering this source. For all monitored sources, the long-term light curves are obtained at 1$-$100\,keV bands. The final results can be accessed on the website of \textit{Insight-HXMT} \footnote{\url{http://hxmtweb.ihep.ac.cn/Transients.jhtml}}, and three examples of long-term light curves are shown in Figure \ref{fig:crab-long-term2}.

The excess variance ($F_{\mathrm{rms}}$) is an important parameter to characterize the flux variability of a source \citep{1984AIPC..115...63S, 2003MNRAS.345.1271V}, so $F_{\mathrm{rms}}$ \footnote{If the flux of a source is stable or its statistical error is larger than its variability, $F_{\mathrm{rms}}$ can be negative, in which case the value of $F_{\mathrm{rms}}$ is meaningless and marked as $F_{\mathrm{rms}}<$0. It appears that a weak source with averaged flux close to zero and no burst in its long-term light curve may be obtained with a very small or large value of $F_{\mathrm{rms}}$, in which case the value may no longer accurately reflect the magnitude of its flux variability.} at each energy band in this work is calculated to quantify the variability of each source using the following equations:
\begin{equation}
    S^2=\frac{1}{N-1}\sum_i^{N}(f_i-\bar{f})^2,
\end{equation}
\begin{equation}
    F_{\mathrm{rms}} = \frac{S^2-\bar{\sigma^2}}{\bar{f}^2} \label{equ:excess},
\end{equation}
and
\begin{equation}
d F_{\mathrm{rms}} = \sqrt{(\sqrt{\frac{2}{N}} \times \frac{\bar{\sigma^2}}{\bar{f}^2})^2+(\sqrt{\frac{\bar{\sigma^2}}{N}}\times\frac{2\sqrt{F_{\mathrm{rms}}}}{\bar{f}})^2},
\end{equation}
where $N$ is the number of points on each light curve, $f_i$ is the flux for each observation, $\bar{\sigma^2}$ is the mean of the square of statistical error, and $d F_{\mathrm{rms}}$ is the error of $F_{\mathrm{rms}}$.  The final $F_{\mathrm{rms}}$ values for monitored sources at different energy bands are listed in our catalog.

Hardness Ratio (HR) is an important parameter that is broadly used to characterize a source's spectral property  \citep{2006ApJ...653.1566J, 2004ApJ...617..262S}. It is defined as:\\
\begin{equation}
    \mathrm{HR} = \frac{H}{S}  \label{equ:hr},
\end{equation}
\noindent where $H$ and $S$ are count rates in hard and soft energy bands, respectively. According to the error propagation formula, the error of HR is calculated as:\\
\begin{equation}
    \sigma_{\mathrm{HR,stat}} = \frac{H}{S}\sqrt{\frac{\sigma^{2}_{\mathrm{S,stat}}}{S^2}+\frac{\sigma^{2}_{\mathrm{H,stat}}}{H^2}} \label{equ:sigma},
\end{equation}
\noindent where $\sigma_{\mathrm{H,stat}}$ and $\sigma_{\mathrm{S,stat}}$ are statistic errors of fluxes at H and S bands, respectively. In this paper, HRs are only calculated for sources with averaged S/N higher than 5 both at H and S bands. The averaged S/N are callculated as follows:
\begin{equation}
    \bar{f} = \frac{\displaystyle\sum_{i}\omega_{i}f_{i}}{\displaystyle\sum_{i}\omega_{i}} \label{equ:average flux},
\end{equation} \\
\begin{equation}
    \omega_{i} = \frac{1}{\sigma_{i}^{2}} \label{equ:weight},
\end{equation}
\begin{equation}
    \sigma^{2}(\bar{f}) = \frac{1}{(\displaystyle\sum_{i}\omega_{i})^2}\displaystyle\sum_{i}\omega_{i}^{2}\sigma_{i}^{2} = \frac{1}{\displaystyle\sum_{i}\frac{1}{\sigma_{i}^{2}}} \label{equ:average sig},
\end{equation}
and
\begin{equation}
    \overline{\mathrm{S/N}} = \frac{\bar{f}}{\bar{\sigma}} \label{equ: average snr},
\end{equation}
where $\bar{f}$ and $f_{i}$ refer to the averaged flux and the best-fit flux of the $i$-th observation, respectively. $\omega_{i}$ and $\sigma_{i}$ are the corresponding weight and statistical error. Equation (\ref{equ: average snr}) is used to calculate the averaged S/N for each source.

We present three \textit{Insight-HXMT} catalogs obtained from the 4 yr GPSS data in the low Galactic latitude sky based on methods 1 - 3, and each catalog contains nine energy bands (1$-$2\,keV, 1$-$6\,keV, 2$-$4\,keV, 2$-$6\,keV, 3$-$5\,keV, 4$-$6\,keV, 5$-$7\,keV, 7$-$40\,keV and 25$-$100\,keV) and the following information: 1) the source names, 2) the source locations, 3) the source types from SIMBAD\footnote{\url{http://simbad.u-strasbg.fr/simbad/}\\It is worth noting that, in this paper, the source type of MAXI\,J1348$-$630 and Swift\,J0243.6$+$6124 are adopted from \cite{2022MNRAS.517L..21C} and \cite{2018AAS...23116004C} rather than the SIMBAD database.}, 4) the averaged fluxes and corresponding errors at each energy band, 5) $F_{\mathrm{rms}}$ at each energy band, and 6) HRs and the corresponding errors. To analyze the source properties more accurately, the results in the catalog that correspond to method 3 are utilized in all statistical analyses of this paper. There are 223 sources with S/N$>$5 at one or more energy bands among 2$-$6\,keV, 7$-$40\,keV, 25$-$100\,keV (hereafter 2$-$100\,keV), including 59 low mass X-ray binaries (LMXBs), 56 high mass X-ray binaries (HMXBs), 13 supernova remnants (SNRs), 11 pulsars, six Seyfert 1 Galaxies, 23 other type sources, and 55 unclassified sources. Their information at the three energy bands is extracted in Tables \ref{tab:catalog1} - \ref{tab:catalog6}, among which the 33 sources marked in blue are bright sources with S/N$>$5 at the three energy bands. Moreover, the compact objects of LMXBs and HMXBs are recorded in column (2) of Tables \ref{tab:catalog1} and \ref{tab:catalog2}, respectively.

\subsection{Properties of \textit{Insight-HXMT} Detectors} \label{sec:property}
Conventionally, the long-term light curve of the Crab nebula (hereafter the `Crab') is used to calculate the systematic errors of instruments, because its spectral characteristics are generally stable. Values of systematic errors at different energy bands can be obtained by solving the following equations, 
\begin{equation}
    \displaystyle\sum_{i}^{N}\frac{(f_{i}-\bar{f})^2}{\sigma_{\mathrm{total},i}^{2}} = N-1 \label{equ:sys},
\end{equation}

\begin{equation}
    \sigma^{2}_{\mathrm{total},i} = \sigma_{\mathrm{sys}}^{2} + \sigma_{\mathrm{stat},i}^2  \label{equ:sigma-total},
\end{equation}
 and
\begin{equation}
    \bar{f} = \frac{\displaystyle\sum_{i}\omega_{i}f_{i}}{\displaystyle\sum_{i}\omega_{i}}, \\ 
    \omega_{i} = \frac{1}{\sigma_{\mathrm{total},i}^{2}} \label{equ:flux and weight},
\end{equation}

\noindent where $N$ is the number of observations, $i$ denotes the $i$-th observation. $\sigma_{\mathrm{total},i}$ is the total error, $\sigma_\mathrm{sys}$ is the systematic error, $\sigma_{\mathrm{stat},i}$ and $f_{i}$ are the statistic error and flux of Crab, $\omega_{i}$ is the weight of $f_{i}$, and $\bar{f}$ is the averaged flux. The systematic errors of all energy bands are listed in Table \ref{tab:Information}.

\begin{figure*}
    \gridline{\fig{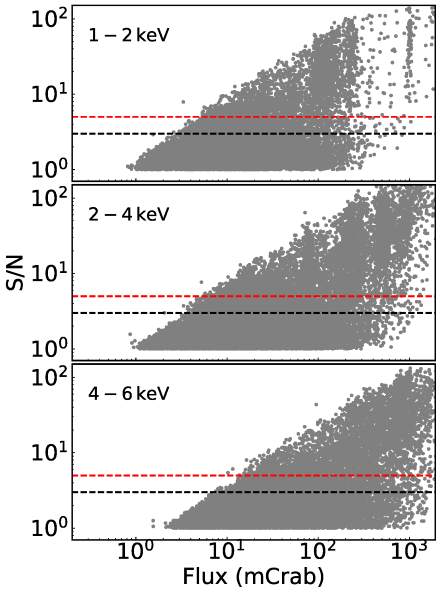}{0.32\textwidth}{(a)}\label{fig:snr-flux1}
              \fig{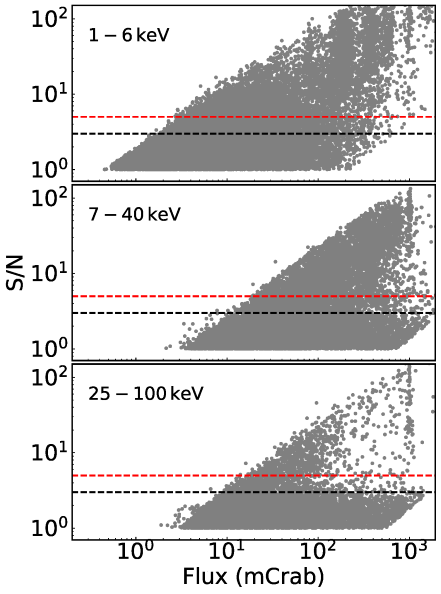}{0.32\textwidth}{(b)}\label{fig:snr-flux2}
              \fig{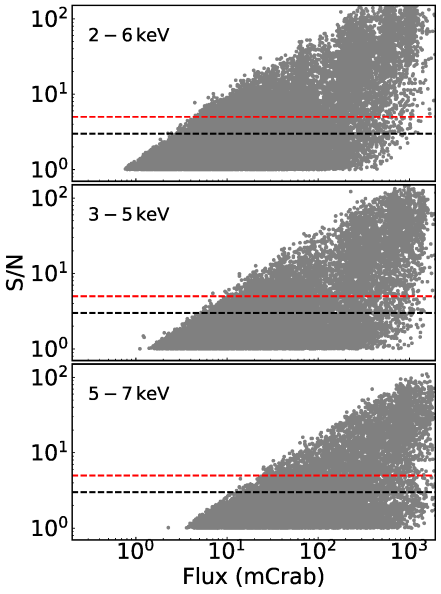}{0.32\textwidth}{(c)}\label{fig:snr-flux3}
              }
    \caption{This figure shows the relationship between flux and S/N at nine energy bands. The black and red dashed lines represent S/N=3 and S/N=5, respectively.}
    \label{fig:snr-flux}
\end{figure*}

The sensitivities at each energy band can be obtained by the S/N-flux relation, which is shown in Figure \ref{fig:snr-flux}. We assume that the spectra of all detected sources show consistent shapes with Crab's, and their fluxes are obtained in Crab's units. That S/N increases with flux according to statistics is often affected by other factors in reality, such as background level, exposure time, and effective area. Therefore, by recording the source flux and S/N obtained in each observation, the distribution in Figure \ref{fig:snr-flux} is obtained, in which the lowest flux when the corresponding S/N $=$ 5 at each energy band is considered as the sensitivity limit. The details are listed in Table \ref{tab:Information}.

\begin{longrotatetable}
\begin{deluxetable*}{L|RRRRRRR|R|R}
    \setcounter{table}{3}
    \tablecaption{Properties of \textit{Insight-HXMT} Detectors in Different Energy Band. \label{tab:Information}}
    \tablewidth{0pt}
    \tablehead{
        \nocolhead{} &
        \multicolumn{7}{|C}{\mathrm{LE}} &
        \multicolumn{1}{|C}{\mathrm{ME}} &
        \multicolumn{1}{|C}{\mathrm{HE}}
        \\
        \nocolhead{} &
        \multicolumn{1}{|C}{1$-$2\,\mathrm{keV}} &
        \colhead{2$-$4\,keV} &
        \colhead{4$-$6\,keV} &
        \colhead{2$-$6\,keV} &
        \colhead{3$-$5\,keV} &
        \colhead{5$-$7\,keV} &
        \colhead{1$-$6\,keV} &
        \multicolumn{1}{|C|}{7$-$40\,\mathrm{keV}} &
        \colhead{25$-$100\,keV}
                }
    \startdata
        \mathrm{$\sigma_{sys}~(\%)$                            } & 2.16     & 3.35      & 2.61   & 3.43  & 3.09      & 3.49     & 3.46      & 1.42    & 0.83    \\
        \mathrm{Source~ Num.$^{a}$                             } & 1337     & 1345      & 1342   & 1343  & 1343      & 1343     & 1336      & 957     & 935     \\
        \mathrm{Bright~ Source~Num.$^{b}$                      } & 261      & 266       & 230    & 294   & 248       & 210      & 268       & 65      & 59     \\
        \mathrm{Sensitivity1~(mCrab)$^{c}$                     } &  4.7     &  4.6      &  13.2  & 4.1   &  8.9      &  21.2    &  2.5      &  17.6   &  12.7 \\
        \mathrm{Sensitivity1~($10^{-10}\,erg\,s^{-1}\,cm^{2}$) } &  0.44    &  0.40     &  0.63  & 0.55  &  0.55     &  0.82    &  0.57     &  3.16   &  1.61  \\
        \mathrm{Sensitivity2~(mCrab)$^{d}$                     } &  0.70    &  0.69     &  1.97  & 0.61  &  1.33     &  3.17    &  0.37     &  2.35   &  2.19 \\
        \mathrm{Sensitivity2~($10^{-11}\,erg\,s^{-1}\,cm^{2}$) } &  0.66    &  0.60     &  0.94  & 0.82  &  0.82     &  1.23    &  0.85     &  4.21   &  2.78  \\
    \enddata
\tablecomments{\\$^{a}$ The total source number that has been covered at each energy band.\\
               $^{b}$ One is considered as a bright source when its averaged S/N is higher than 5 and averaged flux is positive.\\
               $^{c}$ Sensitivity of one scanning observation ($\sim$2.3h). \\
               $^{d}$ Cumulative sensitivity from June 2017 to August 2021. }
\end{deluxetable*}
\end{longrotatetable}

\section{Combine Long-term Light Curves with MAXI} \label{sec:compMAXI}
\begin{figure}
    \centering
    \includegraphics[scale=1.1]{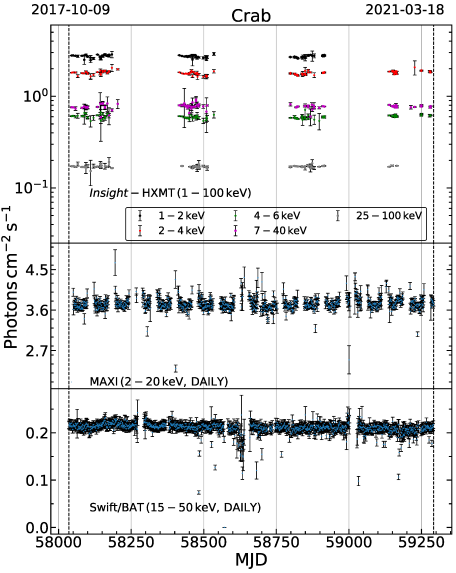}
    \caption{Crab light curves from different missions. (Top) Results monitored by \textit{Insight-HXMT} at 1$-$2\,keV, 2$-$4\,keV, 4$-$6\,keV, 7$-$40\,keV and 25$-$100\,keV.
    (Middle) Results from MAXI at 2$-$20\,keV. 
    (Bottom) Results from Swift/BAT at 15$-$50\,keV. Note that, the Crab light curves of MAXI and Swift/BAT come directly from their webpages, and no additional calibration is done.}
    \label{fig:crab-compare}
\end{figure}

Long-term light curves are essential to reflect the behaviors of X-ray sources, such as their activity duty-cycles and time scales of variability. Providing long-term light curves is one of the main scientific outputs of \textit{Insight-HXMT}. However, the visible duration of each scanned area is limited to a few months per year due to the limitation of the solar angle ($>$ 70$^{\circ}$), which results in gaps in long-term light curves. With the merged data from other surveys, it is helpful to make up for these gaps. Taking Crab as an instance, Figure \ref{fig:crab-compare} shows its long-term light curves that were monitored by \textit{Insight-HXMT}, MAXI \footnote{\url{http://134.160.243.88/star_data/J0534+220/J0534+220_g_lc_1orb_all.dat}}, and Swift/BAT \footnote{\url{https://swift.gsfc.nasa.gov/results/transients/Crab.lc.txt}} from October 2017 to November 2021. It is apparent that \textit{Insight-HXMT} has the advantage of monitoring sources in a wide band, whereas the gaps under its scanning strategy are unavoidable. In comparison, MAXI is an all-sky X-ray monitor and scans the entire sky every 92 minutes, covering the 2$-$30\,keV energy band. MAXI has been monitoring the sky since its launch in 2009 and has accumulated a large amount of continuous observation data. The coded-aperture imager Swift/BAT can cover 88\% of the sky each day and has been scanning the sky since its launch \citep{2013ApJS..209...14K}. Combining the long-term light curves from different missions can reflect more comprehensively the flux variability of X-ray sources in a wider band. In this section, we combine and analyze results at 2$-$4\,keV between \textit{Insight-HXMT} and MAXI, for the reason that the two missions are more sensitive in this energy band and the main radiation energy range of most X-ray sources concentrates in low-energy bands.

\subsection{Calibration Analysis}
\begin{figure}
    \centering
    \includegraphics[scale=1.0]{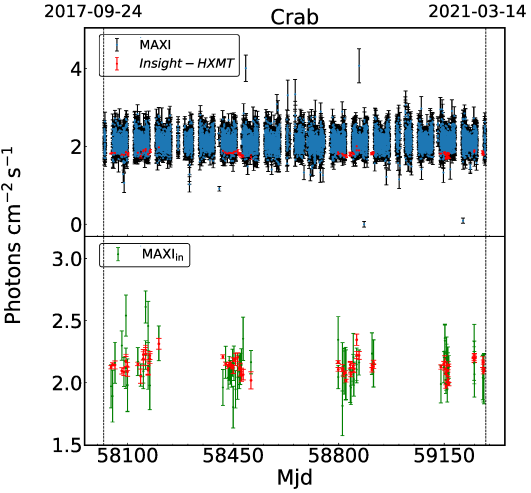}
    \caption{(Top) The long-term light curves of Crab that were monitored by MAXI and \textit{Insight-HXMT} at 2$-$4\,keV band during the same period (from 2017-09-24 to 2021-03-14). The red and blue points are monitored results from \textit{Insight-HXMT} and one-orbit time bin results from MAXI, respectively. The vertical errors are the corresponding 1-$\sigma$ statistical uncertainties. 
    (Bottom) The green points and vertical errors are MAXI interpolation data points and corresponding 1-$\sigma$ statistical uncertainties. The red points are calibrated \textit{Insight-HXMT} data points.}
    \label{fig:crab-inter}
\end{figure}

\begin{figure}
    \centering
    \includegraphics[scale=1.0]{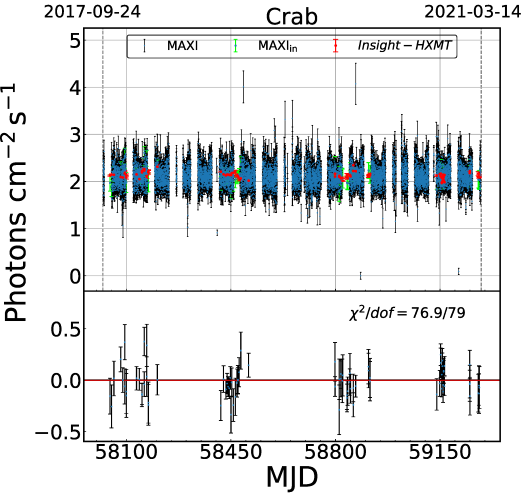}
    \caption{(Top) Crab light curves monitored by MAXI and \textit{Insight-HXMT} at 2$-$4\,keV band. The red, blue and green points are calibrated \textit{Insight-HXMT} data points, one-orbit time bin results from MAXI and MAXI interpolation data points, respectively. The vertical errors are the corresponding 1-$\sigma$ statistical uncertainties. 
    (Bottom) The distribution and corresponding 1-$\sigma$ statistical uncertainties of $F_\mathrm{M}-F_\mathrm{H}$. This is used to check the consistency of two satellites, and the reduced chi-squared is 0.97 (dof=79).}
    \label{fig:contrast crab}
\end{figure}

Before combining the data of the two missions, we analyze and calibrate the results of \textit{Insight-HXMT} by comparing the flux at 2$-$4\,keV of the standard Crab model with the actual detection counts. Then the calibrated result is used to check the consistency with the data of MAXI. The top panel of Figure \ref{fig:crab-inter} shows the long-term light curves of Crab monitored by the two missions during the same period (from 2017$-$09$-$24 to 2021$-$03$-$14). One finds that the monitoring time of the two missions is not strictly consistent. Thus the data points of MAXI are interpolated with those of \textit{Insight-HXMT} based on: 
\begin{equation}
    f_{\mathrm{M,Inter}}=f_{\mathrm{M},i}+\frac{t_{\mathrm{H}}-t_{\mathrm{M},i}}{t_{\mathrm{M},i+1}-t_{\mathrm{M},i}} \times (f_{\mathrm{M},i+1}-f_{\mathrm{M},i}), \label{equ:inter}
\end{equation}
and 
\begin{equation}
    \Delta f_{\mathrm{M,Inter}} =
    \sqrt{(\frac{t_{\mathrm{M},i+1}-t_{\mathrm{H}}}{t_{\mathrm{M},i+1}-t_{\mathrm{M},i}})^2 \sigma_{\mathrm{M},i}^2+(\frac{t_{\mathrm{H}}-t_{\mathrm{M},i}}{t_{\mathrm{M},i+1}-t_{\mathrm{M},i}})^2 \sigma_{\mathrm{M},i+1}^{2}}, \label{equ:inter-err}
\end{equation}
where $f_{\mathrm{M},i}$, $\sigma_{\mathrm{M},i}$ and $t_{\mathrm{M},i}$ are the flux, flux error, and corresponding MJD for the $i$th data point of MAXI, respectively. $t_{\mathrm{H}}$ is the MJD of \textit{Insight-HXMT} between $t_{\mathrm{M},i}$ and $t_{\mathrm{M},i+1}$. In addition, we can notice that the data points of MAXI are overall higher than those of \textit{Insight-HXMT}. We consider the difference is mainly due to the different adoptions of the standard model of Crab that is used in absolute flux calibration. The absorbed power-law model is adopted in \textit{Insight-HXMT} calibration with a photon index of 2.11, a normalization of 8.76\,photon\,$\mathrm{cm^{-2}\,s^{-1}\,keV^{-1}}$, and interstellar absorption $N_\mathrm{H}$ of $\mathrm{3.6} \times \mathrm{10^{21}}\,\mathrm{cm^{-2}}$ \citep{2020JHEAp..27...64L}. The fluxes (2$-$4\,keV) calculated with this model is 1.815\,photon\,$\mathrm{s^{-1}\,cm^{-2}}$ and the averaged detected flux is 1.809 $\pm$ 0.002\,counts\,$\mathrm{s^{-1}\,cm^{-2}}$. In contrast, $\Gamma$ and normalization of the spectrum of Crab are assumed as 2.1 and 10\,photons\,$\mathrm{s^{-1}\,cm^{-2}}$ for MAXI \citep{2018ApJS..235....7H, 2018ApJS..238...32K, 2016PASJ...68S..32T, 2011PASJ...63S.635S}. Furthermore, it is written that 1\,Crab equals to 2.2 $\mathrm{photons\,s^{-1}\,cm^{-2}}$ at 2$-4$\,keV on their webpage \footnote{maxi.riken.jp/top/readme.html}. The difference in the selection of normalization is the main reason for the overall difference of the two Crab light curves in Figure \ref{fig:crab-inter}. Therefore, a correction factor is needed to calibrate the two missions, and it is derived by dividing the averaged value of interpolated data points of MAXI (2.114\,photon\,$\mathrm{cm^{-2}\,s^{-1}}$) and the averaged flux of \textit{Insight-HXMT} (1.809\,counts\,$\mathrm{s^{-1}\,cm^{-2}}$). Consequently, the data points of \textit{Insight-HXMT} are multiplied by 1.17, whose consistency with MAXI is checked based on 
\begin{equation}
    F_{\mathrm{diff}} = f_{\mathrm{M}}-f_{\mathrm{H}}, \label{equ:F-diff}
\end{equation}
and
\begin{equation}
    \chi^{2}_{\mathrm{diff}} = \displaystyle\sum_{i=1}^{N}\frac{(f_{\mathrm{M},i}-f_{\mathrm{H},i})^2}{\sigma_{\mathrm{M},i}^{2}+ \sigma_{\mathrm{H},i}^{2}}, \label{equ:dof}
\end{equation}
where $f_{\mathrm{M}}$ and $f_{\mathrm{H}}$ are interpolated flux of MAXI and flux from \textit{Insight-HXMT}, respectively. $\sigma_{\mathrm{M}}$ and $\sigma_{\mathrm{H}}$ are the corresponding total errors, which are calculated with Equations (\ref{equ:sys}) - (\ref{equ:flux and weight}). The distribution of $F_{\mathrm{diff}}$ is used to calculate the reduced chi-squared based on Equation (\ref{equ:dof}). Figure \ref{fig:contrast crab} shows the Crab light curves and the distribution of $F_{\mathrm{diff}}$. The reduced chi-squared is 0.97 for 79 degrees of freedom (DOF), which indicates that the monitored results of Crab by the two missions are consistent.


\subsection{Data Combining of the Two Missions} \label{sec:DataCom}
\begin{figure*}
    \centering
    \includegraphics[scale=0.8]{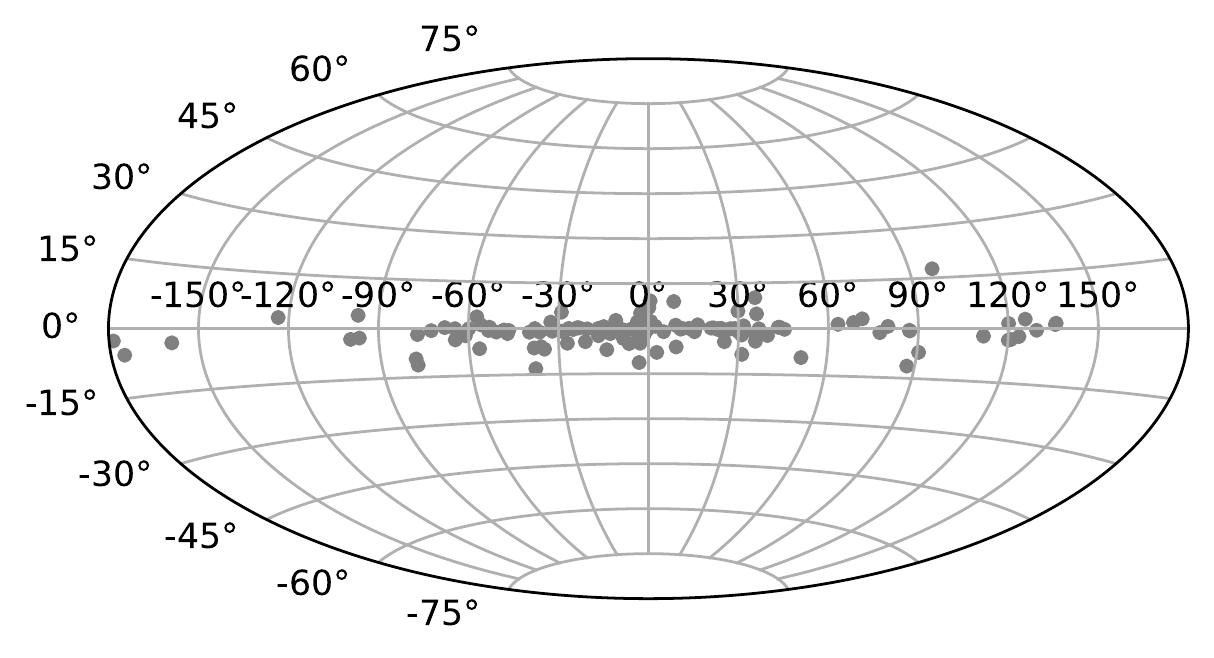}
    \caption{Aitoff projection shows the distribution of sources that are used to combine with MAXI.}
    \label{fig:contrast-source}
\end{figure*}

\begin{figure*}
    \gridline{\fig{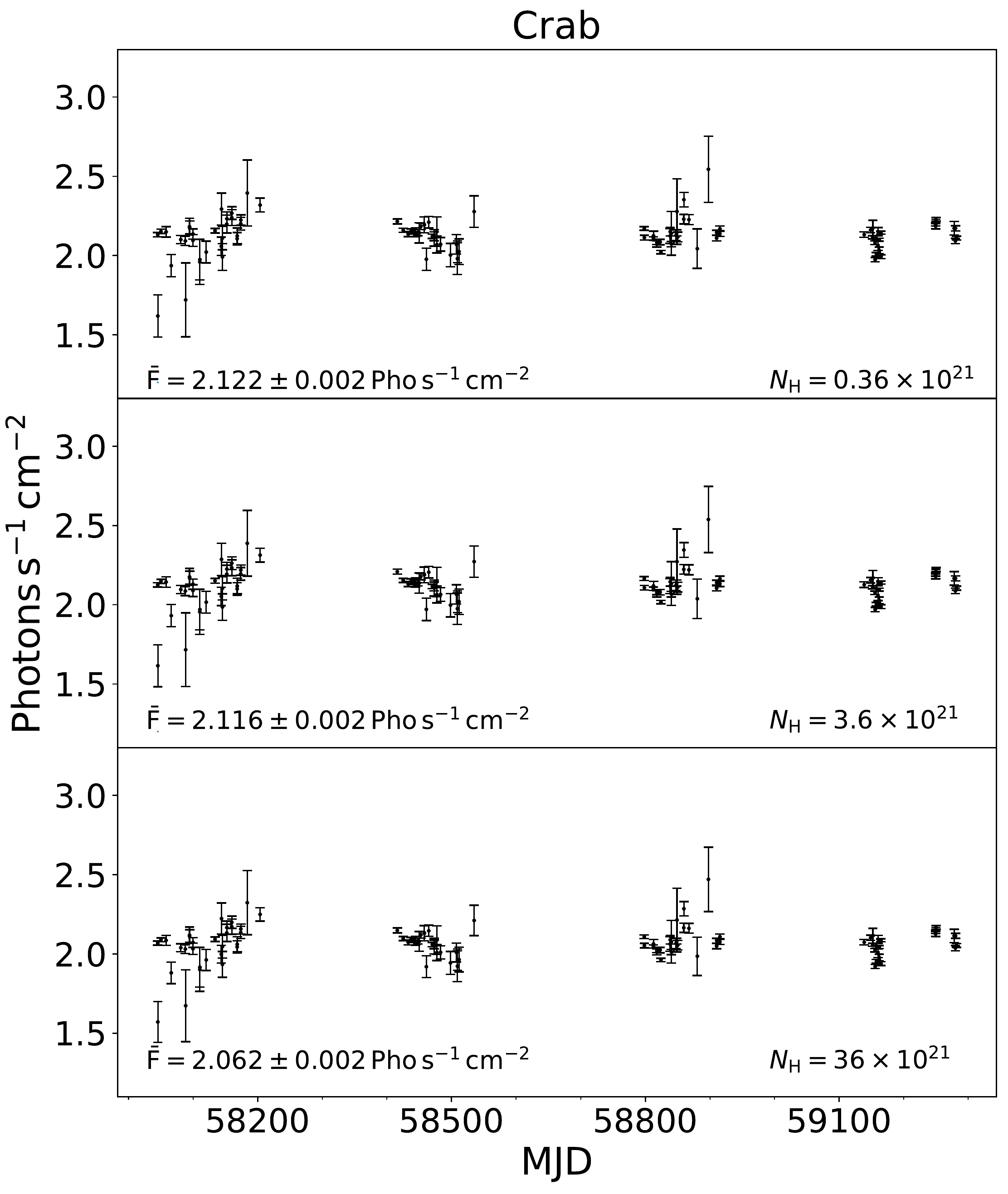}{0.43\textwidth}{(a)}\label{fig:nh1}
              \fig{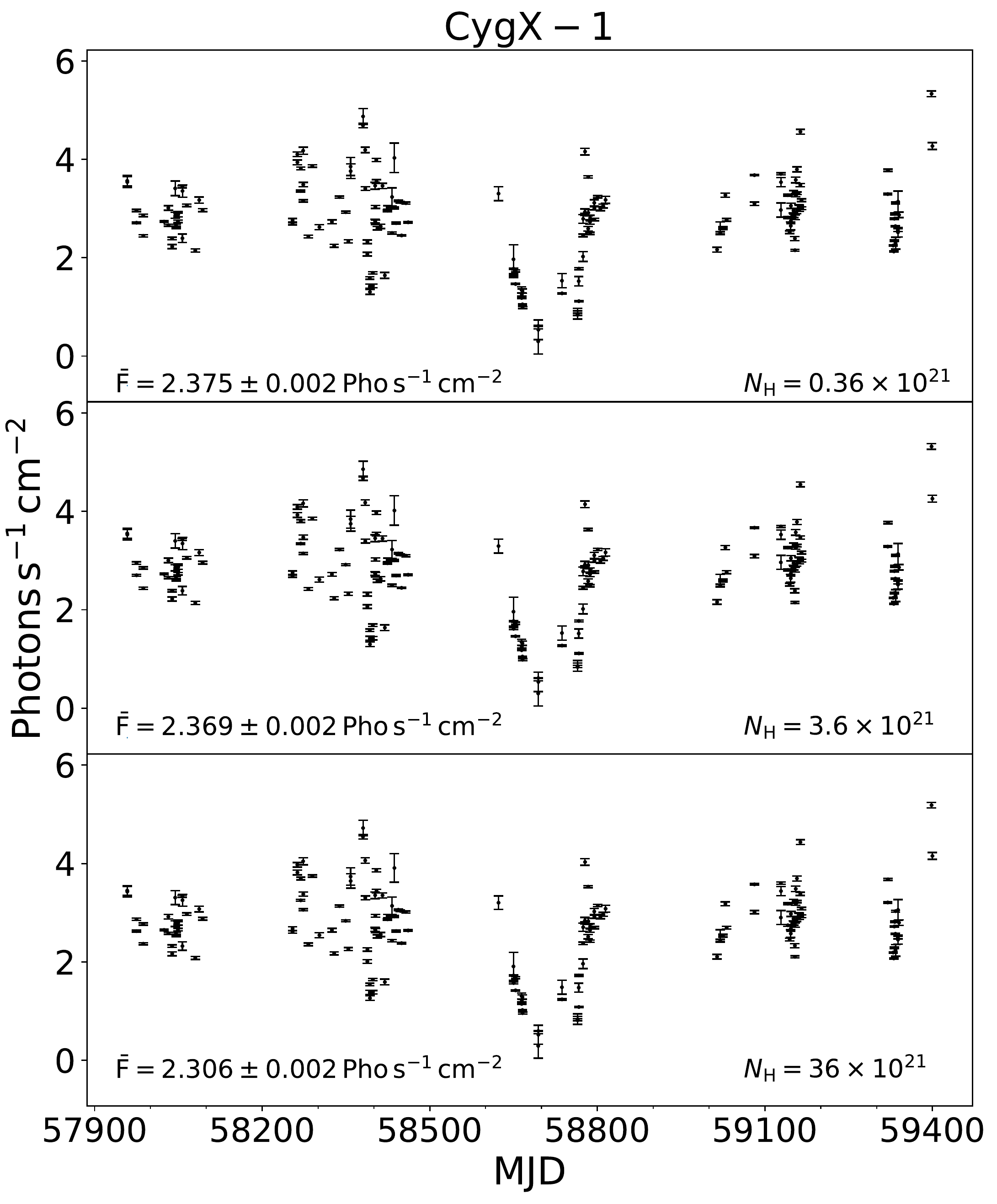}{0.43\textwidth}{(b)}\label{fig:nh2}
              }
    \caption{The long-term light curves of Crab and Cyg\,X$-$1 at 2$-$4\,keV when the interstellar absorption is chosen as $\mathrm{3.6 \times 10^{20}\,cm^{-2}}$, $\mathrm{3.6 \times 10^{21}\,cm^{-2}}$ and $\mathrm{3.6 \times 10^{22}\,cm^{-2}}$, respectively.}
    \label{fig:nh-range}
\end{figure*}

Among the sources with S/N$>$5 (at 2$-$4\,keV) in the 4 yr GPSS catalog of \textit{Insight-HXMT}, a total of 131 sources were monitored by both MAXI and \textit{Insight-HXMT}. Their distribution is shown in Figure \ref{fig:contrast-source}. To obtain more accurate long-term light curves, each sourcce's spectrum is assumed to be an absorbed power-law model with interstellar absorption of $\mathrm{3.6} \times \mathrm{10^{21}}\,\mathrm{cm^{-2}}$. The power-law models (Equation (\ref{equ:powerlaw})) with different $N_0$ and $\mathrm{-\Gamma}$ are used to convolve with the response matrices of \textit{Insight-HXMT} at 2$-$6\,keV, 2$-$4\,keV, and 4$-$6\,keV to obtain the theoretical counts of the corresponding energy band. The calculated results are compared with the detected fluxes at the three energy bands and the model that corresponds to the optimal chi-squared value is determined as the source model. The interstellar absorption is fixed at $\mathrm{3.6} \times \mathrm{10^{21}}\,\mathrm{cm^{-2}}$, because it has little effect on the fitted flux at 2$-$4\,keV when $N_\mathrm{H}$ is fixed with a value that ranges from $\mathrm{3.6} \times \mathrm{10^{20}}\,\mathrm{cm^{-2}}$ to $\mathrm{3.6} \times \mathrm{10^{22}}\,\mathrm{cm^{-2}}$. Here we take Crab and Cyg\,X$-$1 as examples. Their long-term light curves and the averaged fluxes with $N_\mathrm{H}$ selected as $\mathrm{3.6} \times \mathrm{10^{20}}\,\mathrm{cm^{-2}}$, $\mathrm{3.6} \times \mathrm{10^{21}}\,\mathrm{cm^{-2}}$ and $\mathrm{3.6} \times \mathrm{10^{22}}\,\mathrm{cm^{-2}}$ are shown in Figure \ref{fig:nh-range}. We find that the averaged fluxes and long-term light curves vary little with different values of $N_\mathrm{H}$.
\begin{equation}
    N (E) = N_0\,E^{\mathrm{-\Gamma}}\,e^{\mathrm{-\tau}}\,(\mathrm{cts\,cm^{-2}\,s^{-1}\,keV^{-1}}). \label{equ:powerlaw}
\end{equation}

\begin{figure*}
    \gridline{\fig{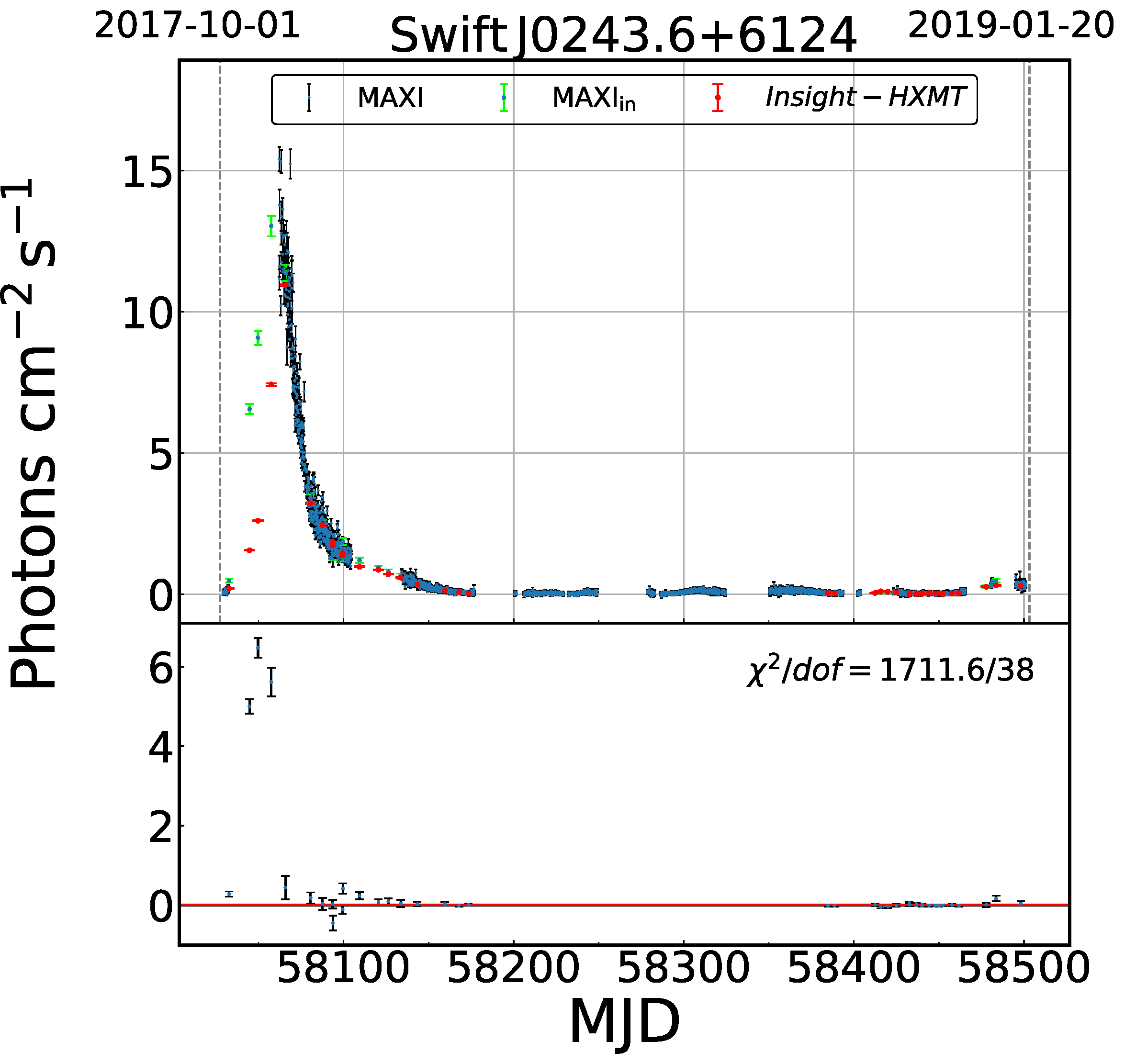}{0.43\textwidth}{(a)}\label{fig:contrast1}
              \fig{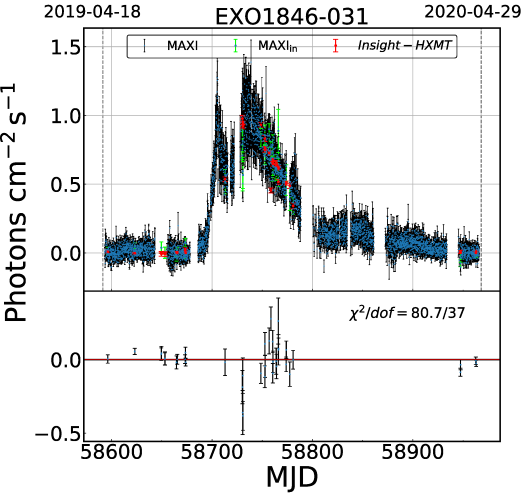}{0.43\textwidth}{(b)}\label{fig:contrast2}
              }
    \gridline{\fig{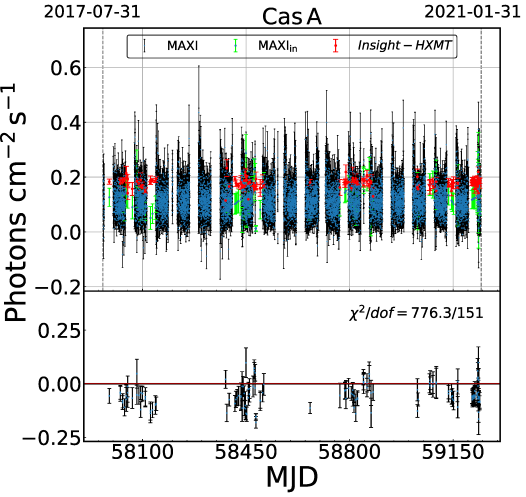}{0.43\textwidth}{(c)}\label{fig:contrast3}   
              \fig{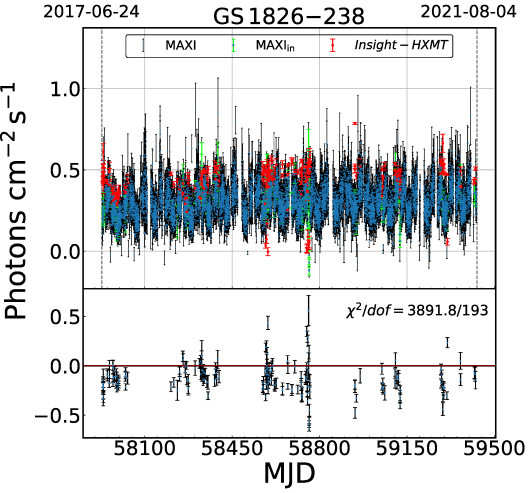}{0.43\textwidth}{(d)}\label{fig:contrast4}
              }
    \caption{Four examples of combined results. The legends for the four panels are the same as those in Figure \ref{fig:contrast crab}.}
    \label{fig:contrast-match}
\end{figure*}

\begin{figure}
    \centering
    \includegraphics[scale=0.32]{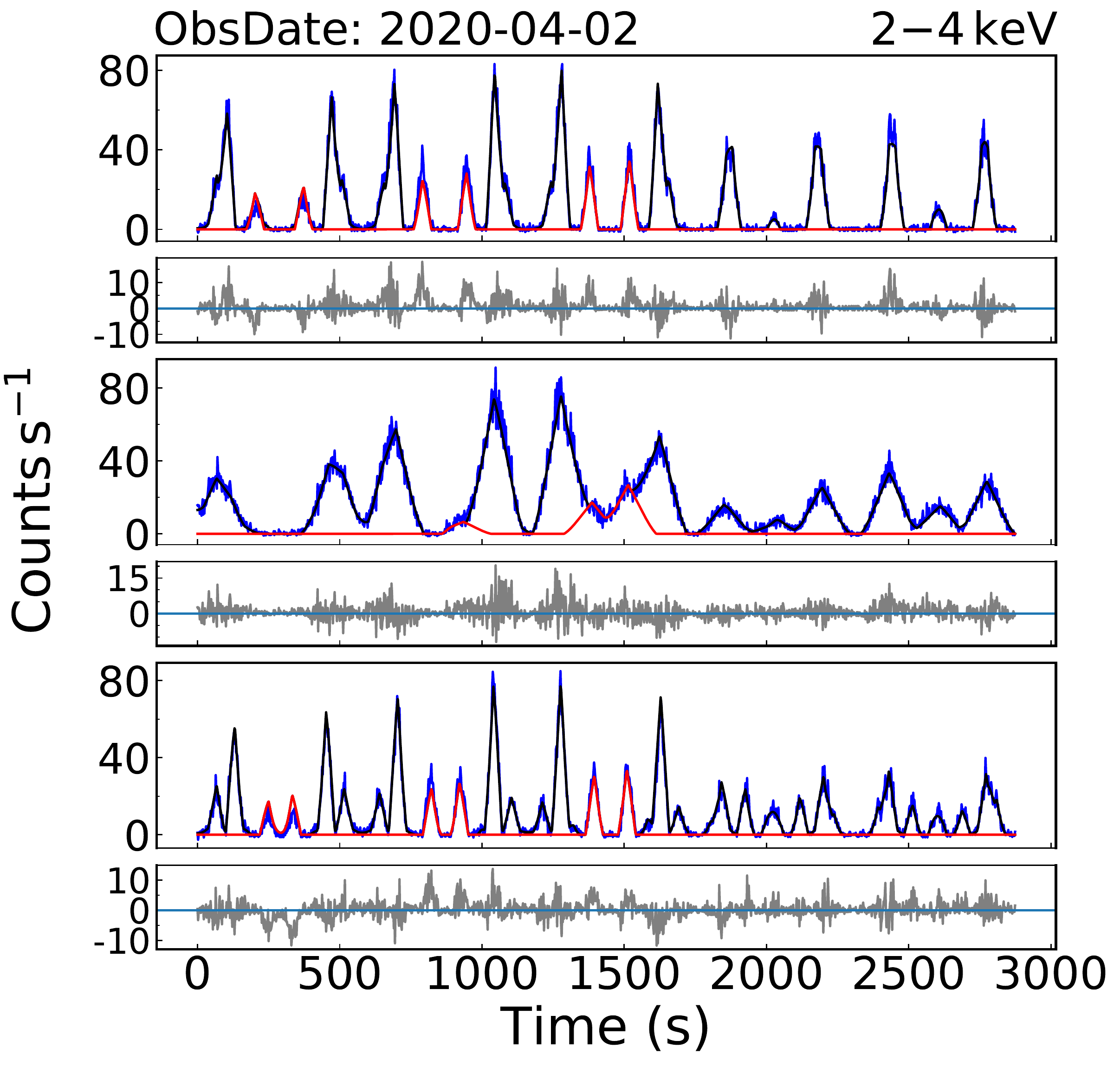}
    \caption{The light curves are from \textit{Insight-HXMT} and observed at MJD 58941.85. The blue and black lines are the light curves at 2$-$4\,keV obtained with LE and best-fit results of bright sources. The red lines represent the contribution of GS\,1826$-$24, and grey lines are residuals.}
    \label{fig:ginga}
\end{figure}

We combine the long-term light curves of the 131 sources and find that $\sim$\,70\% of the sources match well, while others have reduced chi-squared values larger than 2.5. All the combined results are listed in Table \ref{tab:combine-with-maxi}. Four typical combined results are shown in Figure \ref{fig:contrast-match}. One can see that MAXI and \textit{Insight-HXMT} both monitored the outbursts of Swift\,J0243.6$+$6124 and EXO\,1846$-$031, which are shown in panels (a) and (b), respectively. Their results are complementary and the combining can provide more complete outburst histories. The data points of EXO\,1846$-$031 monitored by the two missions are consistent well within the error bars, while the data for Swift\,J0243.6$+$6124 show a large deviation. This is mainly due to a gap in the long-term light curve monitored by MAXI during the outburst phase (MJD 58044 to 58077), while \textit{Insight-HXMT} recorded three flux points during this period. A large deviation is reasonable because the interpolated results cannot accurately restore the outburst lightcurves. It can be seen from panel (c) that the long-term light curve of Cas\,A monitored by \textit{Insight-HXMT} is overall higher than that of MAXI, which may be due to two reasons. The first one is that Cas\,A is an extended source ($\sim$5\,arcmin), and \textit{Insight-HXMT} can record more photons than MAXI due to the larger FOVs. The other possible reason is the complex energy spectrum shape of Cas\,A is far from the power-law model that is assumed by \textit{Insight-HXMT}. In panel (d), the long-term light curves of GS\,1826$-$24 (also called Ginga\,1826$-$24) that are monitored by the two missions are different from each other as well. That GS\,1826$-$24 is a short time-scale variable source is suspected for their different light curves between the two missions. Because MAXI and \textit{Insight-HXMT} both recorded the averaged flux during their exposures, there may be some differences in the recorded results when the source flux is variable. This can be illustrated briefly by the flux at MJD\,58941.85 when \textit{Insight-HXMT} shows the highest flux. The flux monitored by \textit{Insight-HXMT} is 0.78\,$\mathrm{photons\,s^{-1}\,cm^{-2}}$, which is 2.5 times that of by MAXI. Figure \ref{fig:ginga} shows the light curves observed at MJD\,58941.85. In this figure, red solid lines are the best fitting results of GS\,1826$-$24 from \textit{Insight-HXMT}, and grey lines are the corresponding residuals. The residuals change from over-fitting to under-fitting, which means that the source was gradually brighter during this observation. Because the flux of each source is considered constant during the exposure, the fitting results mismatch with the light curves in circumstances where the FOVs cross a source with flux changing. \textit{Insight-HXMT} has the advantage of dealing with such short-timescale variable sources and can obtain a 2-second bin long-term variability of sources based on the light curves and corresponding residuals. The investigation of the short-timescale variable sources by \textit{Insight-HXMT} is beyond the scope of the present work and will be carried out in our following work. 

In conclusion, the long-term light curves of some variable sources recorded by both MAXI and \textit{Insight-HXMT} may show differences, which are understandable and acceptable considering their various time scales and different exposures from the two missions. The data combining of the two missions has the advantage in providing us with more complete long-term light curves of sources.

\section{Statistical Analysis} \label{sec:Analysis}
In the previous sections, the fluxes of the monitored sources were obtained at different energy bands. In this setion, we statistically analyze the flux variabilities based on $F_\mathrm{rms}$ and calculate HRs to investigate the relations between HRs and the source types, as well as those between HRs and the source spatial distributions.

\subsection{Flux Variability}
\begin{figure*}
    \gridline{\fig{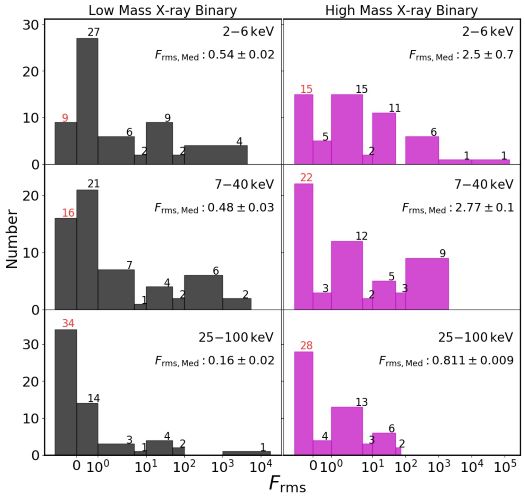}{0.45\textwidth}{(a)}
              \fig{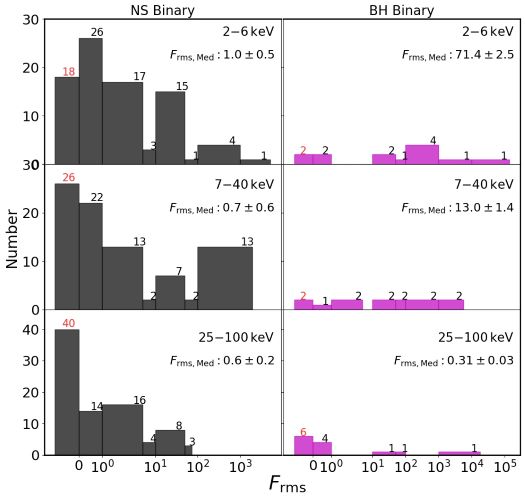}{0.45\textwidth}{(b)}              
             }
    \gridline{
              \fig{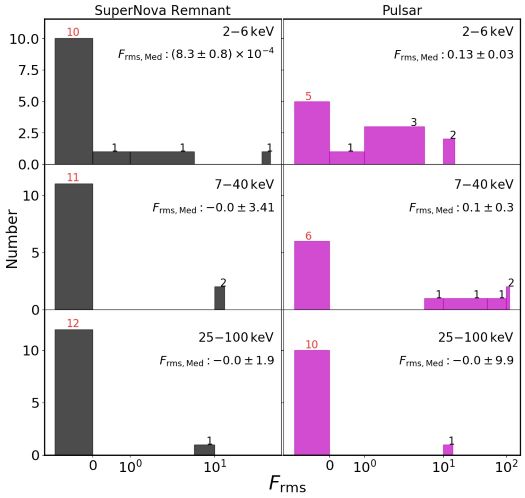}{0.45\textwidth}{(c)}
              \fig{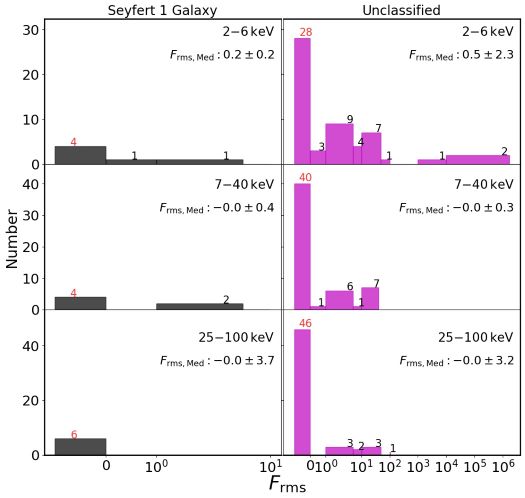}{0.45\textwidth}{(d)}
              }
    \caption{The histogram distribution of $F_\mathrm{rms}$ for the sources that have S/N greater than 5 at 2$-$100\,keV. The X and Y axes are the $F_\mathrm{rms}$ and the source number. The number of sources in each $F_\mathrm{rms}$ interval is labelled above the corresponding bar and the red numbers are the number of flux stable sources.} \label{fig:rms-all}
\end{figure*}

We use $F_{\mathrm{rms}}$ to quantify the flux variability and analyze the properties of the corresponding source types of the 223 sources. When $F_\mathrm{rms}-dF_{\mathrm{rms}}$ of a source is less than or equal to $F_{\mathrm{rms}}$ of Crab at the corresponding energy band, the source is considered as a flux stable one at that band; otherwise it is a flux variable one. Figure \ref{fig:rms-all} shows the distribution of $F_{\mathrm{rms}}$ for these sources at 2$-$6\,keV, 7$-$40\,keV and 25$-$100\,keV and the number of flux stable sources is colored in red in each panel. The X-ray binaries in Tables \ref{tab:catalog1} and \ref{tab:catalog2} are LMXBs and HXMBs, respectively. They are classified into neutron star binaries (NSBs) and black hole binaries (BHBs) based on the compact objects. The distributions of their $F_{\mathrm{rms}}$ are shown in panels (a) and (b). The median $F_{\mathrm{rms}}$\footnote{The median $F_{\mathrm{rms}}$ refers to the middle $F_{\mathrm{rms}}$ if the data count is odd;\\otherwise the weighted average of the middle two values of $F_{\mathrm{rms}}$.}  at different energy bands are utilized to roughly characterize the trend of $F_{\mathrm{rms}}$, and they are labeled in Figure \ref{fig:rms-all}. As can be seen, the fluxes of HMXBs are more variable than those of LMXBs in the three bands and the fluxes of BHBs are more variable than NSBs at 2$-$6\,keV and 7$-$40\,keV. In addition, the values of $F_{\mathrm{rms}}$ of LMXBs, NSB, and BHBs tend to decrease as the energy band increases, which are shown in panels (a) and (b). The fluxes of the SNRs, isolated pulsars, and Seyfert 1 galaxies are more stable than those of X-ray binaries. While several sources in the three types have large $F_{\mathrm{rms}}$, which may be due to the low averaged fluxes or no detected signals above 5\,$\sigma$ (Tables \ref{tab:catalog3} - \ref{tab:catalog5}).

\begin{figure*}
    \gridline{\fig{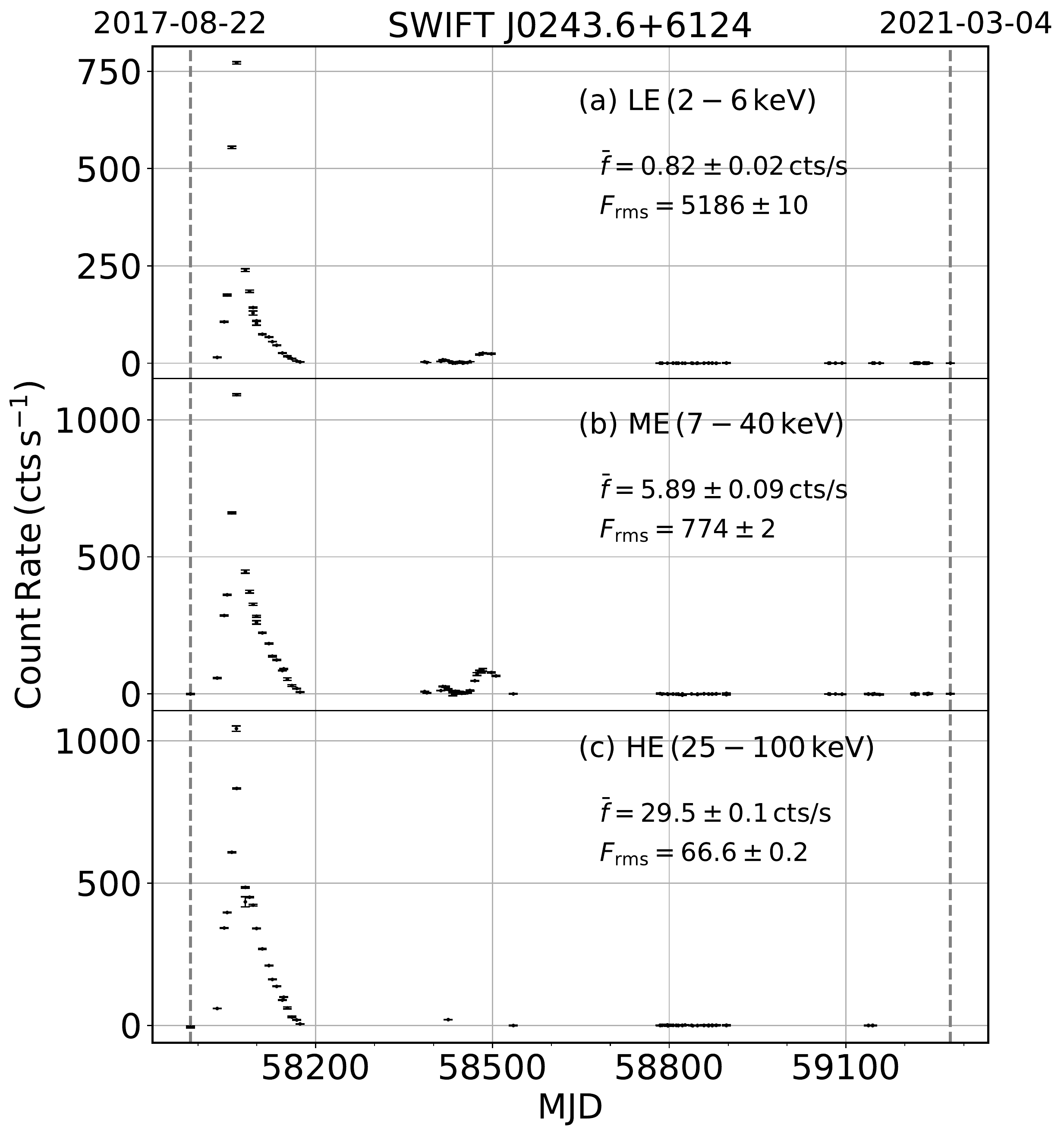}{0.46\textwidth}{(a)}
              \fig{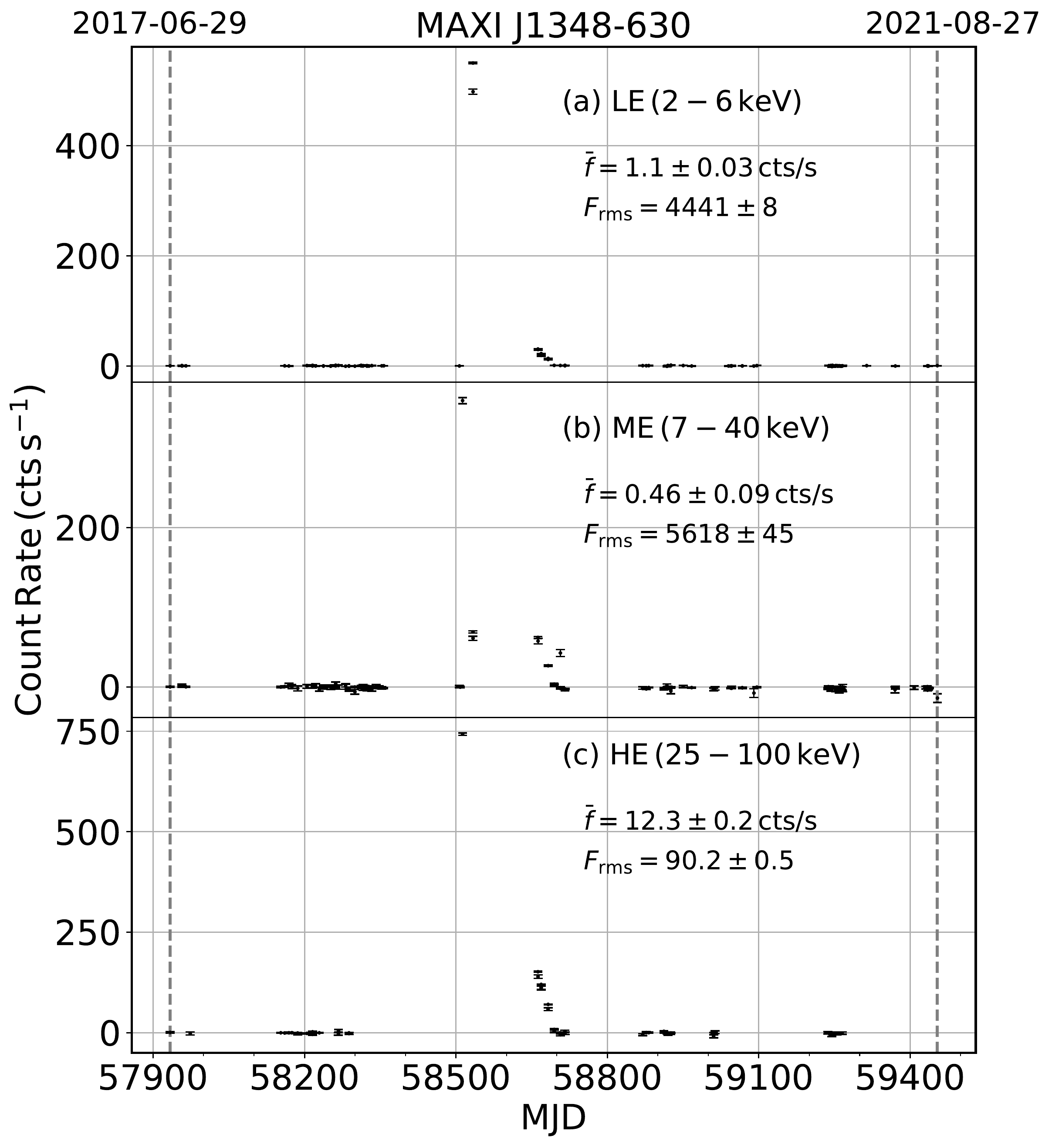}{0.46\textwidth}{(b)}  
             }
    \caption{The long term light curves of Swift\,J0243.6$+$6124 and MAXI\,J1348$-$630.} \label{fig:long-term-twosource}
\end{figure*}

\begin{figure*}
    \gridline{\fig{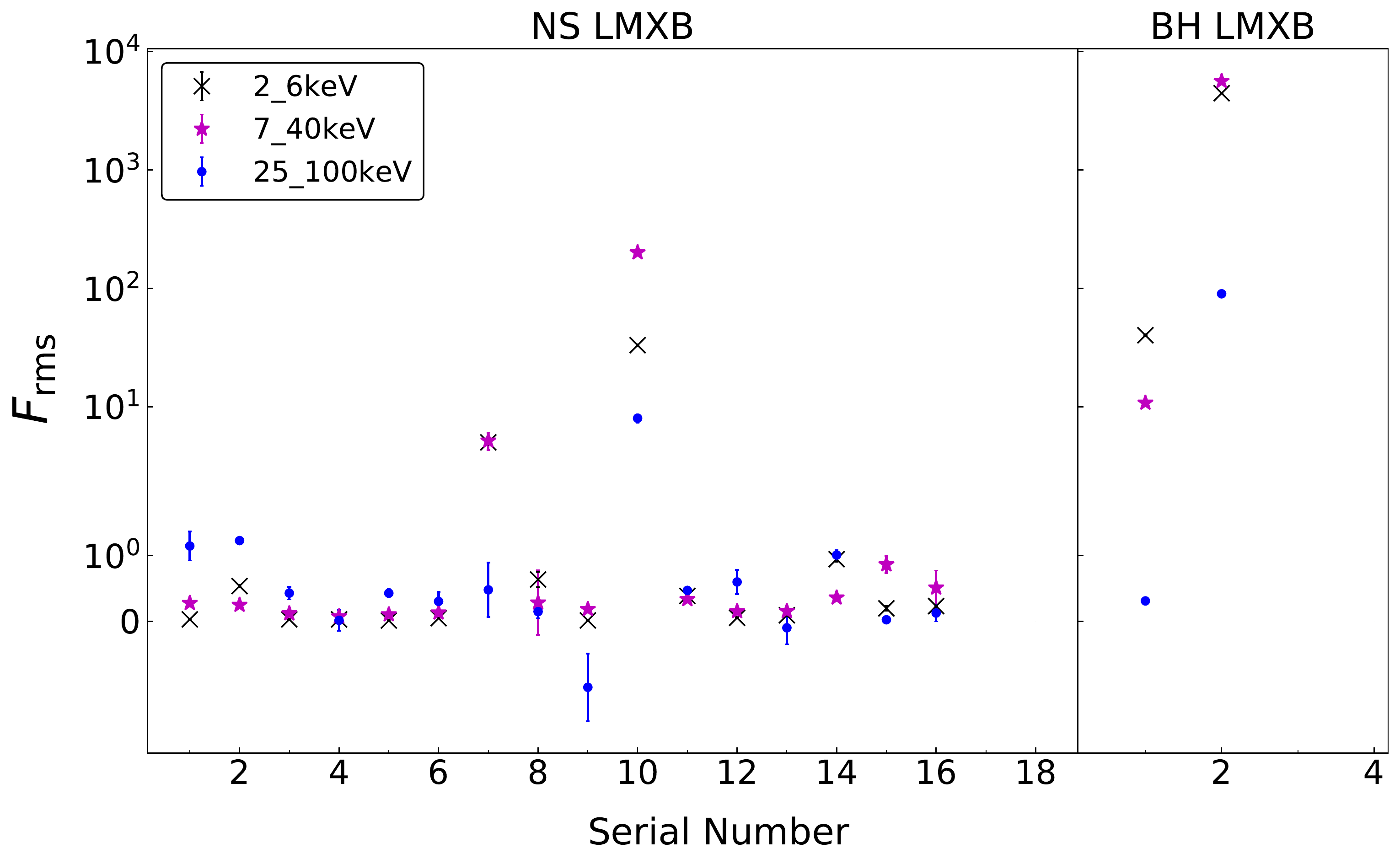}{0.46\textwidth}{(a)}
              \fig{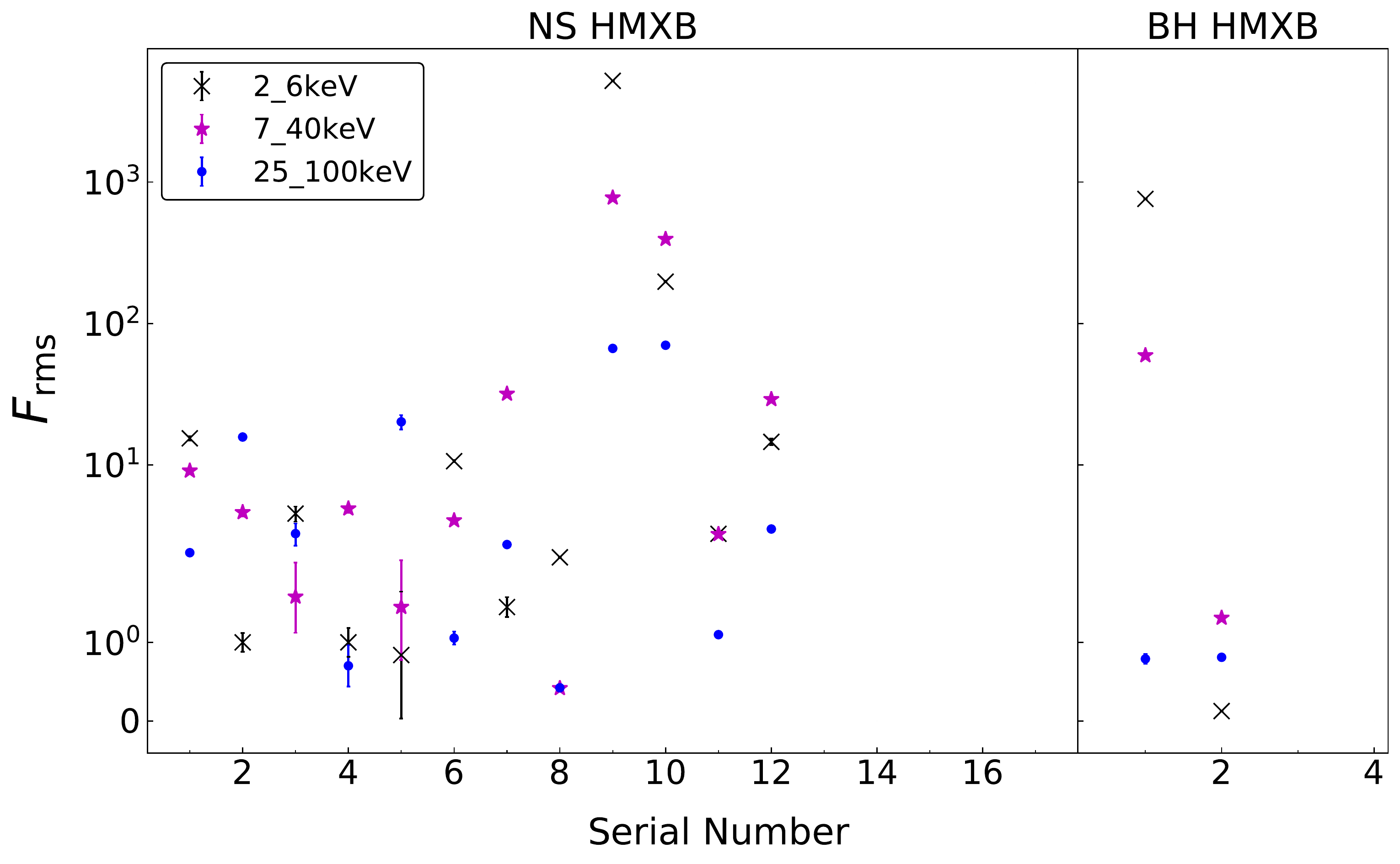}{0.46\textwidth}{(b)}
                }
    \gridline{
              \fig{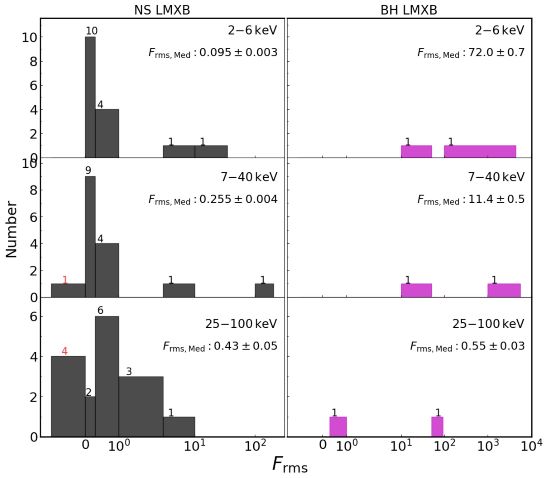}{0.46\textwidth}{(c)}
              \fig{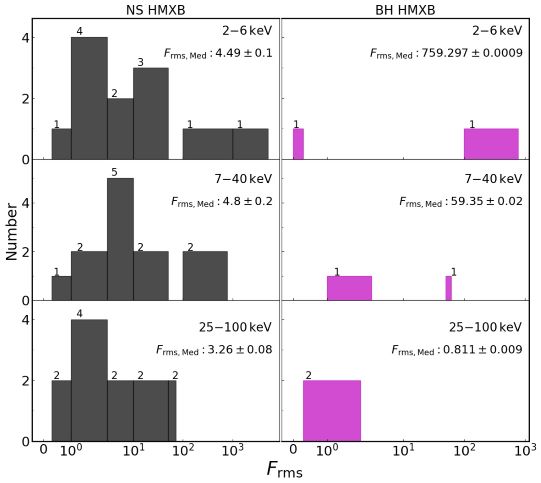}{0.46\textwidth}{(d)}
              }
    \caption{$F_\mathrm{rms}$ distribution for the 32 bright X-ray binaries that have S/N greater than 5 at 2$-$6\,keV, 7$-$40\,keV and 25$-$100\,keV. These sources are classified into LMXB and HMXB based on the mass of donor star and classified into NSB and BHB based on the compact object. The source types and compact objects are labelled on each panel. In panels (a) and (b), the X and Y axes are the serial number and $F_\mathrm{rms}$, respectively. The black cross, magenta pentagram, and blue points represent the $F_\mathrm{rms}$ at 2$-$6\,keV, 7$-$40\,keV, and 25$-$100\,keV for each source, respectively. Vertical error bars represent the error of the $F_\mathrm{rms}$. Panels (c) and (d) are the corresponding histgrams of (a) and (c).}\label{fig:rms}
\end{figure*}

The 33 bright sources metioned in Secction \ref{sec:GPSSresults} include 18 LMXBs (16 NSBs and two BHBs), 14 HMXBs (12 NSBs and 2 BHBs), and one SNR (Crab). Because of the low quantity of SNR, we do not analyze it in detail. The flux variability of the rest 32 X-ray binaries are analyzed carefully. Swift\,J0243.6$+$6124 is a NS HMXB \citep{2018AAS...23116004C}. The outburst was during 2017$-$2018 and recorded by \textit{Insight-HXMT}. Its long-term light curves at three energy bands are shown in panel (a) of Figure \ref{fig:long-term-twosource}, along with the averaged fluxes and $F_{\mathrm{rms}}$ labelled on the corresponding subpanels. MAXI\,J1348$-$630 is a BH LMXB classified recently in 2019 \citep{2022MNRAS.517L..21C} and has large $F_{\mathrm{rms}}$ at the three energy bands. Its long-term light curves recorded by \textit{Insight-HXMT} are shown in panel (b) of Figure \ref{fig:long-term-twosource}, from which it can be seen that the outburst of MAXI\,J1348$-$630 concentrates in 2019. Figure \ref{fig:rms} shows the distribution of $F_{\mathrm{rms}}$ of all the 32 sourcecs at the three energy bands. It is revealed that the values of $F_\mathrm{rms}$ of 15 NS LMXBs at 2$-$6\,keV are lower than those of 7$-$40\,keV or 25$-$100\,keV in panel (a), and the median $F_{\mathrm{rms}}$ of LMXBs tends to increase as the energy band increases in panel (c). This is consistent with \cite{1984PASJ...36..741M}: the energy spectrum of an NS LMXB is composed of a multicolor spectrum from an optically-thick accretion disk and a blackbody spectrum from the neutron star surface, and the former is radiatively stable, while the latter is highly variable. However, the values of $F_\mathrm{rms}$ of most NS LMXBs at 7$-$40\,keV are not larger than those at 25$-$100\,keV, and even one has $F_\mathrm{rms}$ at 2$-$6\,keV larger than that at 7$-$40\,keV and 25$-$100\,keV. It seems that some NS LMXBs are outliers that the $F_\mathrm{rms}$ at higher energy band is larger than that at lower energy band. However, the fact that both 7$-$40\,keV and 25$-$100\,keV contain the 25$-$40\,keV band and the selection range for the 2$-$6\,keV may lead to inaccurate results. In contrast, $F_\mathrm{rms}$ varies more complicatedly at the three energy bands in HMXBs: (a) The flux variation of HMXBs is more active than that of LMXBs at any of the three energy bands: The values of $F_\mathrm{rms}$ of NS LMXBs are lower than those of NS HMXBs at each energy band, and the values of $F_\mathrm{rms}$ of BH LMXBs are lower than those of BH HMXBs at each energy band; (b) $F_\mathrm{rms}$ of the NS HXMBs shows a tendency to get larger initially and then smaller as the energy band increases. These may be influenced by the choice of energy bands, although it cannot be excluded that they are related to the accretion processes of the HXMBs. As a binary stellar system is composed of an accreting compact object and an early type massive star, HMXBs can be subdivided into three subclasses according to the differences of companion stars: (a) Be/X ray binaries, (b) supergiant X-ray binaries, (c) supergiant fast X-ray transients. The compact objects in different classes are powered by mass accretion via Roche-lobe overflow or stellar wind capture \citep{1989PASJ...41....1N, 2006ApJ...646..452S}, which may lead to more diverse outbreaks.

\subsection{Estimation of Hardness Ratio (HR)}
We calclate HRs of the sources with averaged S/N larger than 5 both at H-band and S-band and statistically study the sources with S/N of HRs larger than 3. The source numbers are 49 and 41 when the S-band is fixed as 2$-$6\,keV and H-band is selected as 7$-$40\,keV and 25$-$100\,keV, respectively. Their information can be found in Tables \ref{tab:catalog1} - \ref{tab:catalog6}.

\subsubsection{Relations between HRs and $F_\mathrm{rms}$}
\begin{figure*}
    \gridline{\fig{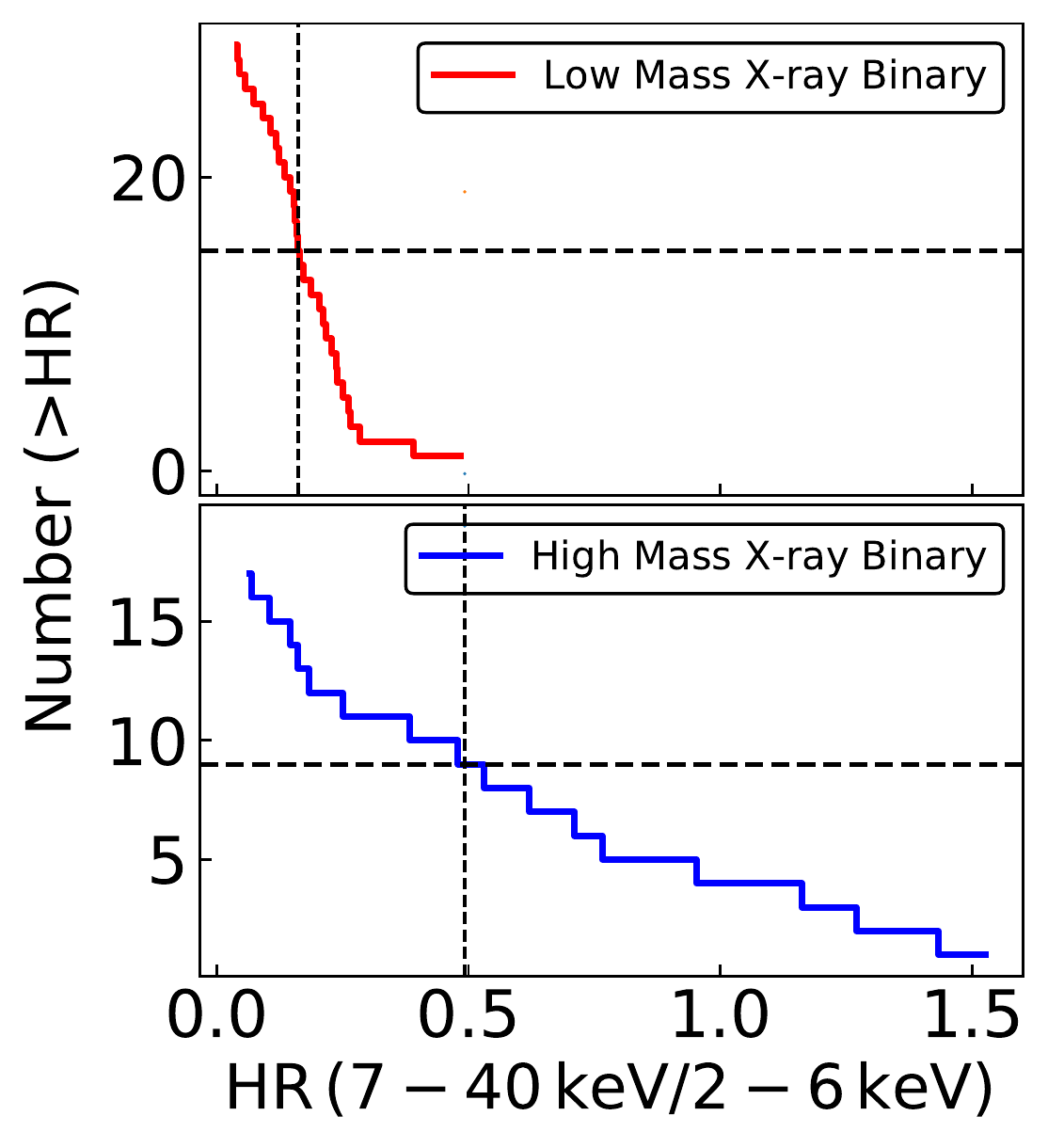}{0.45\textwidth}{(a)}\label{fig:hr-rms1-hh}
              \fig{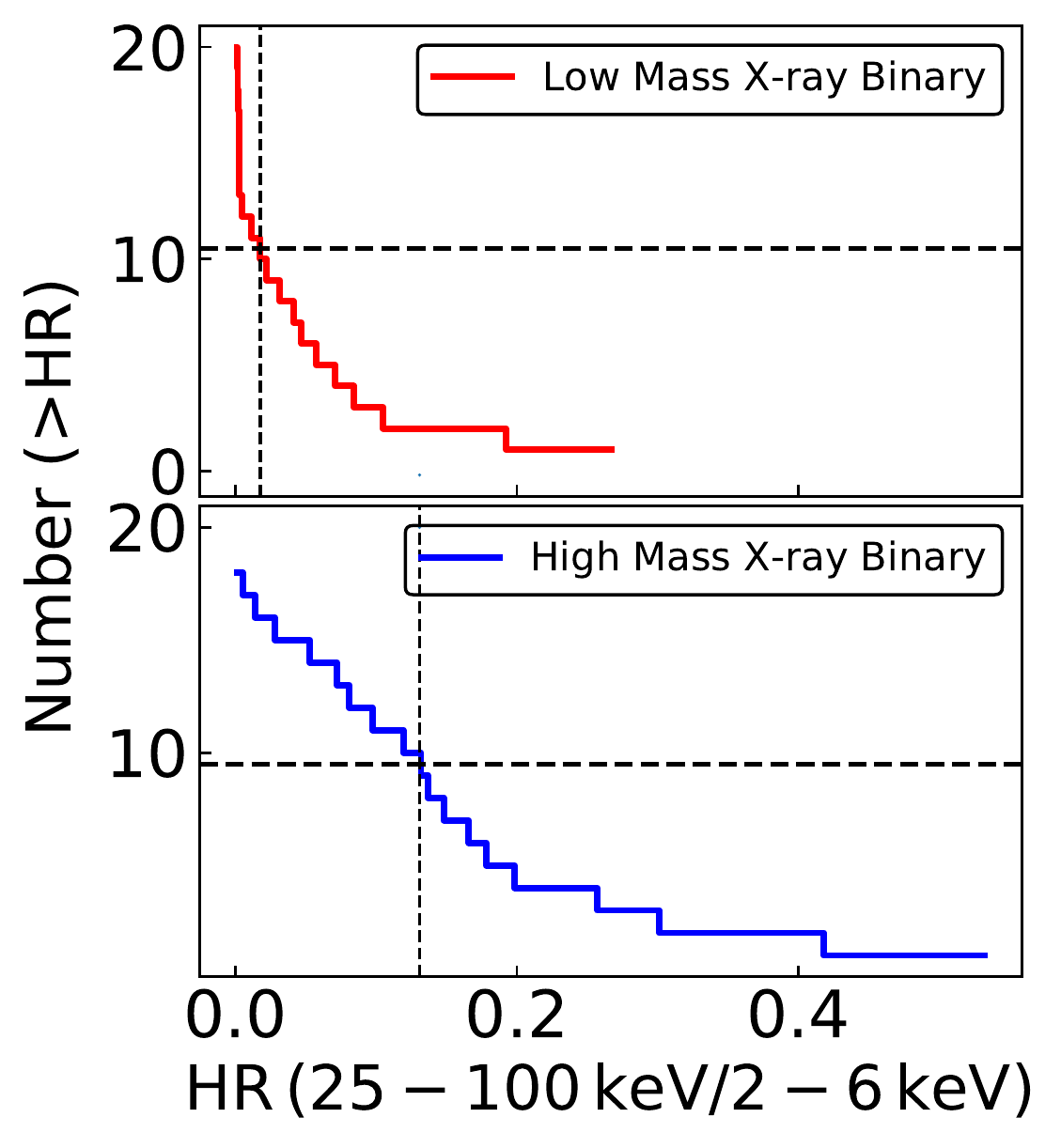}{0.45\textwidth}{(b)}\label{fig:hr-rms2-hh}
              }
    \caption{The relations between HRs and $F_\mathrm{rms}$. H-band is selected as 7$-$40\,keV in panel (a) and 25$-$100\,keV in panel (b).}
    \label{fig:hr-rms}
\end{figure*}

The relationship between HRs and the H-band $F_\mathrm{rms}$ is investigated. Figure \ref{fig:hr-rms} illustrates the HR$-F_\mathrm{rms}$ relationships for different types of sources, and the corresponding median values of HRs and H-band $F_\mathrm{rms}$ are listed in Tabe \ref{tab:HRenergyRuler}. The following can be drawn from Figure \ref{fig:hr-rms} and Table \ref{tab:HRenergyRuler}:
\begin{itemize}
    \item The energy spectra of HMXBs have an overall tendency to be harder than those of LMXBs.
    \item The fluxes of HMXBs are more variable than those of LMXBs.
\end{itemize}
\noindent The sample sizes for other type of sources are too small to do statistical analysis.

\begin{deluxetable*}{L|RRRRRRR}
    \setcounter{table}{4}
    \tablecaption{The Median Value of HR and that of H-band $F_\mathrm{rms}$  (S band: 2$-$6\,keV).} \label{tab:HRenergyRuler}
    \tablewidth{0pt}
    \tablehead{
        \colhead{H-band}   &
        \colhead{LMXB} &
        \colhead{HMXB} &
        \colhead{Pulsar} &
        \colhead{SNR} &
        \colhead{Seyfert1} &
        \colhead{Cluster of Galaxies} &
        \colhead{Unclassified}         
                }
\decimalcolnumbers
    \startdata
        \mathrm{HR_{ 7$-$40\,keV}}   & 0.163 \pm 0.007 &  0.5 \pm 0.2 & $-$   & 0.3193 \pm 0.0007   & $-$   & 0.17 \pm 0.02 & 0.16 \pm 0.02 \\
        F_{\mathrm{rms,7$-$40\,keV}} & 0.35 \pm 0.03   &  4.05 \pm 0.07 &$-$  & 0.0002 \pm 0.0002   & $-$   & 1.5 \pm 0.6 & 1.01 \pm 0.06 \\
        \mathrm{HR_{25$-$100\,keV}}  & 0.017 \pm 0.001 & 0.13 \pm 0.01  & 0.031 \pm 0.005 & 0.072 \pm 10^{-4}   & 0.11 \pm 0.03    & $-$  &$-$   \\
        F_\mathrm{rms,25$-$100\,keV} & 0.43 \pm 0.05    & 2.4 \pm 0.04  & <0    & <0                  & <0    & $-$         & $-$           \\
    \enddata
\tablecomments{\\ $-$ represents there is no this type of source.}
\end{deluxetable*}

\begin{figure*}
    \gridline{\fig{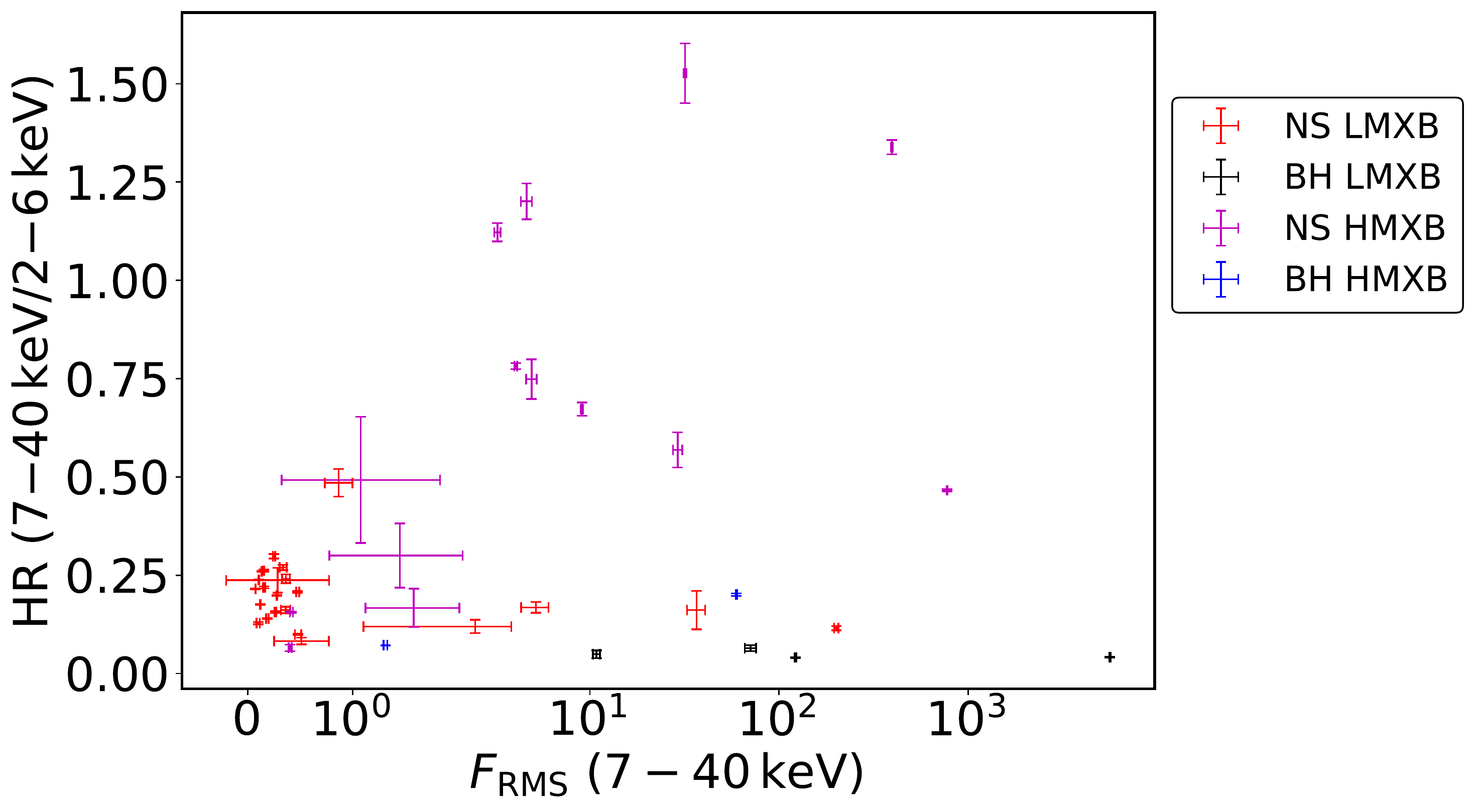}{0.43\textwidth}{(a)}\label{fig:hh}
              \fig{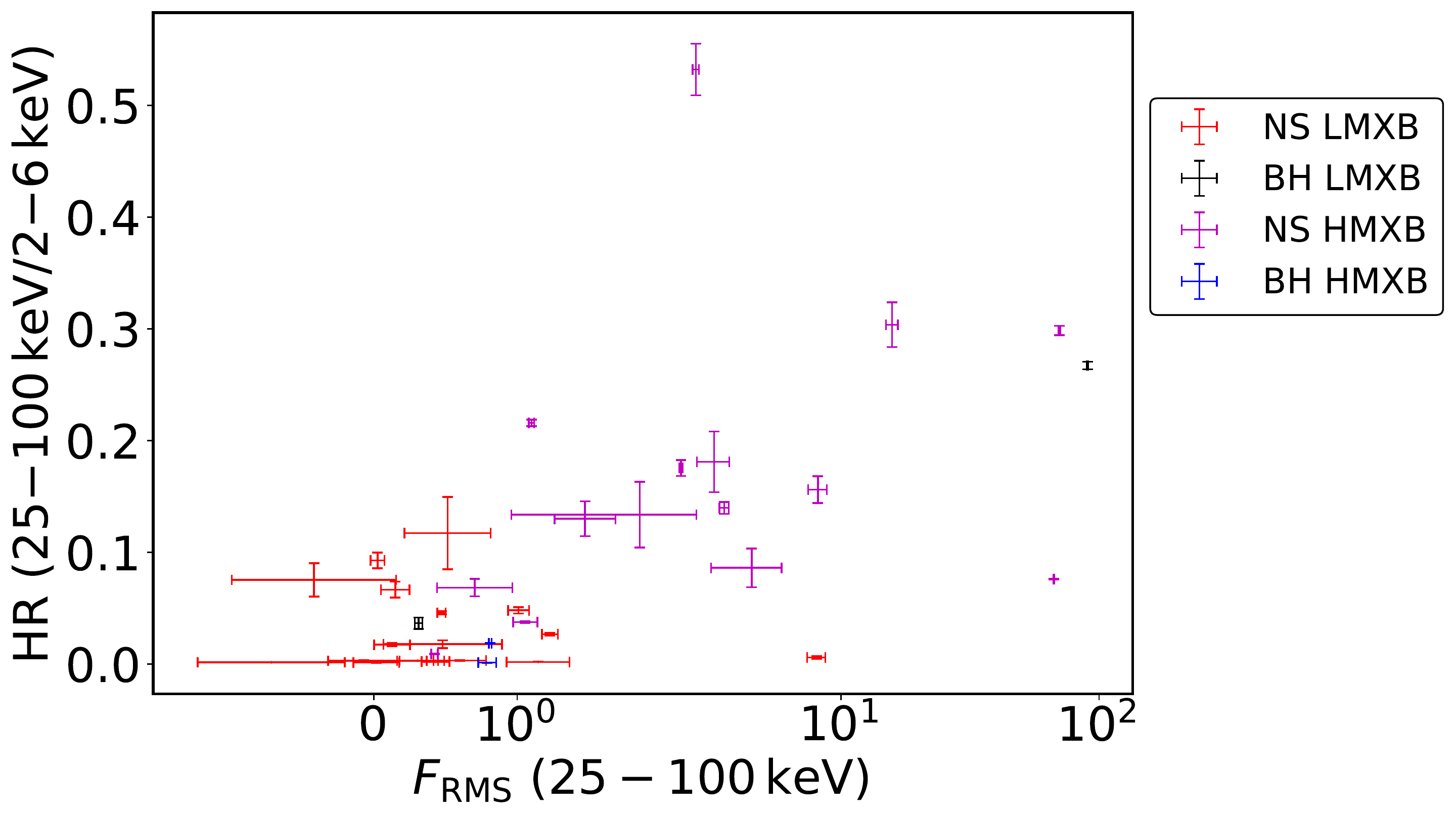}{0.43\textwidth}{(b)}\label{fig:hh2}
              }
    \caption{The relations between HRs and $F_\mathrm{rms}$ for NS and BH binaries.}
    \label{fig:hr-rms-compect}
\end{figure*}

NS and BH binaries are selected for further analysis. Figure \ref{fig:hr-rms-compect} shows the relations between their HRs and $F_\mathrm{rms}$, and the corresponding median values are listed in Table \ref{tab:hr-energy-ruler2}. We can see that:
\begin{itemize}
    \item NS LMXBs are spectrally harder than BH LMXBs when H-band is selected as 7$-$40\,keV, while it is the opposite when 25$-$100\,keV is selected for H-band.
    \item NS HMXBs are spectrally harder than BH HMXBs, no matter which energy band is selected for H-band.
    \item The fluxes of BH LMXBs are more variable than those of NS LMXBs, while the median $F_\mathrm{rms}$ of BH HMXBs are smaller than those of NS HMXBs.
\end{itemize}

\begin{deluxetable*}{L|RRRR}
    \setcounter{table}{5}
    \tablecaption{The Median Value of HR and that of H-band $F_\mathrm{rms}$  (S band: 2$-$6\,keV).} \label{tab:hr-energy-ruler2}
    \tablewidth{0pt}
    \tablehead{
        \colhead{H-band}   &
        \colhead{NS LMXB} &
        \colhead{BH LMXB} &
        \colhead{NS HMXB} &
        \colhead{BH HMXB}        
                }
\decimalcolnumbers
    \startdata
        \mathrm{HR_{ 7$-$40\,keV}}   & 0.1988 \pm 0.0009 &  0.0413 \pm 0.0007  & 0.66 \pm 0.02    & 0.0733 \pm 0.0004              \\
        F_{\mathrm{rms,7$-$40\,keV}} & 0.3 \pm 0.5       &  118 \pm 1          & 4.08 \pm 0.07    & 1.4 \pm 0.02                   \\
        \mathrm{HR_{25$-$100\,keV}}  & 0.011 \pm 0.001   &  0.194 \pm 0.003    & 0.14 \pm 0.005   & 0.01637 \pm 8\times 10^{-5}    \\
        F_\mathrm{rms,25$-$100\,keV} & 0.43 \pm 0.05     &  0.55 \pm 0.03      & 2.74 \pm 0.08    & 0.811 \pm 0.009                \\
    \enddata
\end{deluxetable*}

\subsubsection{Relations between HRs and Source Types}
\begin{figure*}
    \gridline{\fig{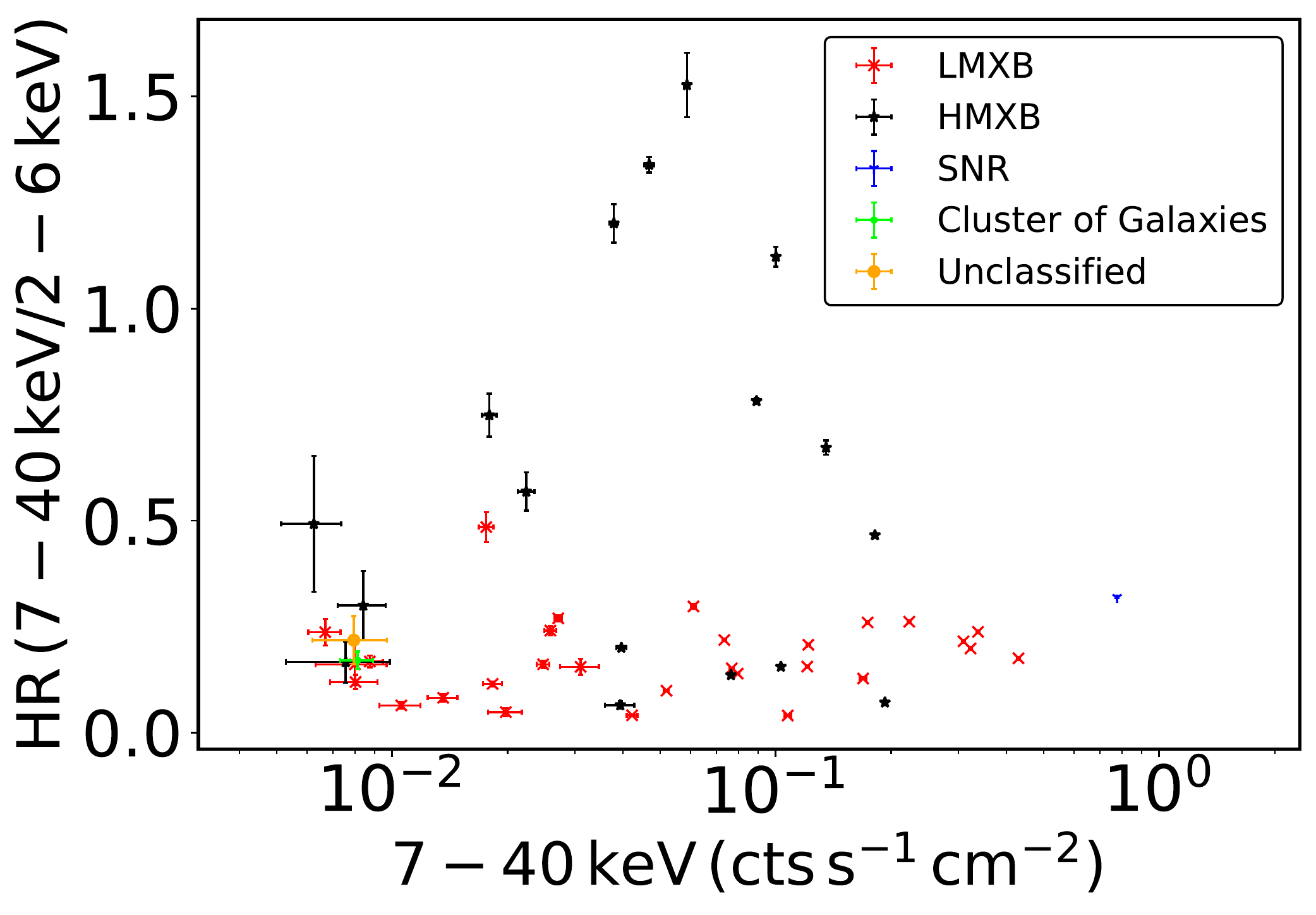}{0.35\textwidth}{(a)}\label{fig:hr-all-num}
              \fig{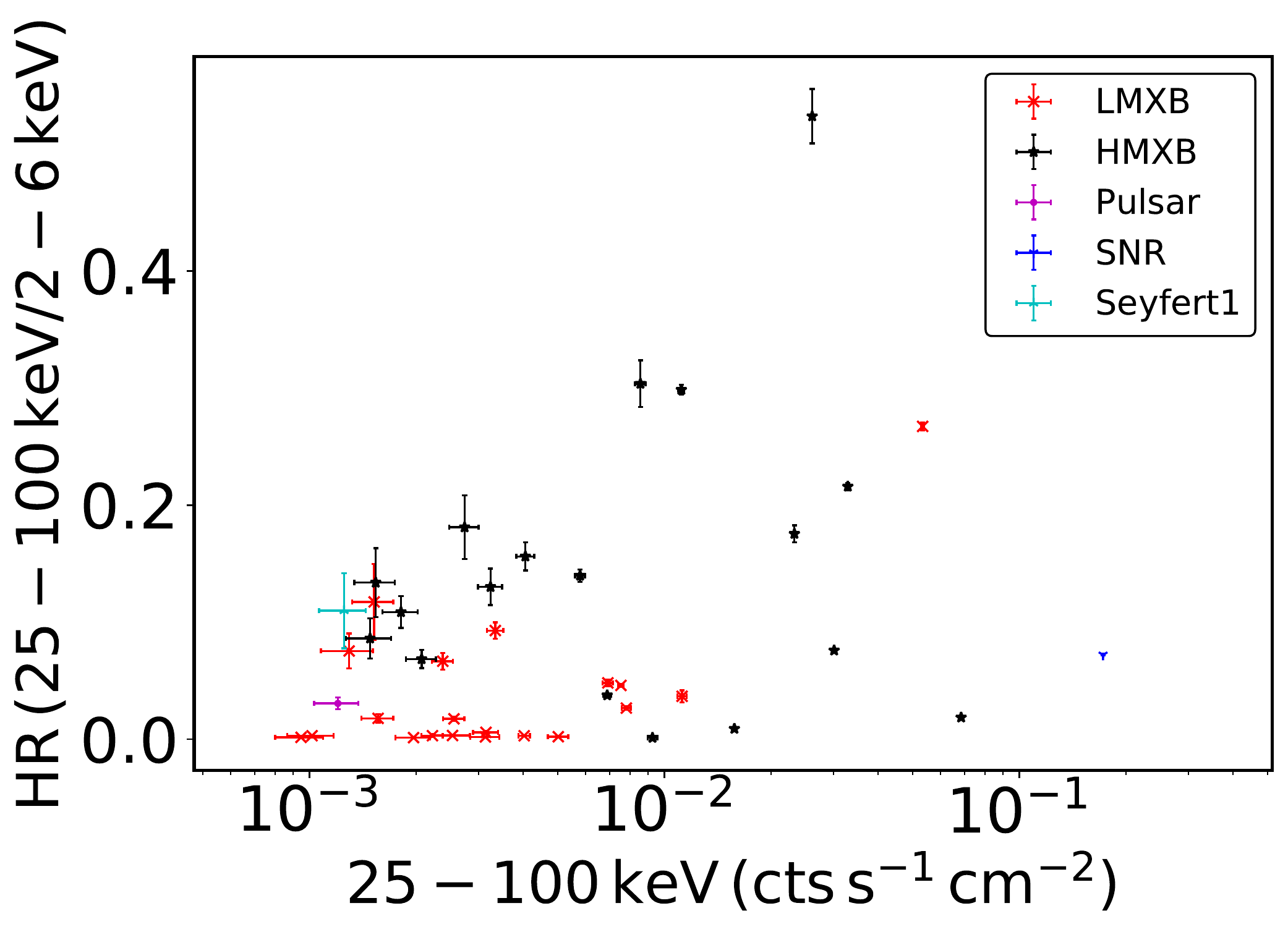}{0.35\textwidth}{(b)}\label{fig:hr-stable-num}
              }
    \caption{The relations between HR and H-band count rates. A color is assigned to each source type.}
    \label{fig:hr-2-6}
\end{figure*}

\begin{figure*}
    \gridline{\fig{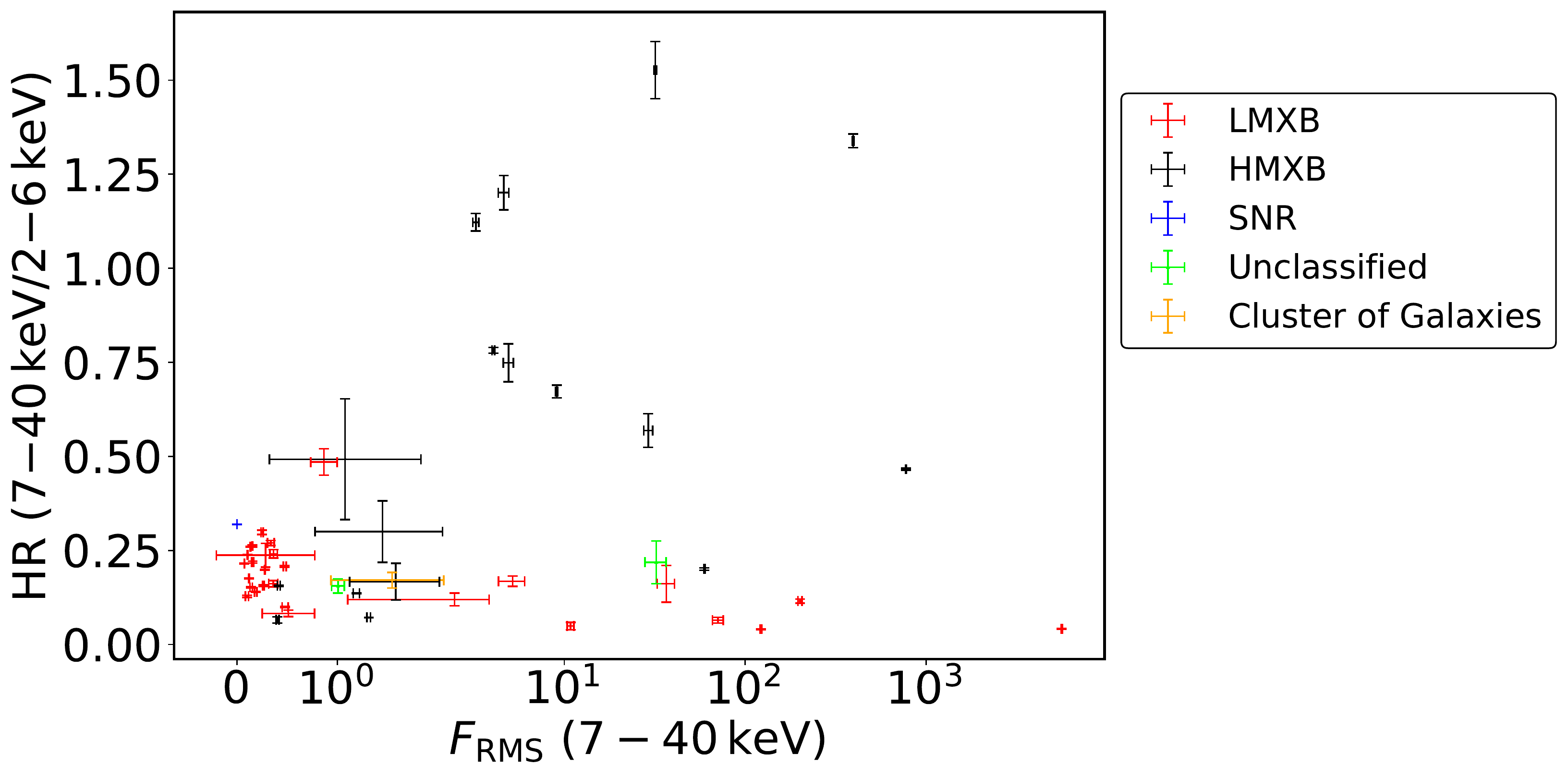}{0.38\textwidth}{(a)}\label{fig:hr-7-40}
			  \fig{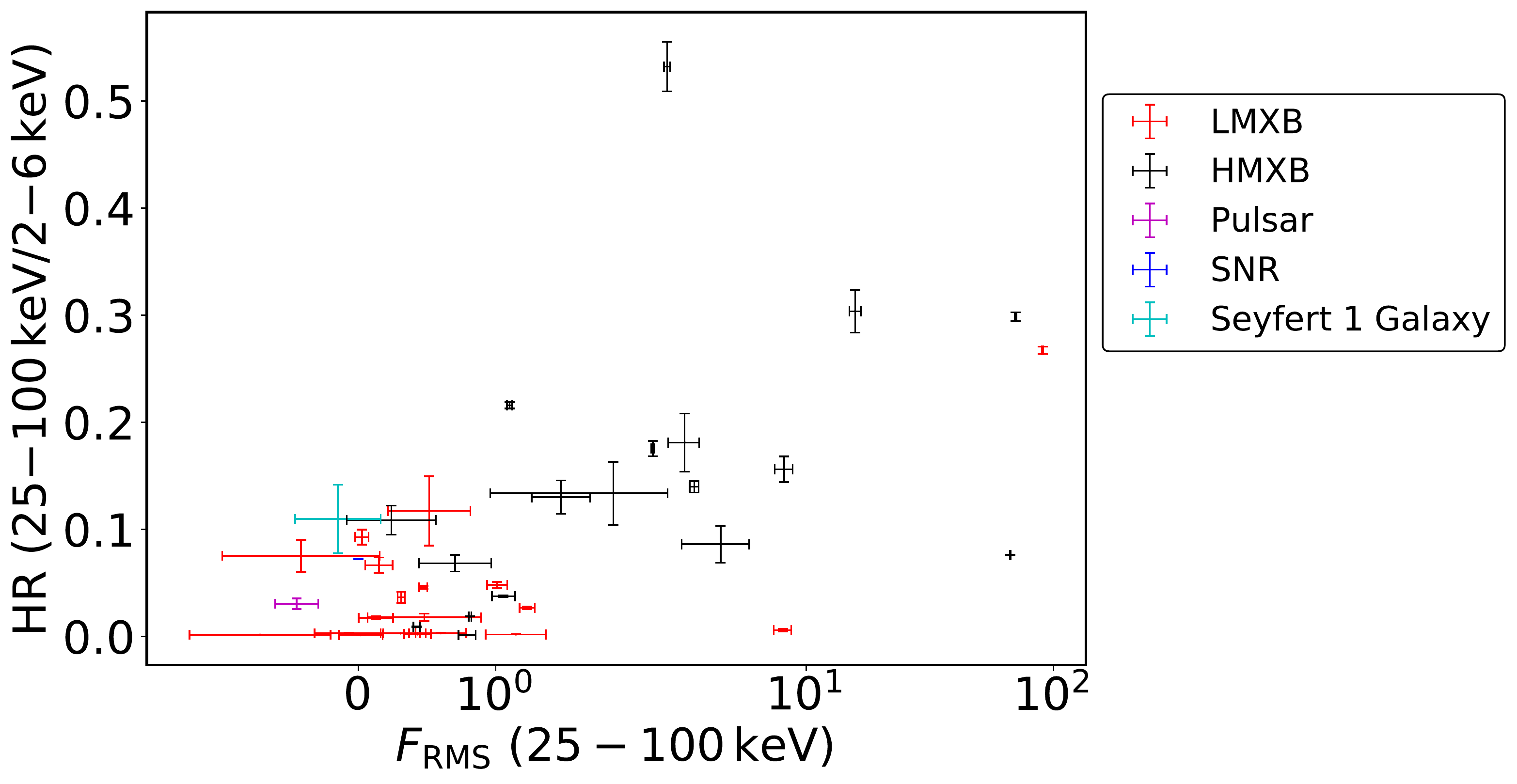}{0.38\textwidth}{(b)}\label{fig:hr-25-100}
              }
    \caption{This figure shows the relations between the cumulative numbers and HRs. X-axis is HR, and Y-axis is the count of sources higher than the value corresponding to X-axis. The black dashed lines are the median values of HRs.}
    \label{fig:hr-2-6-num}
\end{figure*}

In this part, we investigate the possible relations between HRs and source types. Panels (a) and (b) in Figure \ref{fig:hr-2-6} show the relations between HR and H-band count rates when H-band is selected as 7$-$40\,keV and 25$-$100\,keV, respectively. It is clear that HMXBs are overall brighter than LMXBs at 25$-$100\,keV, while it is not obvious at 7$-$40\,keV. X-ray binaries (XRBs) dominate a large portion on the galactic plane \citep{2006ARA&A..44..323F}, which is the same in the two panels. Thus, only the HRs of HMXBs and LMXBs are analyzed. The relation between their HR and source numbers are shown in Figure \ref{fig:hr-2-6-num}. The black dashed lines represent HR medians \footnote{The median refers to the middle HR if data count is odd; otherwise the weighted average of the middle two HRs.}, which are utilized to roughly characterize the trend of HR and are listed in Table \ref{tab:hr-energy-ruler}. Generally, the following conclusions can be drawn:

\begin{enumerate}
    \item HMXBs are overall spectrally harder than LMXBs, no matter which energy range of H-band is selected.
    \item The HR medians of LMXBs and HMXBs become softer as a harder energy range of the H-band is selected.
\end{enumerate}

\begin{deluxetable}{LRR}
    \setcounter{table}{6}
    \tablecaption{The Median Value of HR (S band: 2$-$6\,keV). \label{tab:hr-energy-ruler}}
    \tablewidth{0pt}
    \tablehead{
        \colhead{H band}   &
        \colhead{HMXB} &
        \colhead{LMXB} 
                }
    \startdata
        \mathrm{ 7$-$40\,keV  } & 0.5 \pm 0.2    & 0.161 \pm 0.008      \\
        \mathrm{ 25$-$100\,keV} & 0.13 \pm 0.01  & 0.018 \pm 0.001     \\
    \enddata
\end{deluxetable}

\section{Summery and Conclusion} \label{sec:conclusion}
We have presented the first \textit{Insight-HXMT} catalog at low Galactic latitudes, based on the 4 yr GPSS data mainly at 2$-$6\,keV (LE), $\mathrm{7-40\,keV}$ (ME), and 25$-$100\,keV (HE) bands. The limiting sensitivities of 0.61\,mCrab, 2.35\,mCrab and 2.19\,mCrab are achieved and 1343, 957, and 935 sources have been monitored up to August 2021 at 2$-$6\,keV (LE), 7$-$40\,keV (ME), and 25$-$100\,keV (HE), respectively. 

Combining data from different missions can provide more complete long-term light curves, which are useful to study the nature of these sources. The long-term light curves at 2$-$4\,keV are used to combine with those from MAXI. The light curves of \textit{Insight-HXMT} and MAXI for most sources are consistent well with each other, while some show differences. Two main reasons resulting in those differences are investigated. The first reason comes from the difference in the size of the FOVs of the two missions, which may lead to the different number of photons received from the extended source (such as Cas\,A). The other reason is that an object is a short-time scale variable source, for which \textit{Insight-HXMT} and MAXI both recorded the averaged fluxes. \textit{Insight-HXMT} has advantages in observing short-time scale sources thanks to its large FOVs and scanning strategy.

The 4-year catalog contains 223 sources with S/N greater than 5 at 2$-$100\,keV and 33 sources with S/N greater than 5 at 2$-$6\,keV, 7$-$40\,keV, and 25$-$100\,keV. We have analyzed the flux variabilities for these sources based on their $F_\mathrm{rms}$, and investigated the relations between $F_\mathrm{rms}$ and the source types, and between $F_\mathrm{rms}$ and HRs. The following conclusions are drawn:
\begin{itemize}
    \item During the GPSS of \textit{Insight-HXMT}, most of the SNRs are flux stable. The fluxes of most isolated pulsars and Seyfert 1 Galaxies vary little or are about zero.
    \item The fluxes of BHBs are more variable than those of NSBs.
    \item Most of the values of $F_\mathrm{rms}$ of NS LMXBs at 2$-$6\,keV are lower than those at higher energy bands, which may be related to the composition of the energy spectra of NS LMXBs.
    \item $F_\mathrm{rms}$ of HMXBs varies more complicatedly at different energy bands, which may be due to the complicated accretion processes of HMXBs.
    \item The fluxes of BH LMXBs at 7$-$40\,keV have an overall tendency to be more variable than those of NS LMXBs at the same band, but their HRs (7$-$40\,keV / 2$-$6\,keV) have a softer tendency than NS LMXBs.
    \item The NS HMXBs not only have more variable fluxes than NS LMXBs, but also have harder spectra than NS LMXBs (S-band: 2$-$6\,keV; H-band: 7$-$40\,keV or 25$-$100\,keV).
\end{itemize}

We have studied the relations between HRs and source types. It is found that HMXBs are generally harder than LMXBs, and as a harder energy range of the H-band is selected, the HR of X-ray binaries tends to become softer. 

\begin{acknowledgments}
Chen Wang thank Zeming Zhou and Yufeng Li for their advice and help on improving the manuscript. We thank the referee for her/his pertinent comments and suggestions that helped to improve the quality of our paper. This work made use of the data from the \textit{Insight-HXMT} mission, a project funded by China National Space Administration (CNSA) and the Chinese Academy of Sciences (CAS). The authors thank supports from the National Natural Science Foundation of China under Grants Nos.U1838202, U1838201, U1838105, U1838107, U1838113, the Youth Innovation Promotion Association of the CAS (grant id 2018014), and the National Key R\&D Program of China (Grant No. 2021YFA0718500). This work was partially supported by International Partnership Program of Chinese Academy of Sciences (Grant No.113111KYSB20190020). 
\end{acknowledgments}

\bibliography{main.bib}{}
\bibliographystyle{aasjournal}

\appendix
\section{Catalog of the bright sources and the combined results with MAXI}

\begin{longrotatetable}
\begin{deluxetable*}{LLLRRRRRRRRR}
    \scriptsize
    \setcounter{table}{7}
    \tablecaption{The First Four-year GPSS Catalog of \textit{Insight-HXMT} (Low Mass X-ray Binary). \label{tab:catalog1}}
    \tablewidth{0pt}
    \tabletypesize{}
    \tablehead{
        \colhead{Source name}                                      &
        \colhead{Compact object}                                     &
        \colhead{R.A.}                                             &
        \colhead{Dec.}                                             &
        \colhead{$R_{\mathrm{2-6\,keV}}$}                          &
        \colhead{$R_{\mathrm{7-40\,keV}}$}                         &
        \colhead{$R_{\mathrm{25-100\,keV}}$}                       &
        \colhead{$F_{\mathrm{rms}}1^c$}                            &
        \colhead{$F_{\mathrm{rms}}2^d$}                            &
        \colhead{$F_{\mathrm{rms}}3^e$}                            &
        \colhead{HR1$^a$}                                          &
        \colhead{HR2$^b$}                                                                      
        \\          
        \colhead{ }                                               &
        \colhead{ }                                               &
        \colhead{(deg)}                                           &
        \colhead{(deg)}                                           &
        \colhead{$\mathrm{cts\,s^{-1}}$}                          &
        \colhead{$\mathrm{cts\,s^{-1}}$}                          &
        \colhead{$\mathrm{cts\,s^{-1}}$}                          &
        \colhead{ }                                               &
        \colhead{ }                                               &
        \colhead{ }                                               &
        \colhead{ }                                               &
        \colhead{ }                                 
                }
\decimalcolnumbers
    \startdata
        \textcolor{blue}{\mathrm{GX\,349$+$2}}	&	\mathrm{NS}	&	256.43	&	-36.42	&	89.3\pm0.1	&	54.9\pm0.1	&	2.0\pm0.1	&	0.0315\pm0.0008	&	0.279\pm0.004	&	1.1\pm0.2	&	0.1988\pm0.0009	&	0.0019\pm0.0002	\\
        \mathrm{IGR\,J17091$-$3624}	            &	\mathrm{BH}	&	257.28	&	-36.41	&	1.9\pm0.2	&	<0.5	&	<0.1	&	71\pm2	&	596\pm44	&	<0	&	<0.16	&	<0.11	\\
        \mathrm{SAX\,J1712.6$-$3739}	        &	\mathrm{NS}	&	258.14	&	-37.64	&	0.36\pm0.05	&	<0.3	&	<0.6	&	12\pm1	&	1970\pm316	&	23\pm23	&	<0.24	&	<0.08	\\
        \mathrm{IGR\,J17254$-$3257}	            &	\mathrm{NS}	&	261.35	&	-32.95	&	0.21\pm0.03	&	<-0.22	&	<0.5	&	11\pm2	&	1.7\pm1.0	&	<0	&	<0.43	&	<0.07	\\
        \mathrm{3A\,1728$-$169}	                &	$-$	        &	262.93	&	-16.96	&	28.48\pm0.06	&	12.6\pm0.1	&	<1.9	&	0.0075\pm0.0007	&	0.14\pm0.01	&	<0	&	0.152\pm0.002	&	0.0027\pm0.0005	\\
        \textcolor{blue}{\mathrm{GX\,354$-$0}}	&	\mathrm{NS}	&	262.99	&	-33.83	&	8.1\pm0.07	&	10.97\pm0.08	&	5.82\pm0.1	&	0.54\pm0.02	&	0.25\pm0.01	&	1.23\pm0.06	&	0.298\pm0.006	&	0.027\pm0.001	\\
        \mathrm{1H\,1746$-$370}	                &	$-$	        &	267.55	&	-37.05	&	1.61\pm0.02	&	<0.1	&	<0.4	&	0.58\pm0.05	&	1\pm3	&	147\pm159	&	0.14\pm0.02	&	<0.046	\\
        \mathrm{H\,1755$-$338}	                &	\mathrm{BH}	&	269.67	&	-33.81	&	1.0\pm0.03	&	<-1.11	&	<0.2	&	14.2\pm0.5	&	<0	&	<0	&	<1.4	&	<0.45	\\
\enddata
\tablecomments{\\$^a$ and $^b$ are HR (7$-$40\,keV, 2$-$6\,keV) and HR (25$-$100\,keV, 2$-$6\,keV) calculated with Equation \ref{equ:hr}, respectively.\\
                 $^c$-$^e$ Excess variances calculated with Equation \ref{equ:excess} at 2$-$6keV, 7$-$40\,keV and 25$-$100\,keV, respectively.\\    
                 }
\end{deluxetable*}
\end{longrotatetable}
    
\begin{longrotatetable}
\begin{deluxetable*}{LLLRRRRRRRRR}
    \scriptsize
    \setcounter{table}{8}
    \tablecaption{The First Four-year GPSS Catalog of \textit{Insight-HXMT} (High Mass X-ray Binary). \label{tab:catalog2}}
    \tablewidth{0pt}
    \tabletypesize{}
    \tablehead{
        \colhead{Source name}                                      &
        \colhead{Compact object}                                     &
        \colhead{R.A.}                                             &
        \colhead{Dec.}                                             &
        \colhead{$R_{\mathrm{2-6\,keV}}$}                          &
        \colhead{$R_{\mathrm{7-40\,keV}}$}                         &
        \colhead{$R_{\mathrm{25-100\,keV}}$}                       &
        \colhead{$F_{\mathrm{rms}}1^c$}                            &
        \colhead{$F_{\mathrm{rms}}2^d$}                            &
        \colhead{$F_{\mathrm{rms}}3^e$}                            &
        \colhead{HR1$^a$}                                          &
        \colhead{HR2$^b$}                                                                      
        \\          
        \colhead{ }                                               &
        \colhead{ }                                               &
        \colhead{(deg)}                                           &
        \colhead{(deg)}                                           &
        \colhead{$\mathrm{cts\,s^{-1}}$}                          &
        \colhead{$\mathrm{cts\,s^{-1}}$}                          &
        \colhead{$\mathrm{cts\,s^{-1}}$}                          &
        \colhead{ }                                               &
        \colhead{ }                                               &
        \colhead{ }                                               &
        \colhead{ }                                               &
        \colhead{ }                                 
                }
\decimalcolnumbers
    \startdata
        \mathrm{IGR\,J17252$-$3616}	&	\mathrm{NS}	&	261.30	&	-36.28	&	0.17\pm0.03	&	<0.27	&	<0.7	&	10\pm3	&	582\pm158	&	2\pm5	&	<0.8	&	<0.11	\\
        \mathrm{GRO\,J1750$-$27}	&	\mathrm{NS}	&	267.30	&	-26.65	&	5.4\pm0.2	&	12.08\pm0.09	&	<0.03	&	9.4\pm0.5	&	0.4\pm0.01	&	<0	&	0.065\pm0.008	&	$-$	\\
        \textcolor{blue}{\mathrm{4U\,1700$-$377}}	&	\mathrm{NS}	&	255.99	&	-37.84	&	1.29\pm0.04	&	9.49\pm0.1	&	19.6\pm0.2	&	15.4\pm0.5	&	9.1\pm0.1	&	2.4\pm0.04	&	0.67\pm0.02	&	0.175\pm0.007	\\
        \mathrm{IGR\,J17586$-$2129}	&	$-$	&	269.64	&	-21.39	&	0.38\pm0.05	&	<-0.1	&	<0.3	&	161\pm6	&	<0	&	<0	&	<0.45	&	<0.034	\\
        \mathrm{SAX\,J1818.6$-$1703}	&	\mathrm{NS}	&	274.66	&	-17.05	&	0.68\pm0.06	&	<0.3	&	<0.6	&	23\pm1	&	<0	&	<0	&	<0.26	&	<0.057	\\
        \mathrm{AX\,J1820.5$-$1434}	&	$-$	&	275.12	&	-14.57	&	0.5\pm0.06	&	0.5\pm0.1	&	<0.6	&	23\pm2	&	2\pm1	&	<0	&	<0.14	&	<0.036	\\
        \mathrm{IGR\,J18256$-$1035}	&	$-$	&	276.43	&	-10.58	&	0.32\pm0.03	&	<0.18	&	<0.5	&	2.5\pm0.7	&	<0	&	<0	&	<0.8	&	<0.19	\\
        \mathrm{AX\,1845.0$-$0433}	&	\mathrm{NS}	&	281.92	&	-4.50	&	0.26\pm0.02	&	<-0.08	&	<0.2	&	0.7\pm0.7	&	<0	&	<0	&	<0.7	&	<0.11	\\
\enddata
\tablecomments{\\$^a$ and $^b$ are HR (7$-$40\,keV, 2$-$6\,keV) and HR (25$-$100\,keV, 2$-$6\,keV) calculated with Equation \ref{equ:hr}, respectively.\\
                 $^c$-$^e$ Excess variances calculated with Equation \ref{equ:excess} at 2$-$6keV, 7$-$40\,keV and 25$-$100\,keV, respectively.\\    
                 }
\end{deluxetable*}
\end{longrotatetable}

\begin{longrotatetable}
\begin{deluxetable*}{LLRRRRRRRRR}
    \scriptsize
    \setcounter{table}{9}
    \tablecaption{The First Four-year GPSS Catalog of \textit{Insight-HXMT} (SNR). \label{tab:catalog3}}
    \tablewidth{0pt}
    \tabletypesize{}
    \tablehead{
        \colhead{Source name}                                      &
        \colhead{R.A.}                                             &
        \colhead{Dec.}                                             &
        \colhead{$R_{\mathrm{2-6\,keV}}$}                          &
        \colhead{$R_{\mathrm{7-40\,keV}}$}                         &
        \colhead{$R_{\mathrm{25-100\,keV}}$}                       &
        \colhead{$F_{\mathrm{rms}}1^c$}                            &
        \colhead{$F_{\mathrm{rms}}2^d$}                            &
        \colhead{$F_{\mathrm{rms}}3^e$}                            &
        \colhead{HR1$^a$}                                          &
        \colhead{HR2$^b$}                                                                      
        \\          
        \colhead{ }                                               &
        \colhead{(deg)}                                           &
        \colhead{(deg)}                                           &
        \colhead{$\mathrm{cts\,s^{-1}}$}                          &
        \colhead{$\mathrm{cts\,s^{-1}}$}                          &
        \colhead{$\mathrm{cts\,s^{-1}}$}                          &
        \colhead{ }                                               &
        \colhead{ }                                               &
        \colhead{ }                                               &
        \colhead{ }                                               &
        \colhead{ }                                 
                }
\decimalcolnumbers
    \startdata
        \mathrm{RX\,J1713.7$-$3946}	    &	258.39	&	-39.76	&	1.12\pm0.05	&	<0.87	    &	<0.0	    &	2.8\pm0.3	    &	12\pm3	&	5\pm5	&	<0.19	&	<0.078	\\
        \mathrm{IGR\,J18135$-$1751}	    &	273.40	&	-17.83	&	0.5\pm0.1	&	0.69\pm0.08	&	<0.8	    &	65\pm4	        &	10\pm1	&	<0	&	<0.14	&	<0.5	\\
        \mathrm{SNR\,021.5$-$00.9}	    &	278.40	&	-10.56	&	0.24\pm0.03	&	<0.44	    &	<0.6	    &	0.1\pm0.8	    &	<0	    &	<0	&	<0.59	&	<0.2	\\
        \mathrm{AX\,J1850.2$-$0025}	    &	282.56	&	-0.42	&	0.18\pm0.03	&	<-0.02	    &	<0.5	    &	<0	            &	<0	    &	<0	&	<2.2	&	<0.26	\\
        \mathrm{HESS\,J1834$-$087}	    &	278.73	&	-8.73	&	0.16\pm0.03	&	<0.27	    &	<0.3	    &	<0	            &	<0	    &	<0	&	<0.66	&	<0.29	\\
        \mathrm{AX\,J1838.0$-$0655}	    &	279.51	&	-6.90	&	0.44\pm0.03	&	<0.08	    &	<0.4	    &	0.2\pm0.3	    &	<0	    &	<0	&	<0.44	&	<0.13	\\
        \mathrm{SN\,1006}	            &	225.70	&	-41.91	&	0.28\pm0.06	&	<0.9	    &	<3.5	    &	<0	            &	<0	    &	<0	&	<2.5	&	<1.6	\\
        \mathrm{0FGL\,J1923.0$+$1411}	&	290.83	&	14.16	&	0.25\pm0.03	&	<-0.7	    &	<0.4	    &	<0	            &	<0	    &	<0	&	<2.8	&	<0.44	\\
\enddata
\tablecomments{\\$^a$ and $^b$ are HR (7$-$40\,keV, 2$-$6\,keV) and HR (25$-$100\,keV, 2$-$6\,keV) calculated with Equation \ref{equ:hr}, respectively.\\
                     $^c$-$^e$ Excess variances calculated with Equation \ref{equ:excess} at 2$-$6keV, 7$-$40\,keV and 25$-$100\,keV, respectively.\\    
                     }
\end{deluxetable*}
\end{longrotatetable}
          
\begin{longrotatetable}
\begin{deluxetable*}{LLRRRRRRRRR}
    \scriptsize
    \setcounter{table}{10}
    \tablecaption{The First Four-year GPSS Catalog of \textit{Insight-HXMT} (Pulsar). \label{tab:catalog4}}
    \tablewidth{0pt}
    \tabletypesize{}
    \tablehead{
        \colhead{Source name}                                      &
        \colhead{R.A.}                                             &
        \colhead{Dec.}                                             &
        \colhead{$R_{\mathrm{2-6\,keV}}$}                          &
        \colhead{$R_{\mathrm{7-40\,keV}}$}                         &
        \colhead{$R_{\mathrm{25-100\,keV}}$}                       &
        \colhead{$F_{\mathrm{rms}}1^c$}                            &
        \colhead{$F_{\mathrm{rms}}2^d$}                            &
        \colhead{$F_{\mathrm{rms}}3^e$}                            &
        \colhead{HR1$^a$}                                          &
        \colhead{HR2$^b$}                                                                      
        \\          
        \colhead{ }                                               &
        \colhead{(deg)}                                           &
        \colhead{(deg)}                                           &
        \colhead{$\mathrm{cts\,s^{-1}}$}                          &
        \colhead{$\mathrm{cts\,s^{-1}}$}                          &
        \colhead{$\mathrm{cts\,s^{-1}}$}                          &
        \colhead{ }                                               &
        \colhead{ }                                               &
        \colhead{ }                                               &
        \colhead{ }                                               &
        \colhead{ }                                 
                }
\decimalcolnumbers
    \startdata
        \mathrm{PSR\,B1509$-$58}	&	228.48	&	-59.14	&	0.47\pm0.05	&	<0.54	&	1.4\pm0.2	&	<0	&	96\pm85	&	<0	&	<0.7	&	<0.17	\\
        \mathrm{2EGS\,J1418$-$6049}	&	215.00	&	-60.60	&	0.15\pm0.03	&	<-0.24	&	<0.5	&	<0	&	103\pm4	&	<0	&	<3.5	&	<1.7	\\
        \mathrm{MAXI\,J1409$-$619}	&	212.01	&	-61.98	&	0.18\pm0.03	&	<-0.58	&	<0.5	&	<0	&	0.1\pm0.3	&	<0	&	<2.5	&	<0.4	\\
        \mathrm{HESS\,J1614$-$518}	&	243.62	&	-51.82	&	0.55\pm0.03	&	<0.08	&	<0.4	&	12.4\pm0.8	&	5\pm2	&	<0	&	<0.3	&	<0.07	\\
        \mathrm{PSR\,J1617$-$5055}	&	244.37	&	-50.92	&	0.19\pm0.04	&	<0.15	&	<0.3	&	13\pm3	&	10\pm7	&	<0	&	<0.77	&	<0.11	\\
        \mathrm{IGR\,J17088$-$4008}	&	257.21	&	-40.14	&	0.76\pm0.05	&	<0.59	&	<0.0	&	4.0\pm0.5	&	111\pm14	&	12\pm2	&	<0.24	&	<0.03	\\
        \mathrm{GEV\,J1837$-$0610}	&	279.25	&	-6.17	&	0.11\pm0.02	&	<0.09	&	<0.4	&	<0	&	<0	&	<0	&	<0.7	&	<0.24	\\
        \mathrm{GEV\,J1907$+$0557}	&	286.92	&	5.95	&	0.18\pm0.03	&	<0.0	&	<0.6	&	2\pm1	&	<0	&	<0	&	<1.5	&	<0.18	\\
\enddata
\tablecomments{\\$^a$ and $^b$ are HR (7$-$40\,keV, 2$-$6\,keV) and HR (25$-$100\,keV, 2$-$6\,keV) calculated with Equation \ref{equ:hr}, respectively.\\
                     $^c$-$^e$ Excess variances calculated with Equation \ref{equ:excess} at 2$-$6keV, 7$-$40\,keV and 25$-$100\,keV, respectively.\\    
                     }
\end{deluxetable*}
\end{longrotatetable}

\begin{longrotatetable}
\begin{deluxetable*}{LLRRRRRRRRR}
    \scriptsize
    \setcounter{table}{11}
    \tablecaption{The First Four-year GPSS Catalog of \textit{Insight-HXMT} (Seyfert 1 Galaxy). \label{tab:catalog5}}
    \tablewidth{0pt}
    \tabletypesize{}
    \tablehead{
        \colhead{Source name}                                      &
        \colhead{R.A.}                                             &
        \colhead{Dec.}                                             &
        \colhead{$R_{\mathrm{2-6\,keV}}$}                          &
        \colhead{$R_{\mathrm{7-40\,keV}}$}                         &
        \colhead{$R_{\mathrm{25-100\,keV}}$}                       &
        \colhead{$F_{\mathrm{rms}}1^c$}                            &
        \colhead{$F_{\mathrm{rms}}2^d$}                            &
        \colhead{$F_{\mathrm{rms}}3^e$}                            &
        \colhead{HR1$^a$}                                          &
        \colhead{HR2$^b$}                                                                      
        \\          
        \colhead{ }                                               &
        \colhead{(deg)}                                           &
        \colhead{(deg)}                                           &
        \colhead{$\mathrm{cts\,s^{-1}}$}                          &
        \colhead{$\mathrm{cts\,s^{-1}}$}                          &
        \colhead{$\mathrm{cts\,s^{-1}}$}                          &
        \colhead{ }                                               &
        \colhead{ }                                               &
        \colhead{ }                                               &
        \colhead{ }                                               &
        \colhead{ }                                 
                }
\decimalcolnumbers
    \startdata
        \mathrm{IGR\,J18259$-$0706}	&	276.49	&	-7.17	&	0.21\pm0.02	&	<-0.43	&	<0.5	&	1.9\pm1.2	&	<0	&	<0	&	<2.0	&	<0.21	\\
        \mathrm{IGR\,J18538$-$0102}	&	283.45	&	-1.04	&	0.29\pm0.02	&	<-0.1	&	<0.3	&	<0	&	4\pm1	&	<0	&	<0.7	&	<0.27	\\
        \mathrm{IGR\,J16119$-$6036}	&	242.96	&	-60.63	&	1.15\pm0.04	&	<0.05	&	<1.2	&	0.28\pm0.08	&	<0	&	<0	&	<0.3	&	<0.07	\\
        \mathrm{4U\,1344$-$60}	    &	206.90	&	-60.62	&	0.17\pm0.03	&	<-0.42	&	<1.7	&	0\pm1	&	1.4\pm0.6	&	<0	&	<1.3	&	<0.58	\\
        \mathrm{IGR\,J16056$-$6110}	&	241.46	&	-61.20	&	0.81\pm0.09	&	<0.1	&	<1.1	&	0.0\pm0.2	&	<0	&	<0	&	<0.5	&	<0.22	\\
        \mathrm{IGR\,J21247$+$5058}	&	321.16	&	50.97	&	0.28\pm0.02	&	<0.2	&	1.1\pm0.1	&	<0	&	<0	&	<0	&	<1.9	&	<0.26	\\
\enddata
\tablecomments{\\$^a$ and $^b$ are HR (7$-$40\,keV, 2$-$6\,keV) and HR (25$-$100\,keV, 2$-$6\,keV) calculated with Equation \ref{equ:hr}, respectively.\\
                     $^c$-$^e$ Excess variances calculated with Equation \ref{equ:excess} at 2$-$6keV, 7$-$40\,keV and 25$-$100\,keV, respectively.\\    
                     }
\end{deluxetable*}
\end{longrotatetable}

\begin{longrotatetable}
\begin{deluxetable*}{LLLRRRRRRRRR}
    \scriptsize
    \setcounter{table}{12}
    \tablecaption{The First Four-year GPSS Catalog of \textit{Insight-HXMT} (Unclassified and others). \label{tab:catalog6}}
    \tablewidth{0pt}
    \tabletypesize{}
    \tablehead{
        \colhead{Source name}                                      &
        \colhead{Source type$^{*}$}                                 &
        \colhead{R.A.}                                             &
        \colhead{Dec.}                                             &
        \colhead{$R_{\mathrm{2-6\,keV}}$}                          &
        \colhead{$R_{\mathrm{7-40\,keV}}$}                         &
        \colhead{$R_{\mathrm{25-100\,keV}}$}                       &
        \colhead{$F_{\mathrm{rms}}1^c$}                            &
        \colhead{$F_{\mathrm{rms}}2^d$}                            &
        \colhead{$F_{\mathrm{rms}}3^e$}                            &
        \colhead{HR1$^a$}                                          &
        \colhead{HR2$^b$}                                                                      
        \\          
        \colhead{ }                                               &
        \colhead{ }                                               &
        \colhead{(deg)}                                           &
        \colhead{(deg)}                                           &
        \colhead{$\mathrm{cts\,s^{-1}}$}                          &
        \colhead{$\mathrm{cts\,s^{-1}}$}                          &
        \colhead{$\mathrm{cts\,s^{-1}}$}                          &
        \colhead{ }                                               &
        \colhead{ }                                               &
        \colhead{ }                                               &
        \colhead{ }                                               &
        \colhead{ }                                 
                }
\decimalcolnumbers
    \startdata
        \mathrm{IGR\,J17385$-$3433}	    &\mathrm{Unclassified}	&	264.63	&	-34.55	&	0.23\pm0.03	&	<-0.33	    & <0.56	    &  5\pm1	        &   1.6\pm0.5   & <0	    &	<0.9	&	<0.062	\\
        \mathrm{4PBC\,J1709.7$-$2348}	&\mathrm{Unclassified}	&	257.45	&	-23.75	&	0.38\pm0.05	&	<0.9	    & <1.4	    &  2.2\pm0.8	    &   4\pm2	    & <0	    &	<0.54	&	<20	\\
        \mathrm{IGR\,J17239$-$3143}	    &\mathrm{Unclassified}	&	260.97	&	-31.72	&	0.26\pm0.03	&	<0.03	    & <0.34	    &  2\pm1	        &   20\pm3	    & <0	    &	<0.5	&	<0.13	\\
        \mathrm{HESS\,J1804$-$216}	    &\mathrm{Unclassified}	&	271.17	&	-21.68	&	1.05\pm0.06	&	<0.32	    & <0.4	    &  9.2\pm0.6	    &   <0	        & <0	    &	<0.07	&	<0.014	\\
        \mathrm{IGR\,J18172$-$1944}	    &\mathrm{Unclassified}	&	274.31	&	-19.74	&	0.53\pm0.03	&	<-0.41	    & <0.29	    &  4.4\pm0.5	    &   0.8\pm0.5   & <0	    &	<0.27	&	<0.057	\\
        \mathrm{Swift\,J1728.9$-$3613}	&\mathrm{Unclassified}	&	262.24	&	-36.23	&	0.48\pm0.03	&	<0.23	    & <1.0	    &  25\pm1	        &   30\pm19	    & 4\pm1	    &	<0.09	&	<0.02	\\
        \mathrm{Swift\,J1729.9$-$3437}	&\mathrm{Unclassified}	&	262.54	&	-34.61	&	0.89\pm0.06	&	<0.9	    & <0.56	    &  4.7\pm0.5	    &   39\pm11	    & <0	    &	<0.23	&	<0.06	\\
        \mathrm{AX\,J1825.5$-$1144}	    &\mathrm{Unclassified}	&	276.38	&	-11.75	&	0.14\pm0.02	&	<0.07	    & <0.24	    &  0\pm2	        &   <0	        & <0	    &	<0.7	&	<0.11	\\
\enddata
\tablecomments{\\$^a$ and $^b$ are HR (7$-$40\,keV, 2$-$6\,keV) and HR (25$-$100\,keV, 2$-$6\,keV) calculated with Equation \ref{equ:hr}, respectively.
                     }
\end{deluxetable*}
\end{longrotatetable}

\startlongtable
\begin{deluxetable*}{LRRRRRR}
    \setcounter{table}{13}
    \tablecaption{Combined Results of \textit{Insight-HMXT} and MAXI. \label{tab:combine-with-maxi}}
    \tablewidth{0pt}
    \tablehead{
        \colhead{Source name} &
        \colhead{R.A.} &
        \colhead{Dec} &
        \colhead{Flux Variability$^{a}$} &
        \colhead{$\chi^2$} &
        \colhead{dof} &
        \colhead{Reduced $\chi^2$}
                }
    \startdata
        $\mathrm{IGR\,J00291+5934}$	            &	7.26	&	59.57	&	\mathrm{stable	    }&	26.3	&	49	&	0.5	\\
        $\mathrm{1ES\,0033+595}$	            &	8.97	&	59.83	&	\mathrm{stable	    }&	74.8	&	104	&	0.7	\\
        $\mathrm{GT\,Mus}$	                    &	174.87	&	-65.40	&	\mathrm{stable	    }&	72.9	&	86	&	0.8	\\
        $\mathrm{GRO\,J1008$-$57}$	            &	152.44	&	-58.29	&	\mathrm{variable	}&	69.2	&	73	&	0.9	\\
        $\mathrm{A\,0535+262}$	                &	84.73	&	26.32	&	\mathrm{variable	}&	69.7	&	81	&	0.9	\\
        $\mathrm{XTE\,J1829$-$098}$	            &	277.44	&	-9.90	&	\mathrm{variable	}&	102.0	&	114	&	0.9	\\
        $\mathrm{ClG\,0745$-$1910}$	            &	116.88	&	-19.29	&	\mathrm{stable	    }&	62.6	&	65	&	1.0	\\
        $\mathrm{Crab}$	                        &	83.63	&	22.01	&	\mathrm{stable	    }&	76.9	&	79	&	1.0	\\
\enddata
    \tablecomments{\\$^{a}$ The source is considered as a flux stable one if its $F_\mathrm{rms}-dF_\mathrm{rms}$ is less than or equal to $F_\mathrm{rms}$ of Crab at 2$-$\,4keV band. Otherwise, the source is marked as `variable' in this table.\\ 
    $^{*}$ That many flux variable sources have large reduced $\chi^2$ is understandable and described in Section \ref{sec:compMAXI}. However, in addition to Cas\,A, two flux stable sources (4U\,1323$-$619 and 4U\,1538$-$52) also have large reduced $\chi^2$. This is because some data points given by MAXI may be inaccurate. Their long-term light curves monitored by the two missions are shown in Figure \ref{fig:contrast-4U}. As can be seen in panel (b), the long-term light curve of 4U\,1538$-$52 monitored by MAXI has some data points significantly below zero, which is unreasonable. In panel (a), 4U\,1323$-$619 burst in 2019 (between MJD 58500 and 58600) in the long-term light curve monitored by MAXI. \textit{Insight-HXMT} has two observations (MJD 58533.73 and 58533.86) during this period. However, its data indicate that this source was in a quiet state. The corresponding two observational light curves are shown in Figure \ref{fig:1348}. It can be seen that the results from MAXI are inconsistent with the two observational light curves, and may be influenced by the outbursts of MAXI\,J1348$-$630.
    }
\end{deluxetable*}

\begin{figure*}
    \gridline{\fig{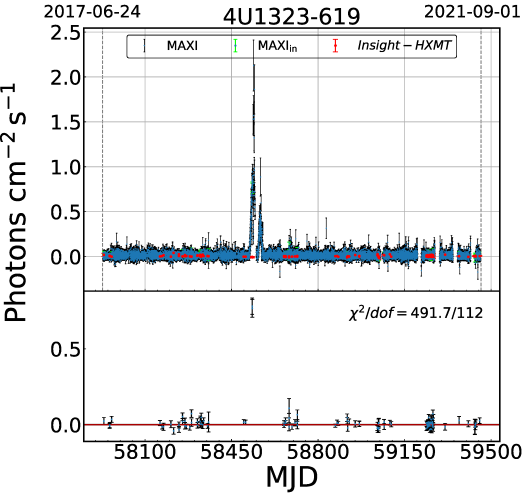}{0.45\textwidth}{(a)}\label{fig:contrast5}
              \fig{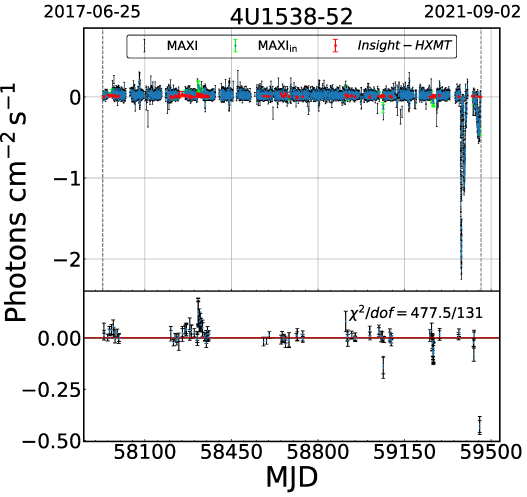}{0.45\textwidth}{(b)}\label{fig:contrast6}
              }
    \caption{The combined results of 4U\,1323$-$619 and 4U\,1538$-$52. The colores and legends are the same as Figure \ref{fig:contrast crab}.}
    \label{fig:contrast-4U}
\end{figure*}

\begin{figure*}
    \gridline{\fig{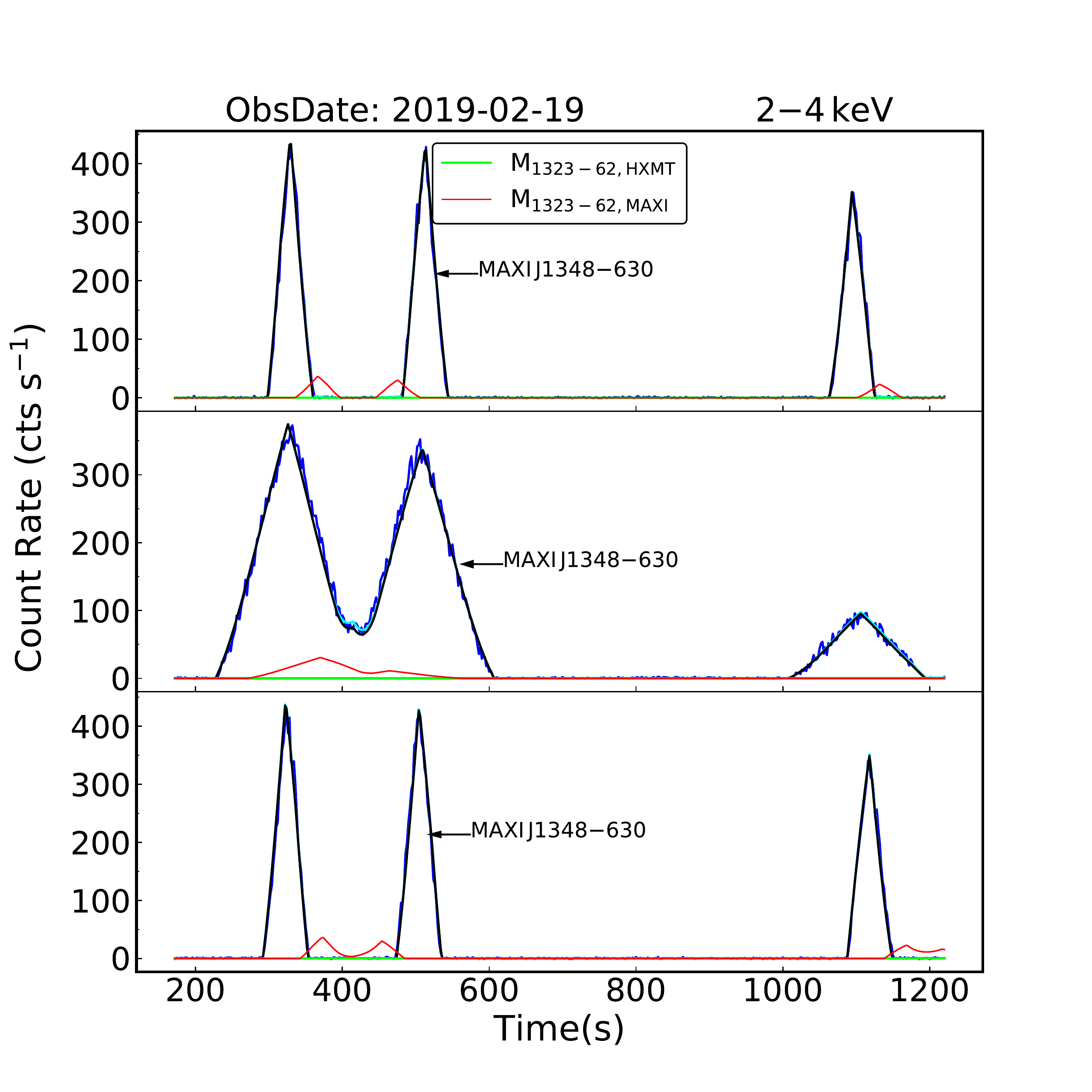}{0.46\textwidth}{(a) MJD 58533.73}\label{fig:contrast7}
              \fig{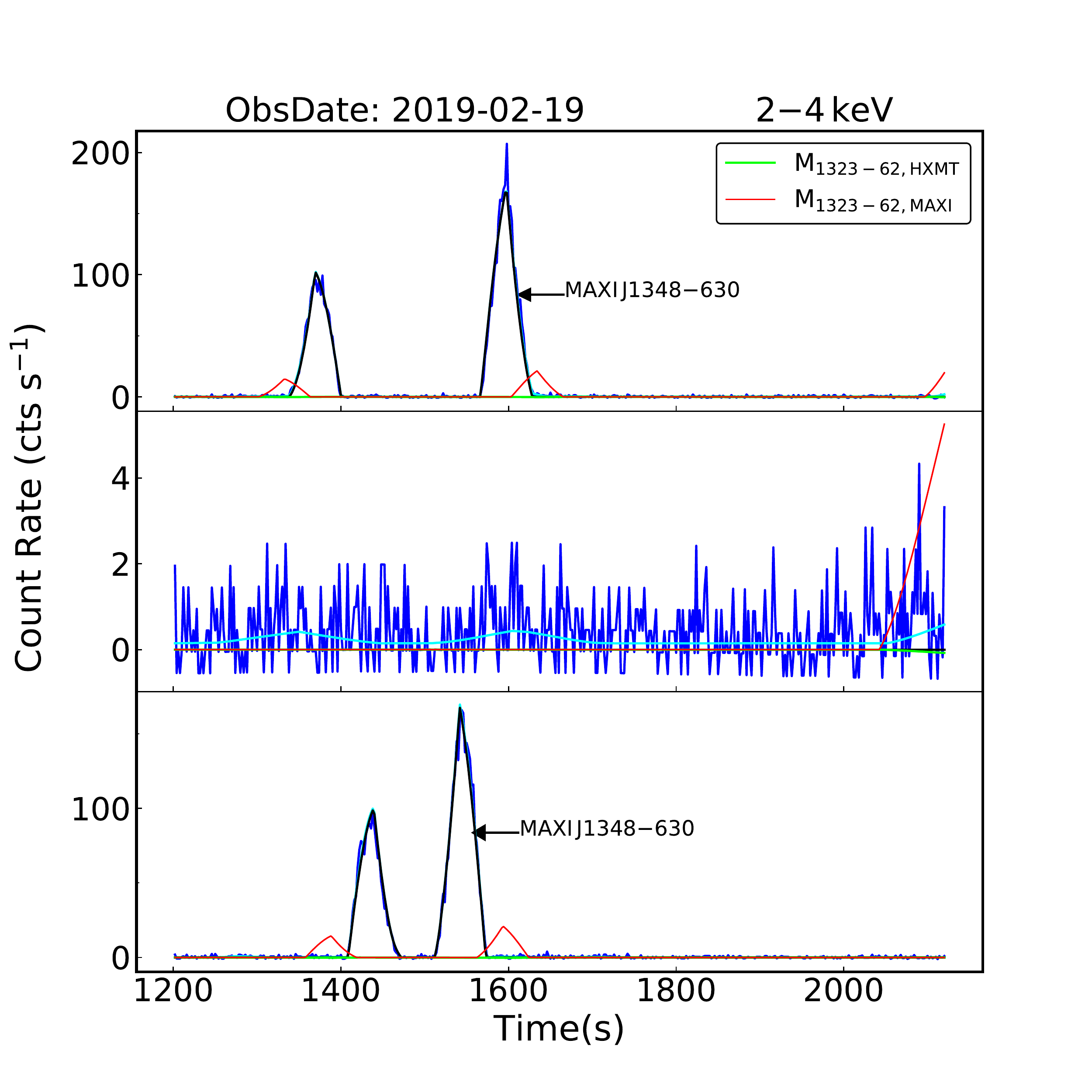}{0.46\textwidth}{(b) MJD 58533.86}\label{fig:contrast8}
              }
    \caption{Panel (a) shows the light curves crossed 4U\,1323$-$619 observed by \textit{Insight-HXMT} at MJD 58533.73, and comparison between MAXI and \textit{Insight-HXMT}. So does panel (b), while MJD is 58533.86. The blue and cyan lines are light curves and the final fitting results of \textit{Insight-HXMT}. MAXI\,1348$-$630 contributes the most peaks on the light curves, and is colored by black. The green and red lines are the comparison of the contribution by 4U\,1323$-$619 between \textit{Insight-HXMT} and MAXI.}
    \label{fig:1348}
\end{figure*}




\end{document}